\documentclass[12pt]{article}
\pdfoutput=1
\usepackage{bbm,latexsym,amsmath,amssymb}		
\usepackage[colorlinks=true,linkcolor=redLinks,citecolor=greenLinks,urlcolor=redLinks, pdfborder={0 0 1}]{hyperref} 
\usepackage{booktabs}
\usepackage{graphicx}
\usepackage{color} 

\definecolor{greenLinks}{rgb}{0, 0.6, 0} 
\definecolor{blueLinks}{rgb}{0, 0, 0.6}
\definecolor{redLinks}{rgb}{0.6, 0, 0} 
\renewcommand{\thefootnote}{\fnsymbol{footnote}}
\newcommand{\zz}{\mathbbm{Z}}
\newcommand{\mnu}{\mathcal{M}_\nu}
\newcommand{\diag}{\mbox{diag}\,}
\newcommand{\bone}{\mathbbm{1}}
\newtheorem{theorem}{Theorem}
\allowdisplaybreaks
\begin{document}
\title{
\normalsize \hfill UWThPh-2014-11 \\
\normalsize \hfill IFIC/14-32 \\*[8mm]
\LARGE Classification of lepton mixing matrices\\
from finite residual symmetries}

\author{
Renato M. Fonseca\,$^{(1)}$\thanks{E-mail: renato.fonseca@ific.uv.es} 
\setcounter{footnote}{2}
and\,
Walter Grimus\,$^{(2)}$\thanks{E-mail: walter.grimus@univie.ac.at} \
\\[5mm]
$^{(1)} \! $
\small AHEP Group, Instituto de F\'isica Corpuscular,
  C.S.I.C./Universitat de Val\`encia \\ 
\small Edificio de Institutos de Paterna, Apartado 22085, E--46071
  Val\`encia, Spain \\ 
\\[2mm]
$^{(2)} \! $
\small University of Vienna, Faculty of Physics \\
\small Boltzmanngasse 5, A--1090 Vienna, Austria
\\[2mm]
}
\date{August 19, 2014}

\maketitle

\begin{abstract}
Assuming that neutrinos are Majorana particles,
we perform a complete classification of all possible mixing matrices which are
fully determined by residual symmetries in the charged-lepton and neutrino
mass matrices. 
The classification is based on the assumption that the residual
symmetries originate from a \emph{finite} flavour symmetry group.
The mathematical tools which allow us to accomplish this
classification are theorems on sums of roots of unity. We find 
$17$ sporadic cases 
plus one infinite series of mixing matrices associated with
three-flavour mixing, 
all
of which have already been discussed in the literature. Only 
the infinite series contains 
mixing matrices which are compatible with the data at the 3~sigma
level.   
\end{abstract}
\newpage
\renewcommand{\thefootnote}{\arabic{footnote}}
\section{Introduction}
Residual symmetries in the mass matrices 
are a means to determine rows and columns
in the mixing matrix $U$ as pure numbers, independent of the values of the
masses, by assuming finiteness of the underlying flavour symmetry
group $G$---for recent reviews on group theory see for
instance~\cite{review-groups,GL-review}. 
This model-independent approach contrasts, for
instance, with the idea of texture 
zeros, where mass ratios are related to mixing angles and CP phases. In recent
years, assuming neutrinos are Majorana particles, much effort has gone into the
discussion of the lepton mixing matrices $U$ from residual symmetries. 
Though a lot of results have been
achieved~\cite{residual,toorop1,toorop2,holthausen,hagedorn,king1,lavoura}, the
present status has not been 
reached in a systematic way.\footnote{Some work has also been done
  in relation to the quark sector~\cite{holthausen1,king} and on $SO(10)$
  GUTs~\cite{lam2014}.} Thus it is by no means clear if the 
cases existing in the literature encompass all relevant mixing matrices.
In the present paper we close this gap by accomplishing a complete
classification of all possible $U$ completely determined by residual
symmetries. It turns out that indeed all relevant mixing
matrices can already be found in the literature. 

Before we delve into the advertised classification, we shortly review the idea
of residual symmetries, which also allows us to introduce our notation.
The mass Lagrangian---obtained through flavour 
symmetry breaking---has the form 
\begin{equation}\label{mm}
\mathcal{L}_\mathrm{mass} = 
-\bar \ell_L M_\ell \ell_R + 
\frac{1}{2} \nu_L^T C^{-1} \mnu \nu_L + \mbox{H.c.},
\end{equation}
where $C$ is the charge-conjugation matrix, $\ell_L$ and $\ell_R$ contain the
three left and right-handed charged-lepton fields, respectively, and
$\nu_L$ consists of the three left-handed neutrino fields.
The neutrino mass matrix $\mnu$ is symmetric, 
due to the assumed Majorana nature of the
neutrinos, but complex in general. 
Residual symmetries in the mass matrices are defined via
\begin{equation}\label{residual}
T^\dagger M_\ell M_\ell^\dagger T = M_\ell M_\ell^\dagger, \quad
S^T \mnu S = \mnu,
\end{equation}
where $T$ and $S$ are unitary matrices.
In the Standard Model, the fields $\nu_L$ and $\ell_L$ are together in
a weak doublet, and we assume that this is also true for any extension
of it.
Therefore, $T$ and $S$ are given in the same weak basis. 
The idea of residual symmetries is that any matrix $T$ and any matrix
$S$ of equation~(\ref{residual}) contributes to the complete flavour
symmetry group $G$ of some unknown extension of the Standard Model.
The group $G$ is broken in the charged-lepton sector to the group
$G_\ell$ with $T \in G_\ell$ and in the neutrino sector to 
$G_\nu$ with $S \in G_\nu$. Because of the specific form of the mass
term in the case of Majorana neutrinos, $G_\nu$ must be a subgroup of
$\zz_2 \times \zz_2 \times \zz_2$~\cite{residual}, or 
\begin{equation}
G_\nu \subseteq \zz_2 \times \zz_2,
\end{equation}
the Klein four-group, if we confine ourselves to matrices
$S$ with $\det S = 1$. 

In this paper we want to discuss all possibilities such that 
the residual symmetry groups $G_\ell$ and $G_\nu$ fully determine the
lepton mixing matrix $U$. Thus we identify $G_\nu$ with the Klein
four-group:
\begin{equation}
G_\nu = \{ \bone,\, S_1,\, S_2,\, S_3 \}
\end{equation}
with three commuting matrices $S_j$ with the properties
$S_j^2 = \bone$ and $S_j S_k = S_l$ for $j \neq k \neq l \neq j$.
As we will see in the following, in most cases $G_\ell$ is
generated by a single matrix $T$. In some cases, however, one needs
two matrices $T_1$, $T_2$ in order to fully determine $U$; in such
instances, one of these matrices alone fixes only a row in $U$.

How can the residual symmetries determine the mixing matrix $U$?
Let us denote the diagonalising matrices of 
$M_\ell M_\ell^\dagger$ and $\mnu$ 
by $U_\ell$ and $U_\nu$, respectively. Then 
\begin{equation}\label{UU}
U_\ell^\dagger M_\ell M_\ell^\dagger U_\ell = 
\diag \left( m_e^2, m_\mu^2, m_\tau^2 \right), \quad
U_\nu^T \mnu U_\nu = \diag \left( m_1, m_2, m_3 \right)
\end{equation}
and the lepton mixing matrix is given by
\begin{equation}\label{U}
U = U_\ell^\dagger U_\nu.
\end{equation}
Since 
\begin{equation}
S_j^T \mnu S_j = \mnu \quad \forall\; j=1,2,3,
\end{equation}
one can show that $U_\nu$ not only diagonalises $\mnu$, but also the matrices
$S_j$. Therefore, the requirement that $U_\nu^\dagger S_j U_\nu$ is
diagonal for $j=1,2,3$ already determines $U_\nu$ and no knowledge of 
$\mnu$ is necessary in this framework. The same argument applies to 
$M_\ell M_\ell^\dagger$ and $T$. The diagonalisation of $T$ determines
$U_\ell$, if $T$ has non-degenerate
eigenvalues. The only other
non-trivial case, differing from the previous one, is that of two
commuting $T_1$, $T_2$, which after diagonalisation have the form
\begin{equation}\label{T1T2}
\hat T_1 = \diag (\lambda'_1, \lambda_1, \lambda_1), \quad
\hat T_2 = \diag (\lambda_2, \lambda'_2, \lambda_2)
\end{equation}
or permutations thereof.
In a nutshell, in the framework of residual symmetries, 
$U_\nu$ and $U_\ell$ are determined by $G_\nu$ and
$G_\ell$, respectively, whence one gets a handle on the lepton mixing matrix
$U$.\footnote{If $U$ is only partially determined by the residual
  symmetries, then either a column or a row is fixed~\cite{lavoura}.} 

However, the framework set forth above is still too general to be
treatable. An important extra ingredient is that the group $G$ generated by
$G_\nu$ and $G_\ell$ is \textit{finite}. This is an \textit{ad hoc}
assumption, but it has the important consequence that the eigenvalues
of all $T \in G_\ell$, the eigenvalues of all products of the
generators of $G$, \textit{e.g.}\ $S_jT$, and the eigenvalues 
of all multiple products of the generators are roots of
unity.\footnote{The eigenvalues of the $S_j$ are $\pm 1$, thus they are
  trivially roots of unity.} 
By taking traces of these matrices we obtain sums of roots of unity, and
therefore known mathematical results concerning these type of sums are
applicable---for a previous application see~\cite{grimus}, which 
form the basis for our classification of all cases of $U$. 
Note that in order for the group $G$ to be finite it is necessary 
but in general not sufficient that the eigenvalues of its generators are 
roots of unity. Nevertheless, 
in the cases discussed in this paper,
it turns out that, if the group
generators and some of their products have finite order, then $G$ is
finite. 

In summary, the classification of possible mixing matrices $U$ is based on
the following premises:
\begin{enumerate}
\renewcommand{\labelenumi}{\roman{enumi}.}
\item
The Standard Model with three families of leptons 
is the low-energy model of some theory of lepton flavour.
\item
Neutrinos have Majorana nature. 
\item 
The residual group $G_\nu$ is the Klein four-group, \textit{i.e.}\ 
$G_\nu = \zz_2 \times \zz_2$.
\item
The group $G$, generated by the elements of $G_\ell$ and $G_\nu$, is
finite.
\item
The mixing matrix $U$ is completely determined by the residual symmetries.
\end{enumerate}

In the following, we will always work in a basis where the $S_j$ are diagonal:
\begin{equation}\label{SSS}
S_1 = \diag (1,-1,-1), \quad
S_2 = \diag (-1,1,-1), \quad
S_3 = \diag (-1,-1,1).
\end{equation}
In this convenient basis, $U_\nu$ is an unknown diagonal matrix of phase
factors. Therefore, in the framework of 
residual symmetries, the
Majorana phases in $U$ are indeterminate and 
the lepton mixing matrix $U$ is simply given by
\begin{equation}\label{U1}
U = U_\ell^\dagger
\end{equation}
up to rephasings.
Moreover, since in our framework the order of the charged-lepton and
neutrino masses is undefined, it is only possible to determine 
$U$ up to independent row and column permutations.

The paper is organized as follows. 
In section~\ref{tools} we discuss some mathematical results. These are  
used later, 
in section~\ref{basic forms}, to
determine the possible forms of $|T|$ which is defined as the matrix
of absolute values of $T$:
\begin{equation}
|T|_{ij} = | T_{ij} | \quad \forall\; i,j = 1,2,3.
\end{equation}
It turns out that there are only five such basic forms
(later on reduced to three by additional considerations), 
modulo independent permutations of rows and columns. 
Then, in section~\ref{equivalent}, we perform a general discussion of
equivalent forms of $T$, which are those forms which lead to trivial
variations of the mixing matrix $U$. 
In particular, we investigate the freedom of permutations
and argue that, without loss of generality, we can confine ourselves to
matrices $T$ which can be written as
\begin{equation}\label{T}
T = \widetilde T \hat\kappa.
\end{equation}
In this formula, $\widetilde T$ contains the ``internal'' or CKM-type
phase of a unitary matrix and $\hat\kappa$ is a diagonal matrix of
phase factors. We identify in section~\ref{permutations} the 
genuinely different forms of $|T|$ which emerge from the five basic forms by
permutations. It remains to investigate the phases of $T$. 
We first determine in section~\ref{internal} the ``internal'' phase of
$T$ for each form of $|T|$. 
In section~\ref{not finite} we demonstrate that forms~1 and~4 do not lead
to finite groups. The extensive  
section~\ref{external} is devoted to the computation of $\hat\kappa$
or ``external'' phases in $T$ for the basics forms~2, 3 and~5. 
At this point we have completely determined $T$ up to equivalent forms, 
so in this section we also provide our main result, 
the possible cases of $U$.
In section~\ref{combining} we discuss the remaining solutions where $G_\ell$
has two generators, each with a twofold degenerate eigenvalue---see
equation~(\ref{T1T2}). 
We conclude with section~\ref{concl}. Lengthy calculational details
are deferred to appendices.

\section{Mathematical tools}
\label{tools}
\paragraph{Vanishing sums of roots of unity:}
We will make use of three theorems related to roots of unity.
The first theorem concerns vanishing sums of roots of unity. 
Some remarks are appropriate
before we reproduce the theorem of Conway
and Jones (theorem~6 in~\cite{conway}).
Formal sums of roots of unity with rational coefficients form
a ring~\cite{conway}. 
A sum of roots of unity $\mathcal{S}'$ is \emph{similar} to 
a sum of roots of unity $\mathcal{S}$
if there is a rational number $q$ and a root of unity
$\delta$ such that $\mathcal{S}' = q \delta \mathcal{S}$.
The length of a sum of roots of unity $\mathcal{S}$ is the number of
distinct roots involved. 
Note, however, that $-\alpha = (-1) \alpha$
for any root of unity $\alpha$, \textit{i.e.}\
$-\alpha$ does \emph{not} count separately for
the length of $\mathcal{S}$;
in other words, $\alpha + (-\alpha)$ has length zero.
The roots occurring in the following theorem are 
\begin{equation}\label{obg}
\omega = e^{2\pi i/3}, \quad 
\beta = e^{2\pi i/5}, \quad
\gamma = e^{2\pi i/7}. 
\end{equation}
Note that
\begin{equation}
\omega = -\frac{1}{2} + i\frac{\sqrt{3}}{2},
\quad
\beta = \frac{\sqrt{5}-1}{4} + i \sqrt{\frac{5 + \sqrt{5}}{8}},
\end{equation}
while $\gamma$ expressed in radicals is too complicated to be shown here.
\begin{theorem}[Conway and Jones]\label{TCJ}
Let $\mathcal{S}$ be a non-empty vanishing sum of length at most~9. 
Then either $\mathcal{S}$ involves $\theta$, $\theta\omega$ and 
$\theta\omega^2$ for some root $\theta$, or $\mathcal{S}$ is similar
to one of 
\begin{enumerate}
\renewcommand{\labelenumi}{\alph{enumi})}
\item
$1 + \beta + \beta^2 +\beta^3 +\beta^4$,
\item 
$- \omega - \omega^2 + \beta + \beta^2 +\beta^3 +\beta^4$, 
\item
$1 + \gamma + \gamma^2 +\gamma^3 +\gamma^4 +\gamma^5 +\gamma^6$, 
\item
$1 + \beta + \beta^4 - (\omega + \omega^2)(\beta^2 + \beta^3)$,
\item
$- \omega - \omega^2 + \gamma + \gamma^2 +\gamma^3 +\gamma^4
  + \gamma^5 + \gamma^6$,
\item 
$\beta + \beta^4 - (\omega + \omega^2)(1 + \beta^2 + \beta^3)$,
\item
$1 + \gamma^2 +\gamma^3 +\gamma^4 +\gamma^5 -( \omega +
  \omega^2)(\gamma + \gamma^6)$,
\item 
$1 - (\omega + \omega^2)(\beta + \beta^2 +\beta^3 +\beta^4)$.
\end{enumerate}
\end{theorem}

\paragraph{Non-vanishing sums of roots of unity:}
The second theorem concerns sums of roots of unity with values on the unit
circle in the complex plane.
\begin{theorem}\label{sum=1}
Let $\zeta$ be an $n$-th root of unity, i.e.\ $\zeta = e^{2\pi i/n}$, and 
let $\mathcal{S} = \sum_{k=0}^{n-1} a_k \zeta^k$ be a sum with integer
coefficients $a_k$. If $|\mathcal{S}| = 1$, then $\mathcal{S}$ is
itself a root of unity. 
\end{theorem}
A proof of this theorem can for instance be deduced from lemma~1.6
in~\cite{washington} and a discussion of this issue can be found 
on~\cite{speyer}. 

The link between theorem~\ref{sum=1} and the problems in the present paper 
is provided by the following theorem. 
\begin{theorem}\label{Troots}
Let $G$ be a finite group with $T \in G_\ell$ and let $c$ be one of the numbers
$1/2$, $(\sqrt{5} + 1)/4$ or $(\sqrt{5} - 1)/4$. 
Moreover, $T_{jj}$ is a diagonal element and $T_{kl}$, $T_{lk}$ 
are off-diagonal elements of $T$. Then the following holds:
\begin{equation}
|T_{jj}| = c \; \Rightarrow \; T_{jj} = c \xi,
\quad
|T_{kl}T_{lk}| = \frac{1}{4} \; \Rightarrow \; T_{kl}T_{lk} = \frac{\xi'}{4}
\end{equation}
with roots of unity $\xi$, $\xi'$.
\end{theorem}
\textbf{Proof:}
It is easy to show that
\begin{equation}
\mathrm{Tr}\, (TS_j) + \mathrm{Tr}\, T = 2T_{jj}.
\end{equation}
Since $T,\, TS_j \in G$ and $G$ is finite, the eigenvalues of
$T$ and $TS_j$ must be roots of unity. Therefore, the traces of these
matrices are sums over three roots of unity and $2T_{jj}$ is a sum
over six roots of unity. Thus, if $2|T_{jj}| = 1$, 
theorem~\ref{sum=1} tells us that $2T_{jj}$ is a root of unity.
Now suppose that 
$2|T_{jj}| = (\sqrt{5} + 1)/2$ or $(\sqrt{5} - 1)/2$. Since
\begin{equation}
\left( \frac{\sqrt{5} + 1}{2} \right)^{-1} = \beta + \beta^4, 
\quad \mbox{and} \quad
\left( \frac{\sqrt{5} - 1}{2} \right)^{-1} = -\beta^2 - \beta^3
\end{equation}
we find that both
\begin{equation}
2T_{jj} \left( \frac{\sqrt{5} + 1}{2} \right)^{-1}
\quad \mbox{and} \quad
2T_{jj} \left( \frac{\sqrt{5} - 1}{2} \right)^{-1}
\end{equation}
are normalized sums over roots of unity. Again, theorem~\ref{sum=1}
applies. This finishes the proof of the first part of the theorem.
For the second part we note that, since $T$ is a unitary matrix,
\begin{equation}
\left( T^{-1} \right)_{jj} = \frac{1}{\det T} 
\left( T_{kk}T_{ll} - T_{kl}T_{lk} \right) = 
\left( T_{jj} \right)^*
\end{equation}
for $j \neq k \neq l \neq j$.
We know that $\det T$ is a product of three roots of unity and that 
$2T_{kk}$, $2T_{ll}$ are sums over roots of unity. Therefore, 
$4T_{kl}T_{lk}$ is a sum over roots of unity as well. Now we apply
once more theorem~\ref{sum=1} and the proof is finished. $\Box$

\paragraph{Root of unity or not, that is the question:}
The next theorem addresses the 
problem of finding out whether 
a complex number $\zeta$ with $|\zeta| = 1$ is a root of unity or not. 
We can answer this question if we know a polynomial $P(x)$ with rational
coefficients such that $P(\zeta) = 0$. Because then from $P(x)$
we can determine the minimal polynomial $m_\zeta(x)$ of $\zeta$, 
which is defined as being irreducible over the rational numbers and
normalized.\footnote{The means that the coefficient of 
   its highest power is~1.}
Since the minimal polynomial is unique the following statement holds.
\begin{theorem}\label{roots}
Let $\zeta$ be a complex number with $|\zeta| = 1$ and $m_\zeta(x)$ 
its minimal polynomial. 
Then $\zeta$ is a root of unity if and only if $m_\zeta(x)$ is a cyclotomic
polynomial. 
\end{theorem}
Note that cyclotomic polynomials are the minimal polynomials of roots of
unity. There is a straightforward corollary to 
theorem~\ref{roots} which makes use of the fact that cyclotomic
polynomials have integer coefficients.
\begin{theorem}\label{roots1}
If there are non-integer coefficients in $m_\zeta(x)$, then $\zeta$ is not a
root of unity.
\end{theorem}

\section{The basic forms of $|T|$}
\label{basic forms}
In this section we determine the basic forms of $|T|$ which follow
from our assumptions listed in the introduction. 

We begin by considering the matrices
\begin{equation}
Y^{(ij)} = T^\dagger S_i T S_j.
\end{equation}
We argue that these matrices 
have eigenvalues 1, $\lambda^{(ij)}$, $(\lambda^{(ij)})^*$. 
\\
\textbf{Proof:} The matrices $Y^{(ij)}$ fulfill
\begin{equation}
\det Y^{(ij)} = 1, \quad
S_j^{-1} Y^{(ij)} S_j  = \left( Y^{(ij)} \right)^\dagger.
\end{equation}
Therefore, the complex conjugate of every eigenvalue of $Y^{(ij)}$
is also an eigenvalue. Moreover, the product of the three
eigenvalues must be one. So the spectrum of $Y^{(ij)}$ contains one 
complex eigenvalue $\lambda^{(ij)}$, its complex conjugate and~1. $\Box$ 
\\[1mm]
Since 
\begin{equation}
\sum_{j=1}^3 S_j = -\bone,
\end{equation}
it follows that
\begin{equation}
\sum_{k=1}^3 \mbox{Tr}\,Y^{(kj)} = 1
\quad \mbox{and} \quad
\sum_{k=1}^3 \mbox{Tr}\,Y^{(ik)} = 1.
\end{equation}
Written in terms of the eigenvalues, these relations are
\begin{equation}\label{lambda}
\sum_{k=1}^3 \left( \lambda^{(kj)} + {\lambda^{(kj)}}^* \right) + 2 = 0
\quad \mbox{and} \quad
\sum_{k=1}^3 \left( \lambda^{(ik)} + {\lambda^{(ik)}}^* \right) + 2 = 0
\end{equation}
for $i,j = 1,2,3$.
These equations can be tackled with theorem~\ref{TCJ}.

Suppose we have obtained the possible $\lambda^{(ij)}$.
Defining 
\begin{equation}
t_{kl} \equiv \left| T_{kl} \right|^2, 
\end{equation}
we observe that
the unitarity of $T$ gives
\begin{equation}
\sum_{k,l} t_{kl} = 3.
\end{equation}
Furthermore, we obtain
\begin{eqnarray}
\mbox{Tr}\, Y^{(ij)} &=& 1 + \lambda^{(ij)} + (\lambda^{(ij)})^*
\nonumber \\
&=& \sum_{k,l} t_{kl} \left( S_i \right)_{kk} \left( S_j \right)_{ll}
\nonumber \\
&=& \sum_{k \neq i} \sum_{l \neq j} t_{kl} + t_{ij} - 
\sum_{k \neq i} t_{kj} - \sum_{l \neq j} t_{il} =
\nonumber \\
&=& 3 - 2\sum_{k \neq i} t_{kj} - 2\sum_{l \neq j} t_{il}
\nonumber \\
&=& -1 + 4 t_{ij}.
\end{eqnarray}
In this way we arrive at the relation 
\begin{equation}\label{t-lambda}
\left| T_{ij} \right|^2 = \frac{1}{2} \left( 1 + \mbox{Re}\,
\lambda^{(ij)} \right).
\end{equation}

Thus 
in order to obtain $\left| T_{ij} \right|$ we are left with the  
task of finding all possible solutions of the generic equation
\begin{equation}\label{lambda2}
\sum_{k=1}^3 \left( \lambda_k + \lambda_k^* \right) + 2 = 0,
\end{equation}
with roots of unity $\lambda_1$, $\lambda_2$, $\lambda_3$.
These solutions are derived in
appendix~\ref{possible eigenvalues-app} through application of
theorem~\ref{TCJ}. 
It turns out that, up to permutations and complex conjugations,  
equation~(\ref{lambda2}) has only three solutions:
\begin{equation}\label{ABC}
\left( \lambda_1,\, \lambda_2,\, \lambda_3 \right) = 
\left\{
\begin{array}{cc}
\left( i,\, \omega,\, \omega \right) & \mbox{(A)}, \\
\left( \omega,\, \beta,\, \beta^2 \right) & \mbox{(B)}, \\
\left( -1,\, \lambda,\, -\lambda \right) & \mbox{(C)},
\end{array} \right.
\end{equation}
where $\lambda = e^{i\vartheta}$ is an arbitrary root of unity.

The details of the tedious manipulations for finding all possible forms of
the matrix $t$ are given in appendix~\ref{basic forms-app}.
Here we mention only that it is important to take into account that
$t$ stems from a unitary matrix---see inequality~(\ref{ineq}), which
rules out many cases. The surprising
result of appendix~\ref{basic forms-app} is that, up to permutations
of rows and columns, there are only five basic forms of $t$. 
In the following, however, we will rather use $|T|$ instead of $t$. Therefore,
we display here the basic forms of $|T|$ 
obtained by inserting the 
solutions~(\ref{ABC}) into equation~(\ref{t-lambda}): 
\begin{subequations}\label{basic}
\begin{eqnarray}
\mbox{Form 1:} &&
\left|T\right| = \left(\begin{array}{ccc}
\frac{1}{\sqrt{2}} & \frac{1}{2} & \frac{1}{2}\\
\frac{1}{2} & \frac{1}{\sqrt{2}} & \frac{1}{2}\\
\frac{1}{2} & \frac{1}{2} & \frac{1}{\sqrt{2}}
\end{array}\right),
\label{basic1}
\\
\mbox{Form 2:} &&
\left|T\right| = \left(\begin{array}{ccc}
0 & \frac{1}{\sqrt{2}} & \frac{1}{\sqrt{2}}\\
\frac{1}{\sqrt{2}} & \frac{1}{2} & \frac{1}{2}\\
\frac{1}{\sqrt{2}} & \frac{1}{2} & \frac{1}{2}
\end{array}\right),
\label{basic2}
\\
\mbox{Form 3:} &&
\left|T\right| = \left(\begin{array}{ccc}
\frac{1}{2} & \frac{\sqrt{5}-1}{4} & \frac{\sqrt{5}+1}{4}\\
\frac{\sqrt{5}+1}{4} & \frac{1}{2} & \frac{\sqrt{5}-1}{4}\\
\frac{\sqrt{5}-1}{4} & \frac{\sqrt{5}+1}{4} & \frac{1}{2}
\end{array}\right),
\label{basic3}
\\
\mbox{Form 4:} && 
\left|T\right| = \left(\begin{array}{ccc}
\frac{1}{\sqrt{2}} & \frac{1}{2} & \frac{1}{2}\\
\frac{1}{2} & \frac{\sqrt{5}-1}{4} & \frac{\sqrt{5}+1}{4}\\
\frac{1}{2} & \frac{\sqrt{5}+1}{4} & \frac{\sqrt{5}-1}{4}
\end{array}\right),
\label{basic4}
\\
\mbox{Form 5:} && 
\left|T\right| = \left(\begin{array}{ccc}
1 & 0 & 0\\
0 & \cos\theta & \sin\theta\\
0 & \sin\theta & \cos\theta
\end{array}\right).
\label{basic5}
\end{eqnarray}
\end{subequations}
Note that form~5 derives from equation~(\ref{basic5-var}) via
$\sin^2\theta = (1-\cos\vartheta)/2$ and, therefore, 
$\theta = \pm \vartheta/2 + k\pi$ 
with an arbitrary integer
$k$. Since we know that $\vartheta$ is a rational angle,
\textit{i.e.}\ a rational multiple of $\pi$, it follows that $\theta$
must be a rational angle too.

\section{Equivalent forms of $T$}
\label{equivalent}
It is expedient to reflect on the ambiguities in the determination of
$U$ with the method used here. 

The first observation is that the choice of basis for the $S_j$ in
equation~(\ref{SSS}) leaves still some freedom for 
basis transformations. 
The reason is that
\begin{equation}\label{V}
S_j \to V^\dagger S_j V
\quad \mbox{with} \quad
V = P \hat \sigma,
\end{equation}
where $P$ is a $3 \times 3$ permutation matrix and $\hat \sigma$ is a diagonal
matrix of phase factors, may transform one $S_j$ into another, but the set of
matrices of equation~(\ref{SSS}) is invariant under this similarity
transformation. On the other hand, the transformation~(\ref{V}) acts also on
$T$ as 
\begin{equation}\label{TV}
T \to V^\dagger T V,
\end{equation}
and we can use this freedom to fix some conventions for $T$. This is an
important issue because it prevents us from over-counting the number of
cases and from studying equivalent cases twice. 
Two matrices 
$T$, $T'$, 
which are connected via 
$T' = V^\dagger T V$, 
are called equivalent in the following. 

The freedom in reordering and rephasing expressed by equation~(\ref{TV}) can
be used for a strategy to determine all inequivalent forms of $T$.
\begin{enumerate}
\item
For each of the five basic forms of equation~(\ref{basic}), we take the matrix
displayed there as the starting point and indicate it by the subscript $A$.
\item
Due to equation~(\ref{TV}) we are allowed to confine ourselves to permutations
from the right in order to find inequivalent matrices 
$|T|_I$ with $I = B,C,\ldots$ 
\item                               
In general there will be less than 
six inequivalent matrices
$|T|$ for each of the basics forms, as 
some matrices which emerge from each other by a permutation from the
right might still be equivalent due to the 
equality of some matrix elements or, in the case
of form~5, because it is possible to make the exchange 
$\cos\theta \leftrightarrow \sin\theta$.
\item
After having found, for each basic form, the inequivalent matrices
$|T|_I = |T|_A P_I$, where $P_I$ is a permutation matrix, we determine the
internal phase associated with each $|T|_I$. Denoting the resulting matrix by 
$\widetilde T_I$, we can choose phase conventions such that 
$\widetilde T_I = \widetilde T_A P_I$ for all $I$.
\item
The matrix $T$ will also have external phases. Applying again 
equation~(\ref{TV}), we can assume that these phases are taken care of by
multiplying $\widetilde T$ by a diagonal matrix of phase factors 
$\hat\kappa$ from the right. In essence, the $\hat\kappa$ will be determined
by the requirement that $T$ has finite order. 
Note, however, that,
given a basic form and one of its possible 
$\widetilde T_I$, 
there can be several solutions of
$\hat\kappa$. 
\item
Finally, since the matrices $S_j$ are real, for every solution $T$ there is
the complex conjugate solution $T^*$.
\end{enumerate}

Eventually we are not interested in the possible $T$ but in the possible mixing
matrices $U$. According to equation~(\ref{U1}) the two matrices are linked via
\begin{equation}\label{eigenvaluesT}
T = U^\dagger \hat T U 
\quad \mbox{with} \quad
\hat T = \diag \left( \lambda^{(0)}_1, \lambda^{(0)}_2, \lambda^{(0)}_3 \right).
\end{equation}
The order of the eigenvalues is indeterminate and $\hat T$ is invariant under
phase transformations. Therefore, $U$ can undergo
rephasing and permutations from the left. Equation~(\ref{TV}) implies that the
same holds from the right, with rephasings and permutations independent from
those on the left. To this indeterminacy one has to add the possibility of
complex conjugation of $U$. Given these trivial phase variations of the 
mixing matrix and the need to factor them out, we will be focusing on $|U|^2$.

The group $G$, determined by the three $S_j$ and by one or two
matrices $T$ which generate $G_\ell$, is not changed by
the above manipulations of $T$. 
Actually, what we have at hand is not directly the group $G$ but its
representation on the three leptonic gauge doublets. In this sense, the change
$T \to T^*$ for the $T \in G_\ell$ corresponds to switching from one
representation to its complex conjugate. We can also remove an overall phase
factor $\xi$ from a $T$, if $\xi$ is root of unity. This will, in general,
change the group, but the group will remain finite.

\section{Permutations of the basic forms}
\label{permutations}
In order to investigate the inequivalent permutations of the
columns of $|T|_A$, it is appropriate to use the representation of the
permutations $p \in S_3$ as permutation matrices: 
\begin{equation}
p \to M(p) = \left( e_{p\left(1\right)}, e_{p\left(2\right)}, e_{p\left(3\right)} \right)
\quad \mbox{with} \quad
e_1 = \left( \begin{array}{c} 1 \\ 0 \\ 0 
\end{array} \right),
\;
e_2 = \left( \begin{array}{c} 0 \\ 1 \\ 0 
\end{array} \right),
\;
e_3 = \left( \begin{array}{c} 0 \\ 0 \\ 1 
\end{array} \right).
\end{equation}
It is easy to check that the representation property 
$M(pp') = M(p) M(p')$ is fulfilled for any two $p,\,p' \in S_3$. 
The basic relation we need is given by
\begin{equation}
M(q)^T |T|_A M(q) = |T|_A \quad \Rightarrow \quad
M(q)^T \Big( |T|_A M(p) \Big) M(q) = 
|T|_A \Big( M(q)^T M(p) M(q) \Big).
\end{equation}
Therefore, invariance of $|T|_A$ under some $q \in S_3$ means that 
$|T|_A M(p_1)$ and $|T|_A M(p_2)$ are equivalent provided that
permutations $p_1$, $p_2$ are related through conjugation by
$q$, \textit{i.e.}\ $p_2 = q^{-1} p_1 q$. 
Now we discuss all five basic forms under this aspect.

It is easy to see that $|T|_A$ of form~1, equation~(\ref{basic1}), is
invariant under \emph{all} $q \in S_3$. Since $S_3$ has three
conjugacy classes, there are three inequivalent forms. 
Apart from $|T|_A$, we choose 
$|T|_B \equiv |T|_A M \left[(23)\right]$ 
and 
$|T|_C \equiv |T|_A M \left[(132)\right]$.

Turning to form~2, equation~(\ref{basic2}), $|T|_A$ is obviously
invariant under $q = (23)$. Given that
$\left|T\right|_{A} M \left[\left(12\right)\right] = 
\left|T\right|_{A} M \left[\left(132\right)\right]$, 
$q^{-1}(12)q = (13)$, and $q^{-1}(132)q = (123)$, 
we can conclude that the only other inequivalent
form is $|T|_B \equiv |T|_A M \left[\left(12\right)\right]$.

Concerning form~3, equation~(\ref{basic3}),
we find that it is invariant under cyclic permutations. 
Since one transposition is transformed into the other two by
conjugation with the two cyclic permutations, 
we can for instance choose 
$|T|_B \equiv |T|_A M \left[(23)\right]$,
as representative of the transpositions. 
Finally, cyclic permutations give the 
inequivalent forms 
$|T|_C \equiv |T|_A M \left[(132)\right]$ 
and
$|T|_D \equiv |T|_A M \left[(123)\right]$.

The matrix $|T|_A$ of form~4, equation~(\ref{basic4}), is invariant
under $q = (23)$. In contrast to form~2, 
all columns are different. Given that $q^{-1} (12) q = (13)$ and
$q^{-1} (123) q = (132)$, we are lead to 
a choice of inequivalent forms
$|T|_B \equiv |T|_A M \left[(23)\right]$, 
$|T|_C \equiv |T|_A M \left[(12)\right]$
and 
$|T|_D \equiv |T|_A M \left[(132)\right]$.

Since in form~5, equation~(\ref{basic5}), the angle $\theta$ is free
apart from being rational, 
the exchange $\cos\theta \leftrightarrow \sin\theta$ relates
equivalent forms. So in effect we are lead back to the argumentation
applied to form~2. For definiteness, we choose 
$|T|_B \equiv |T|_A M \left[(123)\right]$.

In summary, the inequivalent forms obtained by the above 
discussion are given by
\begin{subequations}
\label{basic5-permutations}
\begin{alignat}{3}
&\mbox{Form 1:} 
&\quad
\left|T\right|_A &=
\left(\begin{array}{ccc}
\frac{1}{\sqrt{2}} & \frac{1}{2} & \frac{1}{2}\\
\frac{1}{2} & \frac{1}{\sqrt{2}} & \frac{1}{2}\\
\frac{1}{2} & \frac{1}{2} & \frac{1}{\sqrt{2}}
\end{array}\right),
&\quad
\left|T\right|_B &=
\left(\begin{array}{ccc}
\frac{1}{\sqrt{2}} & \frac{1}{2} & \frac{1}{2}\\
\frac{1}{2} & \frac{1}{2} & \frac{1}{\sqrt{2}}\\
\frac{1}{2} & \frac{1}{\sqrt{2}} & \frac{1}{2}
\end{array}\right),
\\
&
&\quad
\left|T\right|_C &=
\left(\begin{array}{ccc}
\frac{1}{2} & \frac{1}{\sqrt{2}} & \frac{1}{2}\\
\frac{1}{2} & \frac{1}{2} & \frac{1}{\sqrt{2}}\\
\frac{1}{\sqrt{2}} & \frac{1}{2} & \frac{1}{2}
\end{array}\right),
\nonumber
\\
&\mbox{Form 2:} 
&\quad
\left|T\right|_A &=
\left(\begin{array}{ccc}
0 & \frac{1}{\sqrt{2}} & \frac{1}{\sqrt{2}}\\
\frac{1}{\sqrt{2}} & \frac{1}{2} & \frac{1}{2}\\
\frac{1}{\sqrt{2}} & \frac{1}{2} & \frac{1}{2}
\end{array}\right),
&\quad
\left|T\right|_B &=
\left(\begin{array}{ccc}
\frac{1}{\sqrt{2}} & 0 & \frac{1}{\sqrt{2}}\\
\frac{1}{2} & \frac{1}{\sqrt{2}} & \frac{1}{2}\\
\frac{1}{2} & \frac{1}{\sqrt{2}} & \frac{1}{2}
\end{array}\right),
\label{form2AB}
\\
&\mbox{Form 3:}
&\quad
\left|T\right|_A &=
\left(\begin{array}{ccc}
\frac{1}{2} & \frac{\sqrt{5}-1}{4} & \frac{\sqrt{5}+1}{4}\\
\frac{\sqrt{5}+1}{4} & \frac{1}{2} & \frac{\sqrt{5}-1}{4}\\
\frac{\sqrt{5}-1}{4} & \frac{\sqrt{5}+1}{4} & \frac{1}{2}
\end{array}\right),
&\quad
\left|T\right|_B &=
\left(\begin{array}{ccc}
\frac{1}{2} & \frac{\sqrt{5}+1}{4} & \frac{\sqrt{5}-1}{4}\\
\frac{\sqrt{5}+1}{4} & \frac{\sqrt{5}-1}{4} & \frac{1}{2}\\
\frac{\sqrt{5}-1}{4} & \frac{1}{2} & \frac{\sqrt{5}+1}{4}
\end{array}\right),
\label{form3ABCD}
\\
&\hphantom{\mbox{Form 3:}}
&\quad
\left|T\right|_C &=
\left(\begin{array}{ccc}
\frac{\sqrt{5}+1}{4} & \frac{1}{2} & \frac{\sqrt{5}-1}{4}\\
\frac{\sqrt{5}-1}{4} & \frac{\sqrt{5}+1}{4} & \frac{1}{2}\\
\frac{1}{2} & \frac{\sqrt{5}-1}{4} & \frac{\sqrt{5}+1}{4}
\end{array}\right),
&\quad
\left|T\right|_D &=
\left(\begin{array}{ccc}
\frac{\sqrt{5}-1}{4} & \frac{\sqrt{5}+1}{4} & \frac{1}{2}\\
\frac{1}{2} & \frac{\sqrt{5}-1}{4} & \frac{\sqrt{5}+1}{4}\\
\frac{\sqrt{5}+1}{4} & \frac{1}{2} & \frac{\sqrt{5}-1}{4}
\end{array}\right),
\nonumber\\
&\mbox{Form 4:}
&\quad
\left|T\right|_A &=
\left(\begin{array}{ccc}
\frac{1}{\sqrt{2}} & \frac{1}{2} & \frac{1}{2}\\
\frac{1}{2} & \frac{\sqrt{5}-1}{4} & \frac{\sqrt{5}+1}{4}\\
\frac{1}{2} & \frac{\sqrt{5}+1}{4} & \frac{\sqrt{5}-1}{4}
\end{array}\right),
&\quad
\left|T\right|_B &=
\left(\begin{array}{ccc}
\frac{1}{\sqrt{2}} & \frac{1}{2} & \frac{1}{2}\\
\frac{1}{2} & \frac{\sqrt{5}+1}{4} & \frac{\sqrt{5}-1}{4}\\
\frac{1}{2} & \frac{\sqrt{5}-1}{4} & \frac{\sqrt{5}+1}{4}
\end{array}\right),
\\
&\hphantom{\mbox{Form 4:}}
&\quad
\left|T\right|_C &=
\left(\begin{array}{ccc}
\frac{1}{2} & \frac{1}{\sqrt{2}} & \frac{1}{2}\\
\frac{\sqrt{5}-1}{4} & \frac{1}{2} & \frac{\sqrt{5}+1}{4}\\
\frac{\sqrt{5}+1}{4} & \frac{1}{2} & \frac{\sqrt{5}-1}{4}
\end{array}\right),
&\quad
\left|T\right|_D &=
\left(\begin{array}{ccc}
\frac{1}{2} & \frac{1}{\sqrt{2}} & \frac{1}{2}\\
\frac{\sqrt{5}+1}{4} & \frac{1}{2} & \frac{\sqrt{5}-1}{4}\\
\frac{\sqrt{5}-1}{4} & \frac{1}{2} & \frac{\sqrt{5}+1}{4}
\end{array}\right),
\nonumber\\
&\mbox{Form 5:}
&\quad
\left|T\right|_A &=
\left(\begin{array}{ccc}
1 & 0 & 0\\
0 & \cos\theta & \sin\theta\\
0 & \sin\theta & \cos\theta
\end{array}\right),
&\quad
\left|T\right|_B &=
\left(\begin{array}{ccc}
0 & 0 & 1\\
\cos\theta & \sin\theta & 0\\
\sin\theta & \cos\theta & 0
\end{array}\right).
\label{form5AB}
\end{alignat}
\end{subequations}

\section{The internal phase of $T$}
\label{internal}
In order to compute the internal phase of $T$, we can use the formulas
provided in~\cite{branco}:
\begin{eqnarray}
R &=& \mathrm{Re} \left( T_{11} T_{22} T_{12}^* T_{21}^* \right) = 
\frac{1}{2} \left( 1 - t_{11} - t_{22} - t_{12} - t_{21} +
t_{11}t_{22} + t_{12}t_{21} \right), 
\label{R}\\
J &=& \mathrm{Im} \left( T_{11} T_{22} T_{12}^* T_{21}^* \right) = 
\left( t_{11}t_{22} t_{12}t_{21} - R^2 \right)^{1/2}.
\label{J}
\end{eqnarray}
It is well known that the existence of an internal phase of a 
unitary matrix is independent of the phase
convention~\cite{jarlskog}. This can also
be seen from the above equations for $R$ and $J$. 
However, where to place the internal phase in $T$
is, of course, convention-dependent. 
It suffices to find the internal phase for 
subform~A in equation~(\ref{basic5-permutations}) for each of the five basic
forms, since the other subforms emerge from~A by permutation of the
columns.

For each form we choose a suitable set of elements $T_{jk}$ of $T$ which are
real and positive by convention. This set contains three of the elements
occurring in $R$ and $J$. Then we apply equations~(\ref{R})
and~(\ref{J}) and compute the phase of the remaining element. From this,
taking into account that the columns of a unitary matrix form an orthonormal
system, the matrix $\widetilde T$, defined in equation~(\ref{T}), is obtained. 
Since these procedures are standard methods in linear algebra, we only display
the results:
\begin{subequations}
\begin{eqnarray}
\mbox{Form 1:} &&
\widetilde T_A = 
\left(\begin{array}{ccc}
\frac{1}{\sqrt{2}} & \frac{1}{2} & \frac{1}{2}\varphi\\
\frac{1}{2}\varphi^{2} & \frac{1}{\sqrt{2}} & -\frac{1}{2}\varphi^{2}\\
\frac{1}{2}\varphi & -\frac{1}{2} & \frac{1}{\sqrt{2}}
\end{array}\right),
\label{1-tilde}
\\
\mbox{Form 2:} &&
\widetilde T_A = 
\left(\begin{array}{ccc}
0 & \frac{1}{\sqrt{2}} & \frac{1}{\sqrt{2}}\\
\frac{1}{\sqrt{2}} & -\frac{1}{2} & \frac{1}{2}\\
\frac{1}{\sqrt{2}} & \frac{1}{2} & -\frac{1}{2}
\end{array}\right),
\label{2-tilde}
\\
\mbox{Form 3:} &&
\widetilde T_A = 
\left(\begin{array}{ccc}
\frac{1}{2} & -\frac{\sqrt{5}-1}{4} & -\frac{\sqrt{5}+1}{4}\\
\frac{\sqrt{5}+1}{4} & \frac{1}{2} & \frac{\sqrt{5}-1}{4}\\
\frac{\sqrt{5}-1}{4} & -\frac{\sqrt{5}+1}{4} & \frac{1}{2}
\end{array}\right),
\label{3-tilde}
\\
\mbox{Form 4:} &&
\widetilde T_A = 
\left(\begin{array}{ccc}
\frac{1}{\sqrt{2}}\rho_{0} & \frac{1}{2}\omega & \frac{1}{2}\omega^{2}\\
\frac{1}{2} & \frac{\sqrt{5}-1}{4} & \frac{\sqrt{5}+1}{4}\\
\frac{1}{2}\omega^{2} & \frac{\sqrt{5}+1}{4}\omega & \frac{\sqrt{5}-1}{4}
\end{array}\right),
\label{4-tilde}
\\
\mbox{Form 5:} &&
\widetilde T_A = 
\left(\begin{array}{ccc}
1 & 0 & 0\\
0 & \cos\theta & \sin\theta\\
0 & -\sin\theta & \cos\theta
\end{array}\right).
\label{5-tilde}
\end{eqnarray}
\end{subequations}
Of the phase factors occurring in these formulas, $\omega$ is defined in
equation~(\ref{obg}) and
\begin{equation}\label{phi-rho0}
\varphi = \frac{ 1 - i\sqrt{7}}{\sqrt{8}},
\quad
\rho_0 = \frac{\sqrt{5} + i\sqrt{3}}{\sqrt{8}}.
\end{equation}
As discussed in section~\ref{equivalent}, $\widetilde T$ only
needs to be computed modulo complex conjugation.

\section{Forms~1 and~4 do not lead to finite groups}
\label{not finite}
\subsection{Form 1}
We begin with 
\begin{equation}\label{1A}
\mbox{Form~1A:} \quad
T =
\left(\begin{array}{ccc}
\frac{1}{\sqrt{2}} & \frac{1}{2} & \frac{1}{2}\varphi\\
\frac{1}{2}\varphi^{2} & \frac{1}{\sqrt{2}} & -\frac{1}{2}\varphi^{2}\\
\frac{1}{2}\varphi & -\frac{1}{2} & \frac{1}{\sqrt{2}}
\end{array}\right)\textrm{diag}\left(\kappa_{1},\kappa_{2},\kappa_{3}\right).
\end{equation}
Theorem~\ref{Troots} requires that the phase factors of $T_{12}T_{21}$,
$T_{13}T_{31}$ and $T_{23}T_{32}$ are roots of unity. Therefore,
these phases are $\kappa_1 \kappa_2 \varphi^2 \equiv \xi'_{12}$, 
$\kappa_1 \kappa_3 \varphi^2 \equiv \xi'_{13}$ and
$\kappa_2 \kappa_3 \varphi^2 \equiv \xi'_{23}$, respectively, 
with roots of unity $\xi'_{jk}$ ($j < k$). From these equations we derive
$\kappa_1^2 = \varphi^{-2} \xi'_{12}\xi'_{13}/\xi'_{23}$, etc. Hence it follows
that $\kappa_j = \varphi^{-1} \xi_j$ with roots of unity $\xi_j$.
Furthermore, 
\begin{equation}\label{det1}
\left( \det T \right)^2 = x \left( \xi_1\xi_2\xi_3 \right)^2
\quad \mbox{with} \quad 
x = \frac{1+3i\sqrt{7}}{8}.
\end{equation}
Since $T$ has finite order, $x$ has to be a root of unity. But one can easily
check that $x$ fulfills 
\begin{equation}
x^2 - \frac{1}{4}x + 1 = 0.
\end{equation}
Therefore, according to theorem~\ref{roots1}, $x$ is not a root of unity and $T$
of equation~(\ref{1A}) does not belong to a finite group.

Next we consider
\begin{equation}
\mbox{Form~1B:} \quad
T =
\left(\begin{array}{ccc}
\frac{1}{\sqrt{2}} & \frac{1}{2}\varphi & \frac{1}{2}\\
\frac{1}{2}\varphi^{2} & -\frac{1}{2}\varphi^{2} & \frac{1}{\sqrt{2}}\\
\frac{1}{2}\varphi & \frac{1}{\sqrt{2}} & -\frac{1}{2}
\end{array}\right)\textrm{diag}\left(\kappa_{1},\kappa_{2},\kappa_{3}\right).
\end{equation}
Applying again theorem~\ref{Troots}, we find that 
the phase factors of $T_{12}T_{21}$, $T_{22}$ and $T_{33}$ are roots of
unity. From this it is easy to show that $\kappa_1 = \varphi^{-1} \xi_1$, 
$\kappa_2 = \varphi^{-2} \xi_2$ and $\kappa_3 = \xi_3$ with roots of unity
$\xi_j$. Now we proceed as with form~1A and obtain again
equation~(\ref{det1}), which excludes form~1B as well.

Finally we discuss
\begin{equation}
\mbox{Form 1C:} \quad
T =
\left(\begin{array}{ccc}
\frac{1}{2}\varphi & \frac{1}{\sqrt{2}} & \frac{1}{2}\\
-\frac{1}{2}\varphi^{2} & \frac{1}{2}\varphi^{2} & \frac{1}{\sqrt{2}}\\
\frac{1}{\sqrt{2}} & \frac{1}{2}\varphi & -\frac{1}{2}
\end{array}\right)\textrm{diag}\left(\kappa_{1},\kappa_{2},\kappa_{3}\right).
\end{equation}
Here theorem~\ref{Troots} tells us that the phase factors of
$T_{11}$, $T_{22}$ and $T_{33}$ are roots of unity and one obtains the same
relations for $\kappa_j$ as for form~1B. Computing the determinant of $T$, 
we are again lead to equation~(\ref{det1}). Hence form~1C is excluded.

\subsection{Form 4}
We proceed analogously to form~1. We first consider
\begin{equation}
\mbox{Form~4A:} \quad
T = 
\left(\begin{array}{ccc}
\frac{1}{\sqrt{2}}\rho_{0} & \frac{1}{2}\omega & \frac{1}{2}\omega^{2}\\
\frac{1}{2} & \frac{\sqrt{5}-1}{4} & \frac{\sqrt{5}+1}{4}\\
\frac{1}{2}\omega^{2} & \frac{\sqrt{5}+1}{4}\omega & \frac{\sqrt{5}-1}{4}
\end{array}\right)\textrm{diag}\left(\kappa_{1},\kappa_{2},\kappa_{3}\right).
\end{equation}
Theorem~\ref{Troots} requires that the phase factors of
$T_{12} T_{21}$, $T_{22}$ and $T_{33}$ are roots of unity, therefore, the
$\kappa_j$  must all be roots of unity. Computing the determinant of $T$, we
obtain 
\begin{equation}
\det T = \chi \kappa_1 \kappa_2 \kappa_3 
\quad \mbox{with} \quad
\chi = \frac{1}{8}
\left[ 1+3\sqrt{5}-i\sqrt{3}\left(\sqrt{5}-1\right) \right].
\end{equation}
Since $T$ has finite order, its determinant is a root of unity and so is
$\chi$. Then also
\begin{equation}
\chi \omega = \frac{1}{4} \left( -1 + i \sqrt{15} \right)
\end{equation}
must be root of unity. But $\chi \omega$ is a root of the equation
\begin{equation}
x^2 + \frac{1}{2}x + 1 = 0.
\end{equation}
Therefore, according to theorem~\ref{roots1}, it cannot be a root of unity,
which is a contradiction. Hence form~4A does not lead to a finite group.

Now we consider
\begin{equation}
\mbox{Form~4B:} \quad
T = 
\left(\begin{array}{ccc}
\frac{1}{\sqrt{2}}\rho_{0} & \frac{1}{2}\omega^{2} & \frac{1}{2}\omega\\
\frac{1}{2} & \frac{\sqrt{5}+1}{4} & \frac{\sqrt{5}-1}{4}\\
\frac{1}{2}\omega^{2} & \frac{\sqrt{5}-1}{4} & \frac{\sqrt{5}+1}{4}\omega
\end{array}\right)\textrm{diag}\left(\kappa_{1},\kappa_{2},\kappa_{3}\right).
\end{equation}
In the same way as for form~1A, 
we find that all $\kappa_j$ are roots of unity. In this case
\begin{equation}
\det T = -\chi \kappa_1 \kappa_2 \kappa_3 
\end{equation}
and the argument excluding this form goes through as before.

Next we discuss
\begin{equation}
\mbox{Form~4C:} \quad
T = 
\left(\begin{array}{ccc}
\frac{1}{2}\omega & \frac{1}{\sqrt{2}}\rho_{0} & \frac{1}{2}\omega^{2}\\
\frac{\sqrt{5}-1}{4} & \frac{1}{2} & \frac{\sqrt{5}+1}{4}\\
\frac{\sqrt{5}+1}{4}\omega & \frac{1}{2}\omega^{2} & \frac{\sqrt{5}-1}{4}
\end{array}\right)\textrm{diag}\left(\kappa_{1},\kappa_{2},\kappa_{3}\right).
\end{equation}
The determinant of this form is equal to the one of form~4B, and the 
phases $\kappa_j$ have to be roots of unity as before. 
So one can exclude this form as well.

The final case is
\begin{equation}
\mbox{Form~4D:} \quad
T = 
\left(\begin{array}{ccc}
\frac{1}{2}\omega^{2} & \frac{1}{\sqrt{2}}\rho_{0} & \frac{1}{2}\omega\\
\frac{\sqrt{5}+1}{4} & \frac{1}{2} & \frac{\sqrt{5}-1}{4}\\
\frac{\sqrt{5}-1}{4} & \frac{1}{2}\omega^{2} & \frac{\sqrt{5}+1}{4}\omega
\end{array}\right)\textrm{diag}\left(\kappa_{1},\kappa_{2},\kappa_{3}\right).
\end{equation}
Again, all $\kappa_j$ must be roots of unity and the determinant is the same as
for form~4A.

It is interesting to note that forms~1 and~4, the only two of the basic five
forms which have a non-trivial internal phase, do not lead to finite groups.
Thus we do not discuss them further.

\section{The external phases of $T$ and the resulting mixing matrices}
\label{external}
It remains to investigate the remaining forms~2, 3 and~5.
We still have to determine the external phases 
$\hat\kappa = \diag (\kappa_1, \kappa_2, \kappa_3 )$ of 
$T = \widetilde T \hat\kappa$ 
for a total of eight subforms of $\widetilde T$:
\begin{subequations}\label{subforms}
\begin{alignat}{5}
&\mbox{Form 2:} &\quad &\widetilde T_A, 
&\quad \widetilde T_B &= \widetilde T_A\, M[(12)], &&&&
\\
&\mbox{Form 3:} &\quad &\widetilde T_A,  
&\quad \widetilde T_B &= \widetilde T_A\, M[(23)],
&\quad \widetilde T_C &= \widetilde T_A\, M[(132)],
&\quad \widetilde T_D &= \widetilde T_A\, M[(123)],
\\
&\mbox{Form 5:} &\quad &\widetilde T_A,  
&\quad \widetilde T_B &= \widetilde T_A\, M[(123)], &&&&
\end{alignat}
\end{subequations}
which are read off from 
equations~(\ref{form2AB}) and~(\ref{2-tilde}),
equations~(\ref{form3ABCD}) and~(\ref{3-tilde}),
and
equations~(\ref{form5AB}) and~(\ref{5-tilde}),
for forms~2, 3 and~5, respectively. 
All these forms have trivial internal phases.

The basic idea to determine the possible $\hat\kappa$ for each subform is
the observation that, for any element $T' \in G_\ell$, the matrix
$|T'|$ must be of one of the basics forms~2, 3 or~5, after having
excluded forms~1 and~4. In particular, this holds for $T' = T^2$.
Therefore, for any of the eight subforms under consideration the elements
\begin{equation}
\left| \left( T^2 \right)_{jk} \right| = 
\left| \sum_{l=1}^3 \widetilde T_{jl} \kappa_l \widetilde T_{lk} \right| 
\end{equation}
must belong to a matrix $P_1 |T|_A P_2$ where $P_1$ and $P_2$ are
arbitrary permutation matrices. Two remarks are in order.
Let us assume that we discuss for instance $T$ of form~2.
\begin{enumerate}
\renewcommand{\labelenumi}{\roman{enumi}.}
\item
Then $T^2$ can belong to form~2, 3 or~5.
\item
Since we have used up already the freedom of permutations for the
subform we begin with, we have to admit \textit{all} possible
permutations $P_1$ and $P_2$.
\end{enumerate}
In the following we will present, for each subform in an own
subsection, the results for the phases $\hat\kappa$, 
the eigenvalues $\lambda^{(0)}_j$ ($j=1,2,3$) of $T$ and the matrix $|U|^2$. 
Every triple consisting of $T$, including its external phases, 
the eigenvalues of $T$ and $|U|^2$ is called solution. 
The latter matrices, whose entries are the $|U_{jk}|^2$, 
represent the main result of this paper. The
details of the computations for each subform are deferred to
appendix~\ref{details}. Every solution 
obtains a tag $\mathcal{C}_i$ in the case of a \textit{complete} 
determination of $|U|^2$ by a single $T$. 
In the case that $T$ is degenerate and determines
only one row of $|U|^2$, then the tag is $\mathcal{P}_i$ which stands
for \textit{partial}.

As we will see, in general, there are several solutions of external
phases $\hat\kappa$ for each of the eight subforms of
equation~(\ref{subforms}). Sometimes it occurs that for one 
$\widetilde T$ there are two matrices $\hat\kappa_a$ and
$\hat\kappa_b$ 
of external phase such that the corresponding matrices
$T_a$ and $T_b$ have the same $|U|^2$ and 
eigenvalues which are related by
\begin{equation}\label{lambda-related}
\left( \lambda^{(0)}_{a1},\lambda^{(0)}_{a2},\lambda^{(0)}_{a3} \right) 
\propto 
\left( \lambda^{(0)}_{b1},\lambda^{(0)}_{b2},\lambda^{(0)}_{b3} \right)
\quad \mbox{or} \quad
\left( \lambda^{(0)}_{b1},\lambda^{(0)}_{b2},\lambda^{(0)}_{b3} \right)^*.
\end{equation}
In this case, we assign only one solution tag.

\subsection{Form 2A}
In appendix~\ref{form 2A} it is shown that form~2A 
\begin{equation}\label{T2A}
T = \left(\begin{array}{ccc}
0 & \frac{1}{\sqrt{2}} & \frac{1}{\sqrt{2}}\\
\frac{1}{\sqrt{2}} & -\frac{1}{2} & \frac{1}{2}\\
\frac{1}{\sqrt{2}} & \frac{1}{2} & -\frac{1}{2}
\end{array}\right)
\textrm{diag}\left(\kappa_{1},\kappa_{2},\kappa_{3}\right)\end{equation}
requires 
$\kappa_2 = \pm \kappa_3$ and thus leads to two subcases.

In the first subcase with solution tag $\mathcal{C}_{1}$, we have
\begin{equation}
\kappa_2 = \kappa_3, 
\quad
\left(\lambda_{1}^{(0)},\lambda_{2}^{(0)},\lambda_{3}^{(0)}\right) =
\left(\sqrt{\kappa_1 \kappa_2},-\sqrt{\kappa_1 \kappa_2}, -\kappa_2 \right)
\end{equation}
and the mixing matrix
\begin{equation}
\mathcal{C}_{1}: \quad |U|^2 = 
\left(\begin{array}{ccc}
\frac{1}{2} & \frac{1}{4} & \frac{1}{4}\\
\frac{1}{2} & \frac{1}{4} & \frac{1}{4}\\
0 & \frac{1}{2} & \frac{1}{2}
\end{array}\right).
\end{equation}

The second subcase is given by
\begin{equation}
\kappa_2 = -\kappa_3, 
\quad
\left(\lambda_{1}^{(0)},\lambda_{2}^{(0)},\lambda_{3}^{(0)}\right) =
\kappa\left(1,\omega,\omega^{2}\right),
\end{equation}
\begin{equation}
\mathcal{C}_{2}: \quad |U|^2 = 
\frac{1}{3}\left(\begin{array}{ccc}
1 & 1+\textrm{Re}\,\sigma & 1-\textrm{Re}\,\sigma\\
1 & 1+\textrm{Re}\left(\omega\sigma\right) &
1-\textrm{Re}\left(\omega\sigma\right)\\ 
1 & 1+\textrm{Re}\left(\omega^{2}\sigma\right) &
1-\textrm{Re}\left(\omega^{2}\sigma\right) 
\end{array}\right).
\end{equation}
For $\mathcal{C}_2$, the quantities $\kappa$ and $\sigma$ are roots of
unity related to $\kappa_1$ and $\kappa_2$ by
\begin{equation}
\kappa^3 = -\kappa_1 \kappa^2_2
\quad \mbox{and} \quad
\sigma = -\kappa\kappa_2^*,			
\end{equation}
respectively. 
Note that the transformations 
$\sigma\rightarrow\left(-\omega\right)^{x}\sigma$
and
$\sigma\rightarrow\left(-\omega\right)^{x}\sigma^{*}$ with
$x=0,\ldots,5$
lead to a permutation of the mixing pattern in $|U|^2$,
which accounts for 12 of
the 36 possible row and column permutations.
Once this permutation freedom is taken into account, it becomes
clear that two roots of unity $\sigma$, $\sigma'$ will yield the
same mixing angles if and only if 
$\textrm{Re}\left(\sigma^{6}\right) =
\textrm{Re}\left({\sigma'}^{ 6}\right)$.

While $\mathcal{C}_1$ is known as bimaximal mixing,
$\mathcal{C}_{2}$ corresponds to trimaximal mixing; 
for the specific choice of $\sigma$ such that $\sigma^6=1$, 
tribimaximal mixing is obtained. 
It will turn out that $\mathcal{C}_{2}$
is the only series of mixing matrices 
genuinely involving the three flavours. 
From residual symmetries, $\mathcal{C}_{2}$ has for
instance been derived in~\cite{toorop2}, while it was obtained
in~\cite{holthausen} from GAP~\cite{GAP} and the
Small Groups Library~\cite{SGL}. 
Recently, this series has been accommodated in a model~\cite{GL2013}.
We note that the $|U|^2$ of $\mathcal{C}_{2}$ has a trivial CKM-type phase. 

If $\kappa_1 = \kappa_2$ in $\mathcal{C}_1$, then 
$\lambda^{(0)}_2 = \lambda^{(0)}_3$ and the residual symmetries fix
only one row in the mixing matrix $U$:
\begin{equation}
\kappa_1 = \kappa_2 = \kappa_3,
\quad
 \left(\lambda_{1}^{(0)},\lambda_{2}^{(0)},\lambda_{3}^{(0)}\right) =
\kappa_1\left(1,-1,-1\right),
\end{equation}
with the mixing matrix
\begin{equation}
\mathcal{P}_{1}: \quad |U|^{2} = 
\left(\begin{array}{ccc}
\frac{1}{2} & \frac{1}{4} & \frac{1}{4}\\
\times & \times & \times \\
\times & \times & \times
\end{array}\right).
\end{equation}
The symbol ``$\times$'' indicates the positions in $|U|^2$ which are not
fully determined by the $T$ of equation~(\ref{T2A}) with
$\hat\kappa \propto \bone$.

\subsection{Form 2B}
According to the analysis carried out in appendix~\ref{form 2B}, 
the external phases allowed for form~2B are 
\begin{equation}
\hat\kappa = \diag \left( x, \pm x^*, 1 \right) \kappa_3
\quad \mbox{with} \quad
x = \pm\varphi, \; \pm \rho_0,
\end{equation}
where $\varphi$ and $\rho_0$ are defined in
equation~(\ref{phi-rho0}) and $\kappa_3$ is an arbitrary root of
unity. Complex conjugation of $\hat\kappa$ leads
to further allowed cases. However, as discussed earlier, this
is a trivial variation of the solutions and we ignore it.
Thus there are eight subcases of form~2B.
\paragraph{First subcase:}
\begin{align}
T & =\left(\begin{array}{ccc}
\frac{1}{\sqrt{2}} & 0 & \frac{1}{\sqrt{2}}\\
-\frac{1}{2} & \frac{1}{\sqrt{2}} & \frac{1}{2}\\
\frac{1}{2} & \frac{1}{\sqrt{2}} & -\frac{1}{2}
\end{array}\right)
\textrm{diag}\left(-\varphi,-\varphi^{*},1\right)\kappa_{3}
\end{align}
with eigenvalues
\begin{equation}
\left(\lambda^{(0)}_1, \lambda^{(0)}_2, \lambda^{(0)}_3 \right) = 
\kappa_3 ( -1,i,-i).
\end{equation}
We denote this solution by $\mathcal{C}_3$. It gives the mixing matrix
\begin{equation}
\mathcal{C}_3: \quad |U|^2 = 
\left(
\begin{array}{ccc}
 \frac{1}{4} & \frac{1}{4} & \frac{1}{2} \\
 \frac{1}{8} \left(3+\sqrt{7}\right) & \frac{1}{8}
   \left(3-\sqrt{7}\right) & \frac{1}{4} \\
 \frac{1}{8} \left(3-\sqrt{7}\right) & \frac{1}{8}
   \left(3+\sqrt{7}\right) & \frac{1}{4} \\
\end{array}
\right).
\end{equation}
\paragraph{Second subcase:}
\begin{align}
T & =\left(\begin{array}{ccc}
\frac{1}{\sqrt{2}} & 0 & \frac{1}{\sqrt{2}}\\
-\frac{1}{2} & \frac{1}{\sqrt{2}} & \frac{1}{2}\\
\frac{1}{2} & \frac{1}{\sqrt{2}} & -\frac{1}{2}
\end{array}\right)
\textrm{diag}\left(-\varphi,\varphi^{*},1\right)\kappa_{3}
\end{align}
with eigenvalues
\begin{equation}
\left(\lambda^{(0)}_1, \lambda^{(0)}_2, \lambda^{(0)}_3 \right) = 
\kappa_3 ( \gamma^2,\gamma^4,\gamma).
\end{equation}
In this case with the solution tag $\mathcal{C}_4$,
the seventh root of unity, $\gamma = e^{2\pi i/7}$,
occurs in the eigenvalues and the mixing matrix reads 
\begin{equation}
\mathcal{C}_4: \quad |U|^2 = 
\left(
\begin{array}{ccc}
r_1 & r_2 & r_3 \\
r_2 & r_3 & r_1 \\
r_3 & r_1 & r_2
\end{array} \right),
\end{equation}
where the $r_i$ are the roots of the equation
\begin{equation}
-1 + 14x -56x^2 + 56x^3 = 0.
\end{equation}
Their approximate numerical values are 
$r_1 = 0.664$, $r_2 = 0.204$, $r_3 = 0.132$.
\paragraph{Third subcase:}
\begin{align}
T & =\left(\begin{array}{ccc}
\frac{1}{\sqrt{2}} & 0 & \frac{1}{\sqrt{2}}\\
-\frac{1}{2} & \frac{1}{\sqrt{2}} & \frac{1}{2}\\
\frac{1}{2} & \frac{1}{\sqrt{2}} & -\frac{1}{2}
\end{array}\right)\textrm{diag}\left(\varphi,\varphi^{*},1\right)\kappa_{3}
\end{align}
with eigenvalues
\begin{equation}
\left(\lambda^{(0)}_1, \lambda^{(0)}_2, \lambda^{(0)}_3 \right) = 
-\kappa_3 ( 1, \omega^2,\omega) 
\end{equation}
leads to solution $\mathcal{C}_5$ with the mixing matrix
\begin{equation}
\mathcal{C}_5: \quad |U|^2 = 
\left(
\begin{array}{ccc}
\frac{1}{6} & \frac{1}{6} & \frac{2}{3} \\
\frac{1}{12} \left(5-\sqrt{21}\right) & 
\frac{1}{12} \left(5+\sqrt{21}\right) & 
\frac{1}{6} \\
\frac{1}{12} \left(5+\sqrt{21}\right) & 
\frac{1}{12} \left(5-\sqrt{21}\right) &
\frac{1}{6} \\
\end{array}
\right).
\end{equation}
\paragraph{Fourth subcase:}
\begin{align}
T & =\left(\begin{array}{ccc}
\frac{1}{\sqrt{2}} & 0 & \frac{1}{\sqrt{2}}\\
-\frac{1}{2} & \frac{1}{\sqrt{2}} & \frac{1}{2}\\
\frac{1}{2} & \frac{1}{\sqrt{2}} & -\frac{1}{2}
\end{array}\right)\textrm{diag}\left(\varphi,-\varphi^{*},1\right)\kappa_{3}
\end{align}
with eigenvalues
\begin{equation}
\left(\lambda^{(0)}_1, \lambda^{(0)}_2, \lambda^{(0)}_3 \right) = 
\kappa_3 ( \gamma^5,\gamma^3,\gamma^6)
\end{equation}
does not produce a new mixing matrix, but repeats the mixing
matrix of solution~$\mathcal{C}_4$ of the second subcase. Moreover, the
eigenvalues of $T$ of the second and fourth subcases are related by
complex conjugation. Therefore, according to the philosophy put forward in the
beginning of section~\ref{external}, we do not assign a new solution
tag here.
\paragraph{Fifth subcase:}
\begin{align}
T & =\left(\begin{array}{ccc}
\frac{1}{\sqrt{2}} & 0 & \frac{1}{\sqrt{2}}\\
-\frac{1}{2} & \frac{1}{\sqrt{2}} & \frac{1}{2}\\
\frac{1}{2} & \frac{1}{\sqrt{2}} & -\frac{1}{2}
\end{array}\right)\textrm{diag}\left(\rho_{0},\rho_{0}^{*},1\right)\kappa_{3}
\end{align}
with eigenvalues
\begin{equation}
\left(\lambda^{(0)}_1, \lambda^{(0)}_2, \lambda^{(0)}_3 \right) = 
-\kappa_3 (\beta^3,\beta^2,1). 
\end{equation}
In this case, the fifth root of unity, $\beta = e^{2\pi i/5}$, appears.
The mixing matrix of this solution is given by
\begin{equation}
\mathcal{C}_6: \quad |U|^2 = 
\left(
\begin{array}{ccc}
\frac{1}{8} \left(3+\frac{1}{\sqrt{5}} + \sqrt{6+\frac{6}{\sqrt{5}}}\right) & 
\frac{1}{8} \left(3+\frac{1}{\sqrt{5}} - \sqrt{6+\frac{6}{\sqrt{5}}}\right) & 
\frac{1}{20} \left(5-\sqrt{5}\right) \\
\frac{1}{8} \left(3+\frac{1}{\sqrt{5}} - \sqrt{6+\frac{6}{\sqrt{5}}}\right) &
\frac{1}{8} \left(3+\frac{1}{\sqrt{5}} + \sqrt{6+\frac{6}{\sqrt{5}}}\right) &
\frac{1}{20} \left(5-\sqrt{5}\right) \\
\frac{1}{20} \left(5-\sqrt{5}\right) & 
\frac{1}{20} \left(5-\sqrt{5}\right) &
\frac{1}{10} \left(5+\sqrt{5}\right) 
\end{array}
\right).
\end{equation}
\paragraph{Sixth subcase:}
\begin{align}
T & =\left(\begin{array}{ccc}
\frac{1}{\sqrt{2}} & 0 & \frac{1}{\sqrt{2}}\\
-\frac{1}{2} & \frac{1}{\sqrt{2}} & \frac{1}{2}\\
\frac{1}{2} & \frac{1}{\sqrt{2}} & -\frac{1}{2}
\end{array}\right)\textrm{diag}\left(\rho_{0},-\rho_{0}^{*},1\right)\kappa_{3}
\end{align}
with eigenvalues
\begin{equation}
\left(\lambda^{(0)}_1, \lambda^{(0)}_2, \lambda^{(0)}_3 \right) = 
\kappa_3 \omega\,( i,1,-i).
\end{equation}
Here we obtain the mixing matrix
\begin{equation}
\mathcal{C}_7: \quad |U|^2 = 
\left(
\begin{array}{ccc}
\frac{1}{16} \left(5-\sqrt{3}+\sqrt{5}-\sqrt{15}\right) & 
\frac{1}{16} \left(5-\sqrt{3}-\sqrt{5}+\sqrt{15}\right) 
& 
\frac{1}{8} \left(3+\sqrt{3}\right) \\
 \frac{1}{8} \left(3-\sqrt{5}\right) & \frac{1}{8} \left(3+\sqrt{5}\right) &
   \frac{1}{4} \\
 \frac{1}{16} \left(5+\sqrt{3}+\sqrt{5}+\sqrt{15}\right) 
&
   \frac{1}{16} \left(5+\sqrt{3}-\sqrt{5}-\sqrt{15}\right) &
   \frac{1}{8} \left(3-\sqrt{3}\right) \\
\end{array}
\right).
\end{equation}
\paragraph{Seventh subcase:}
\begin{align}
T & =\left(\begin{array}{ccc}
\frac{1}{\sqrt{2}} & 0 & \frac{1}{\sqrt{2}}\\
-\frac{1}{2} & \frac{1}{\sqrt{2}} & \frac{1}{2}\\
\frac{1}{2} & \frac{1}{\sqrt{2}} & -\frac{1}{2}
\end{array}\right)\textrm{diag}\left(-\rho_{0},-\rho_{0}^{*},1\right)\kappa_{3}
\end{align}
with eigenvalues
\begin{equation}
\left(\lambda^{(0)}_1, \lambda^{(0)}_2, \lambda^{(0)}_3 \right) = 
-\kappa_3 (\beta^4,\beta,1).
\end{equation}
Here again the fifth root of unity, $\beta$, occurs. The corresponding mixing
matrix is
\begin{equation}
\mathcal{C}_8: \quad |U|^2 = 
\left(
\begin{array}{ccc}
 \frac{1}{8} \left(3-\frac{1}{\sqrt{5}}-\sqrt{6-\frac{6}{\sqrt{5}}}\right) &
   \frac{1}{8} \left(3-\frac{1}{\sqrt{5}}+\sqrt{6-\frac{6}{\sqrt{5}}}\right) &
   \frac{1}{20} \left(5+\sqrt{5}\right) \\
 \frac{1}{8} \left(3-\frac{1}{\sqrt{5}}+\sqrt{6-\frac{6}{\sqrt{5}}}\right) &
   \frac{1}{8} \left(3-\frac{1}{\sqrt{5}}-\sqrt{6-\frac{6}{\sqrt{5}}}\right) &
   \frac{1}{20} \left(5+\sqrt{5}\right) \\
 \frac{1}{20} \left(5+\sqrt{5}\right) & \frac{1}{20} \left(5+\sqrt{5}\right) &
   \frac{1}{10} \left(5-\sqrt{5}\right) \\
\end{array}
\right).
\end{equation}
\paragraph{Eighth subcase:}
\begin{align}
T & =\left(\begin{array}{ccc}
\frac{1}{\sqrt{2}} & 0 & \frac{1}{\sqrt{2}}\\
-\frac{1}{2} & \frac{1}{\sqrt{2}} & \frac{1}{2}\\
\frac{1}{2} & \frac{1}{\sqrt{2}} & -\frac{1}{2}
\end{array}\right)\textrm{diag}\left(-\rho_{0},\rho_{0}^{*},1\right)\kappa_{3}
\end{align}
with eigenvalues
\begin{equation}
\left(\lambda^{(0)}_1, \lambda^{(0)}_2, \lambda^{(0)}_3 \right) = 
\kappa_3\, \omega^2\,( -i,1,i).
\end{equation}
These eigenvalues are proportional to the complex conjugate ones of
solution~$\mathcal{C}_7$, sixth subcase, and the mixing matrix is 
also the same. Thus we do not assign a new solution tag.

\subsection{Form 3A}
According to the analysis in appendix~\ref{form 3A}, the possible external
phases for form~3A are 
\begin{equation}\label{kappa-3A}
\hat\kappa = 
\diag\left(1,\omega,\omega^2 \right)\kappa_1, \;
\diag\left(-\omega^2,\omega,1\right)\kappa_3, \;
\diag\left(\omega,-\omega^2,1\right)\kappa_3, \;
\diag\left(\omega,1,-\omega^2\right)\kappa_2
\end{equation}
with arbitrary roots of unity $\kappa_1$, $\kappa_2$ and $\kappa_3$.
\paragraph{First subcase:}
\begin{align}
T & =\left(\begin{array}{ccc}
\frac{1}{2} & -\frac{\sqrt{5}-1}{4} & -\frac{\sqrt{5}+1}{4}\\
\frac{\sqrt{5}+1}{4} & \frac{1}{2} & \frac{\sqrt{5}-1}{4}\\
\frac{\sqrt{5}-1}{4} & -\frac{\sqrt{5}+1}{4} & \frac{1}{2}
\end{array}\right)\textrm{diag}\left(1,\omega,\omega^{2}\right)\kappa_{1}
\end{align}
with eigenvalues
\begin{equation}
\left(\lambda^{(0)}_1, \lambda^{(0)}_2, \lambda^{(0)}_3 \right) = 
\kappa_1 \,( 1,\omega,\omega^2).
\end{equation}
The corresponding mixing matrix is
\begin{equation}
\mathcal{C}_{9}: \quad |U|^2 =
\left(\begin{array}{ccc}
\frac{2}{3} & \frac{1}{6} & \frac{1}{6}\\
\frac{1}{6} & \frac{2}{3} & \frac{1}{6}\\
\frac{1}{6} & \frac{1}{6} & \frac{2}{3}
\end{array}\right).
\end{equation}
\paragraph{Second subcase:}
Next we consider
\begin{align}
T & =\left(\begin{array}{ccc}
\frac{1}{2} & -\frac{\sqrt{5}-1}{4} & -\frac{\sqrt{5}+1}{4}\\
\frac{\sqrt{5}+1}{4} & \frac{1}{2} & \frac{\sqrt{5}-1}{4}\\
\frac{\sqrt{5}-1}{4} & -\frac{\sqrt{5}+1}{4} & \frac{1}{2}
\end{array}\right)
\textrm{diag}\left(-\omega^{2},\omega,1\right)\kappa_{3}
\end{align}
with eigenvalues
\begin{equation}
\left(\lambda^{(0)}_1, \lambda^{(0)}_2, \lambda^{(0)}_3 \right) = 
\kappa_3\, \omega^2\,( -i,-1,i).
\end{equation}
This subcase provides the mixing matrix
\begin{equation}
\mathcal{C}_{10}: \quad |U|^2 = 
\left(\begin{array}{ccc}
\frac{1}{4} & \frac{1}{8}\left(3+\sqrt{3}\right)  & 
\frac{1}{8}\left(3-\sqrt{3}\right) \\
\frac{1}{2} & \frac{1}{4} & \frac{1}{4}\\
\frac{1}{4} & \frac{1}{8}\left(3-\sqrt{3}\right) & 
\frac{1}{8}\left(3+\sqrt{3}\right)
\end{array}\right).
\end{equation}
According to equation~(\ref{kappa-3A}) there are two more subcases to
consider, however, it turns out that these again lead to the 
eigenvalues and the mixing matrix of
$\mathcal{C}_{10}$. 

\subsection{Form 3B}
According to the analysis in appendix~\ref{form 3B}, there are four different
cases of external phases:
\begin{equation}
\hat\kappa = 
\diag\left(1,1,-1\right)\kappa_{1},\;
\diag\left(1,1,1\right)\kappa_{1},\;
\diag\left(1,-1,1\right)\kappa_{1},\;
\diag\left(-1,1,1\right)\kappa_{3},
\end{equation}
with arbitrary roots of unity $\kappa_1$, $\kappa_3$.
\paragraph{First subcase:}
\begin{align}
T & =\left(\begin{array}{ccc}
\frac{1}{2} & -\frac{\sqrt{5}+1}{4} & -\frac{\sqrt{5}-1}{4}\\
\frac{\sqrt{5}+1}{4} & \frac{\sqrt{5}-1}{4} & \frac{1}{2}\\
\frac{\sqrt{5}-1}{4} & \frac{1}{2} & -\frac{\sqrt{5}+1}{4}
\end{array}\right)\textrm{diag}\left(1,1,-1\right)\kappa_{1}
\end{align}
has the eigenvalues
\begin{equation}
\left(\lambda^{(0)}_1, \lambda^{(0)}_2, \lambda^{(0)}_3 \right) = 
\kappa_1 \,\left( 1,\beta,\beta^4 \right)
\end{equation}
and provides the mixing matrix
\begin{equation}
\mathcal{C}_{11}: \quad |U|^2 = 
\left(
\begin{array}{ccc}
 \frac{1}{10} \left(5-\sqrt{5}\right) & 0 & \frac{1}{10}
   \left(5+\sqrt{5}\right) \\
 \frac{1}{20} \left(5+\sqrt{5}\right) & \frac{1}{2} &
   \frac{1}{20}\left(5-\sqrt{5}\right) \\
 \frac{1}{20} \left(5+\sqrt{5}\right) & \frac{1}{2} &
   \frac{1}{20}\left(5-\sqrt{5}\right) \\
\end{array}
\right).
\end{equation}
\paragraph{Second subcase:}
The next subcase is
\begin{align}
T & =\left(\begin{array}{ccc}
\frac{1}{2} & -\frac{\sqrt{5}+1}{4} & -\frac{\sqrt{5}-1}{4}\\
\frac{\sqrt{5}+1}{4} & \frac{\sqrt{5}-1}{4} & \frac{1}{2}\\
\frac{\sqrt{5}-1}{4} & \frac{1}{2} & -\frac{\sqrt{5}+1}{4}
\end{array}\right)\textrm{diag}\left(1,1,1\right)\kappa_{1}.
\end{align}
The eigenvalues are
\begin{equation}
\left(\lambda^{(0)}_1, \lambda^{(0)}_2, \lambda^{(0)}_3 \right) = 
-\kappa_1 \,\left( \omega,\omega^2,1 \right)
\end{equation}
and the mixing matrix is given by
\begin{equation}
\mathcal{C}_{12}: \quad |U|^2 =
\left(
\begin{array}{ccc}
 \frac{1}{2} & \frac{1}{12} \left(3+\sqrt{5}\right) & \frac{1}{12}
   \left(3-\sqrt{5}\right) \\
 \frac{1}{2} & \frac{1}{12} \left(3+\sqrt{5}\right) & \frac{1}{12}
   \left(3-\sqrt{5}\right) \\
 0 & \frac{1}{6} \left(3-\sqrt{5}\right) & \frac{1}{6}
   \left(3+\sqrt{5}\right) \\
\end{array}
\right).
\end{equation}
\paragraph{Third subcase:}
\begin{align}
T & =\left(\begin{array}{ccc}
\frac{1}{2} & -\frac{\sqrt{5}+1}{4} & -\frac{\sqrt{5}-1}{4}\\
\frac{\sqrt{5}+1}{4} & \frac{\sqrt{5}-1}{4} & \frac{1}{2}\\
\frac{\sqrt{5}-1}{4} & \frac{1}{2} & -\frac{\sqrt{5}+1}{4}
\end{array}\right)\textrm{diag}\left(1,-1,1\right)\kappa_{1}
\end{align}
with eigenvalues
\begin{equation}
\left(\lambda^{(0)}_1, \lambda^{(0)}_2, \lambda^{(0)}_3 \right) = 
\kappa_1 \,\left( \beta^2,\beta^3,1 \right)
\end{equation}
gives the mixing matrix
\begin{equation}
\mathcal{C}_{13}: \quad |U|^2 =
\left(
\begin{array}{ccc}
 \frac{1}{20} \left(5-\sqrt{5}\right) & \frac{1}{20}
   \left(5+\sqrt{5}\right) & \frac{1}{2} \\
 \frac{1}{20} \left(5-\sqrt{5}\right) & \frac{1}{20}
   \left(5+\sqrt{5}\right) & \frac{1}{2} \\
 \frac{1}{10} \left(5+\sqrt{5}\right) & \frac{1}{10} \left(5-\sqrt{5}\right) &
 0 \\ 
\end{array}
\right).
\end{equation}
Solution $\mathcal{C}_{13}$ has the same mixing matrix as
$\mathcal{C}_{11}$, however, 
its eigenvalues are not related by way of equation~(\ref{lambda-related}).
Therefore, we count it as a separate solution. 
\paragraph{Fourth subcase:}
The last subcase
\begin{align}
T & =\left(\begin{array}{ccc}
\frac{1}{2} & -\frac{\sqrt{5}+1}{4} & -\frac{\sqrt{5}-1}{4}\\
\frac{\sqrt{5}+1}{4} & \frac{\sqrt{5}-1}{4} & \frac{1}{2}\\
\frac{\sqrt{5}-1}{4} & \frac{1}{2} & -\frac{\sqrt{5}+1}{4}
\end{array}\right)\textrm{diag}\left(-1,1,1\right)\kappa_{3}
\end{align}
has two degenerate eigenvalues:
\begin{equation}
\left(\lambda^{(0)}_1, \lambda^{(0)}_2, \lambda^{(0)}_3 \right) = 
\kappa_3 \left( 1,-1,-1 \right).
\end{equation}
Therefore, the mixing matrix is only partially determined:
\begin{equation}
\mathcal{P}_2: \quad |U|^2 =
\left( \begin{array}{ccc}
\frac{1}{4} & \frac{1}{8} \left(3+\sqrt{5}\right) & 
\frac{1}{8} \left(3-\sqrt{5}\right) \\
\times & \times & \times \\
\times & \times & \times 
\end{array} \right).
\end{equation}

\subsection{Form 3C}
According to the analysis in appendix~\ref{form 3C} there are four
solutions for the external phases:
\begin{equation}\label{external-3C}
\hat\kappa = 
\diag \left( \omega,-\omega^2, 1 \right) \kappa_3, \;
\diag\left(-\omega^2,\omega,1\right)\kappa_3, \;
\diag\left(\omega,\omega^2,1\right)\kappa_3, \;
\diag\left(\omega,1,-\omega^2\right)\kappa_2,
\end{equation}
with arbitrary roots of unity $\kappa_2$, $\kappa_3$.
\paragraph{First subcase:}
We begin with
\begin{equation}
T=
\left(\begin{array}{ccc}
-\frac{\sqrt{5}+1}{4} & \frac{1}{2} & -\frac{\sqrt{5}-1}{4}\\
\frac{\sqrt{5}-1}{4} & \frac{\sqrt{5}+1}{4} & \frac{1}{2}\\
\frac{1}{2} & \frac{\sqrt{5}-1}{4} & -\frac{\sqrt{5}+1}{4}
\end{array}\right)
\textrm{diag} \left( \omega, -\omega^2, 1 \right) \kappa_{3},
\end{equation}
which has the eigenvalues
\begin{equation}
\left(\lambda^{(0)}_1, \lambda^{(0)}_2, \lambda^{(0)}_3 \right) = 
-\kappa_3 \,\left( 1,\omega^2,\omega \right)
\end{equation}
and leads to the mixing matrix
\begin{equation}
\mathcal{C}_{14}: \quad |U|^2 =
\left(
\begin{array}{ccc}
 \frac{1}{12} \left(3-\sqrt{5}\right) & \frac{1}{12}
   \left(3-\sqrt{5}\right) & \frac{1}{6} \left(3+\sqrt{5}\right) \\
 \frac{1}{12} \left(3-\sqrt{5}\right) & \frac{1}{6}
   \left(3+\sqrt{5}\right) & \frac{1}{12} \left(3-\sqrt{5}\right) \\
 \frac{1}{6} \left(3+\sqrt{5}\right) & 
 \frac{1}{12} \left(3-\sqrt{5}\right) &
 \frac{1}{12} \left(3-\sqrt{5}\right)
\end{array}
\right).
\end{equation}
\paragraph{Second subcase:}
Next we tackle 
\begin{align}
T & =\left(\begin{array}{ccc}
-\frac{\sqrt{5}+1}{4} & \frac{1}{2} & -\frac{\sqrt{5}-1}{4}\\
\frac{\sqrt{5}-1}{4} & \frac{\sqrt{5}+1}{4} & \frac{1}{2}\\
\frac{1}{2} & \frac{\sqrt{5}-1}{4} & -\frac{\sqrt{5}+1}{4}
\end{array}\right)
\textrm{diag}\left(-\omega^{2},\omega,1\right)\kappa_{3}.
\end{align}
This matrix has the eigenvalues
\begin{equation}\label{eigen-3C-2}
\left(\lambda^{(0)}_1, \lambda^{(0)}_2, \lambda^{(0)}_3 \right) = 
-\kappa_3 \left( \beta^4,\beta,1 \right)
\end{equation}
and provides the mixing matrix
\begin{equation}
\mathcal{C}_{15}: \quad
|U|^2 =
\left(
\begin{array}{ccc}
 \frac{1}{8} \left(3+\frac{1}{\sqrt{5}}-\sqrt{6
   +\frac{6}{\sqrt{5}}}\right) & \frac{1}{8}
   \left(3+\frac{1}{\sqrt{5}}+\sqrt{6
   +\frac{6}{\sqrt{5}}}\right) & \frac{1}{20}
   \left(5-\sqrt{5}\right) \\
 \frac{1}{8} \left(3+\frac{1}{\sqrt{5}}+\sqrt{6
   +\frac{6}{\sqrt{5}}}\right) &
 \frac{1}{8} \left(3+\frac{1}{\sqrt{5}}-\sqrt{6
   +\frac{6}{\sqrt{5}}}\right) &
   \frac{1}{20} \left(5-\sqrt{5}\right) \\
 \frac{1}{20} \left(5-\sqrt{5}\right) & \frac{1}{20}
   \left(5-\sqrt{5}\right) & \frac{1}{10} \left(5+\sqrt{5}\right)
\end{array}
\right).
\end{equation}

According to equation~(\ref{external-3C}) there are two more subcases
with eigenvalues 
\begin{equation}
\kappa_3\, \omega^2 \left( \beta^4,\beta,1 \right)
\quad \mbox{and} \quad
-\kappa_2\,\omega \left( \beta^4,\beta,1 \right),
\end{equation}
respectively. It turns out
that in both subcases the mixing matrix is that of
$\mathcal{C}_{15}$. Since by choice of $\kappa_3$ and $\kappa_2$,
respectively, we can even match the eigenvalues of $T$ in
equation~(\ref{eigen-3C-2}), we do not count these two subcases as
separate solutions.

\subsection{Form 3D}
According to appendix~\ref{form 3D} there are the four solutions
\begin{equation}
\hat\kappa = 
\left( -\omega,\omega^2, 1 \right) \kappa_3, \;
\diag\left(\omega,-\omega^2,1\right)\kappa_3, \;
\diag\left(\omega,\omega^2,1\right)\kappa_3, \;
\diag\left(\omega,1,-\omega^2\right)\kappa_2,
\end{equation}
with arbitrary roots of unity $\kappa_2$, $\kappa_3$,
for the external phases.
\paragraph{First subcase:}
We begin with
\begin{align}
T & =\left(\begin{array}{ccc}
-\frac{\sqrt{5}-1}{4} & -\frac{\sqrt{5}+1}{4} & \frac{1}{2}\\
\frac{1}{2} & \frac{\sqrt{5}-1}{4} & \frac{\sqrt{5}+1}{4}\\
-\frac{\sqrt{5}+1}{4} & \frac{1}{2} & \frac{\sqrt{5}-1}{4}
\end{array}\right)
\textrm{diag}\left(-\omega,\omega^{2},1\right)\kappa_{3},
\end{align}
which has the eigenvalues 
\begin{equation}
\left(\lambda^{(0)}_1, \lambda^{(0)}_2, \lambda^{(0)}_3 \right) = 
-\kappa_3 \left( \omega, 1, \omega^2 \right)
\end{equation}
and provides the mixing matrix
\begin{equation}
\mathcal{C}_{16}: \quad
|U|^2 =
\left(
\begin{array}{ccc}
 \frac{1}{6} \left(3-\sqrt{5}\right) & \frac{1}{12}
   \left(3+\sqrt{5}\right) & \frac{1}{12} \left(3+\sqrt{5}\right)
   \\
 \frac{1}{12} \left(3+\sqrt{5}\right) & \frac{1}{12}
   \left(3+\sqrt{5}\right) & \frac{1}{6} \left(3-\sqrt{5}\right) \\
 \frac{1}{12} \left(3+\sqrt{5}\right) & \frac{1}{6}
   \left(3-\sqrt{5}\right) & \frac{1}{12} \left(3+\sqrt{5}\right)
   \\
\end{array}
\right).
\end{equation}
\paragraph{Second subcase:}
Next we consider
\begin{align}
T & =\left(\begin{array}{ccc}
-\frac{\sqrt{5}-1}{4} & -\frac{\sqrt{5}+1}{4} & \frac{1}{2}\\
\frac{1}{2} & \frac{\sqrt{5}-1}{4} & \frac{\sqrt{5}+1}{4}\\
-\frac{\sqrt{5}+1}{4} & \frac{1}{2} & \frac{\sqrt{5}-1}{4}
\end{array}\right)
\textrm{diag}\left(\omega,-\omega^{2},1\right)\kappa_{3}.
\end{align}
Its eigenvalues are given by
\begin{equation}
\left(\lambda^{(0)}_1, \lambda^{(0)}_2, \lambda^{(0)}_3 \right) = 
-\kappa_3 \left( \beta^3, \beta^2, 1 \right)
\end{equation}
and the corresponding mixing matrix is
\begin{equation}
\mathcal{C}_{17}: \quad |U|^2 =
\left(
\begin{array}{ccc}
 \frac{1}{8}
   \left(3-\frac{1}{\sqrt{5}}-\sqrt{6-\frac{6}{\sqrt{5}}}\right) &
   \frac{1}{8}
   \left(3-\frac{1}{\sqrt{5}}+\sqrt{6-\frac{6}{\sqrt{5}}}\right) &
   \frac{1}{20} \left(5+\sqrt{5}\right) \\
 \frac{1}{8}
   \left(3-\frac{1}{\sqrt{5}}+\sqrt{6-\frac{6}{\sqrt{5}}}\right) &
   \frac{1}{8}
   \left(3-\frac{1}{\sqrt{5}}-\sqrt{6-\frac{6}{\sqrt{5}}}\right) &
   \frac{1}{20} \left(5+\sqrt{5}\right) \\
 \frac{1}{20} \left(5+\sqrt{5}\right) & \frac{1}{20}
   \left(5+\sqrt{5}\right) & \frac{1}{10} \left(5-\sqrt{5}\right)
   \\
\end{array}
\right).
\end{equation}

The third and fourth subcase have the eigenvalues
\begin{equation}
\kappa_3\,\omega \left( \beta^3, \beta^2, 1 \right)
\quad \mbox{and} \quad
-\kappa_2 \left( \beta^3, \beta^2, 1 \right),
\end{equation}
respectively. Thus they are those of the second subcase multiplied by
a phase factor. Since they also reproduce the mixing matrix of the
second subcase, we do not consider them further.

\subsection{Form 5A}
\label{two-flavour}
If $T$ is of form~5A, then obviously we have two-flavour mixing,
which is completely off from realistic lepton mixing. 
There is a rather large number of such two-flavour mixing solutions, namely
complete solutions $\mathcal{C}_{18}$--$\mathcal{C}_{29}$, and 
partial solutions 
$\mathcal{P}_3$--$\mathcal{P}_{14}$ where one row of $|U|^2$ is determined.
Since these solutions are not relevant from the physics point of view,
we defer them to appendix~\ref{2f}, where we list them for completeness.

\subsection{Form 5B}
We are considering here the matrix
\begin{equation}
T =
\left(\begin{array}{ccc}
0 & 0 & 1\\
\cos\theta & \sin\theta & 0\\
-\sin\theta & \cos\theta & 0
\end{array}\right)
\hat\kappa,
\end{equation}
where $e^{i\theta}$ is a 
root of unity. 
Denoting the eigenvalues of $T$ by
$\lambda_j$ and those of $TS_1$, with $S_1$ defined in equation~(\ref{SSS}), 
by $\lambda'_j$, we find, by taking the determinants and traces,
\begin{equation}\label{system}
\lambda_1\lambda_2\lambda_3 = \lambda'_1\lambda'_2\lambda'_3,
\quad
\sin\theta\, \kappa_2 = \lambda_1+\lambda_2+\lambda_3,
\quad
-\sin\theta\, \kappa_2 = \lambda'_1+\lambda'_2+\lambda'_3.
\end{equation}
Thus we obtain a vanishing sum of six roots of unity
\begin{equation}
\lambda_1 + \lambda_2 + \lambda_3 + 
\lambda'_1 + \lambda'_2 + \lambda'_3 = 0.
\end{equation}
According to theorem~\ref{TCJ}, the vanishing sum over six roots of unity
could be similar to sum~b), but this case can be excluded by the first
relation of equation~(\ref{system}). Then, theorem~\ref{TCJ} allows for 
the following two solutions of this equality:
\begin{subequations}
\begin{align}
\label{5B-1}
&\delta_1(1-1)+ \delta_2(1-1) + \delta_3(1-1) = 0,\\
\label{5B-2}
&\sigma_1(1+\omega+\omega^2) + \sigma_2(1+\omega+\omega^2) = 0,
\end{align}
\end{subequations}
with arbitrary roots of unity 
$\delta_k$ ($k=1,2,3$) and $\sigma_l$ ($l=1,2$).
\paragraph{First subcase:}
First we discuss solution~(\ref{5B-1}). The assignment
\begin{equation}
(\lambda_1, \lambda_2, \lambda_3) = (\delta_1,\delta_2,\delta_3),
\quad
(\lambda'_1, \lambda'_2, \lambda'_3) = (-\delta_1,-\delta_2,-\delta_3)
\end{equation}
leads to a contradiction with the first relation in equation~(\ref{system}).
So without loss of generality we are left with
\begin{equation}
(\lambda_1, \lambda_2, \lambda_3) = (\delta_1,\delta_2,-\delta_2),
\quad
(\lambda'_1, \lambda'_2, \lambda'_3) = (-\delta_1,\delta_3,-\delta_3).
\end{equation}
Since 
$\sum_j \lambda_j = -\sum_j \lambda'_j = \delta_1$, 
this first subcase yields $\sin^2\theta = 1$. 
Choosing without loss of generality
$\sin\theta = 1$, we end up with the two-flavour mixing case
\begin{equation}
T = \left( \begin{array}{ccc}
0 & 0 & \kappa_3 \\ 0 & \kappa_2 & 0 \\ -\kappa_1 & 0 & 0 
\end{array} \right)
\end{equation}
with a mixing angle of $45^\circ$.
Note that $\kappa_2$ and $\kappa_1\kappa_3$ are roots of unity. Since
two-flavour mixing is dealt with in subsection~\ref{two-flavour},
we do not assign an extra solution tag to this.
\paragraph{Second subcase:}
Now we discuss solution~(\ref{5B-2}). If we make the assignment
\begin{equation}\label{as}
(\lambda_1, \lambda_2, \lambda_3) = 
\sigma_1( 1,\omega,\omega^2),
\quad
(\lambda'_1, \lambda'_2, \lambda'_3) = 
\sigma_2( 1,\omega,\omega^2),
\end{equation}
it immediately follows that 
$\sum_j \lambda_j = \sum_j \lambda'_j  = 0$ and, therefore,
$\sin\theta = 0$. Taking without loss of
generality $\cos\theta = 1$, we obtain
\begin{equation}
T = \left( \begin{array}{ccc}
0 & 0 & \kappa_3 \\ \kappa_1 & 0 & 0 \\ 0 & \kappa_2 & 0 
\end{array} \right)
\end{equation}
and the mixing matrix
\begin{equation}
\mathcal{C}_{30}: \quad |U|^2 = 
\frac{1}{3} \left( \begin{array}{ccc}
1 & 1 & 1 \\ 1 & 1 & 1 \\1 & 1 & 1 
\end{array} \right).
\end{equation}
Note that here the product $\kappa_1\kappa_2\kappa_3$ must be a root of
unity. 

It remains to discuss possibilities of assignment other than that of
equation~(\ref{as}). Without loss of generality, we can confine ourselves to
\begin{equation}\label{as1}
(\lambda_1, \lambda_2, \lambda_3) = 
\left( \sigma_1\omega^i, \sigma_1\omega^k, \sigma_2\omega^p \right),
\quad
(\lambda'_1, \lambda'_2, \lambda'_3) = 
\left( \sigma_2\omega^m, \sigma_2\omega^n, \sigma_1\omega^l \right),
\end{equation}
with $i=1$, $k=2$, $l=3$ and permutations thereof and
$m=1$, $n=2$, $p=3$ and permutations thereof.
With this assignment, 
the first relation in equation~(\ref{system}) requires 
\begin{equation}
\sigma_1^2 \sigma_2 \omega^{i+k+p} = 
\sigma_1 \sigma_2^2 \omega^{m+n+l}.
\end{equation}
Because of $\omega^{i+k+l} = \omega^{m+n+p} = 1$,
$\omega^{2l} = \omega^{-l}$ and $\omega^{2p} = \omega^{-p}$, we arrive at
\begin{equation}
\sigma_1 \omega^l = \sigma_2 \omega^p
\quad \Rightarrow \quad
\left\{
\begin{array}{ccc}
(\lambda_1, \lambda_2, \lambda_3) &=&
\sigma_1 \left( \omega^i, \omega^k, \omega^l \right),
\\
(\lambda'_1, \lambda'_2, \lambda'_3) &=& 
\sigma_2 \left( \omega^m, \omega^n, \omega^p \right).
\end{array}
\right.
\end{equation}
Therefore, also for the general assignment in equation~(\ref{as1}) we obtain
$\sum_j \lambda_j = \sum_j \lambda'_j  = 0$ and we are lead back to
solution $\mathcal{C}_{30}$. 

\section{Combining two $\mathcal{P}$-type solutions}
\label{combining}
Here we discuss the cases where two generators $T_1,\,T_2 \in G_\ell$ are
necessary in order to fully determine the mixing matrix $U$. This 
happens if each $T_j$ has a twofold degenerate eigenvalue---see
equation~(\ref{T1T2}):
\begin{equation}
T_1 = U^\dagger \hat T_1 U,
\quad
T_2 = U^\dagger \hat T_2 U
\quad \mbox{with} \quad 
\hat T_1 = \diag\left(\lambda'_1,\lambda_1,\lambda_1 \right),
\quad
\hat T_2 = \diag\left(\lambda_2,\lambda'_2,\lambda_2 \right)
\end{equation}
and roots of unity $\lambda_1$, $\lambda'_1$, $\lambda_2$, $\lambda'_2$
fulfilling
\begin{equation}  
\lambda_1 \neq \lambda'_1, 
\quad
\lambda_2 \neq \lambda'_2.
\end{equation}

Actually, we can only hope to discover a new $|U|^2$, not already covered by
the previous section, if \textit{none} of the elements of $G_\ell$ is
non-degenerate. Let us examine this point further. The product $T_1T_2$
is degenerate if and only if  
$\lambda_1\lambda'_2 = \lambda'_1\lambda_2$ whence if follows that
\begin{equation}
\hat T_2 = \frac{\lambda_2}{\lambda_1}\,
\diag \left( \lambda_1, \lambda'_1, \lambda_1 \right). 
\end{equation}
Next we consider
\begin{equation}
\left( \hat T_1 \right)^2 \hat T_2 = \lambda_1^2 \lambda_2 \,
\diag\left( \left( \frac{\lambda'_1}{\lambda_1} \right)^2,
\frac{\lambda'_1}{\lambda_1}, 1 \right).
\end{equation}
The only possibility to avoid non-degeneracy in this matrix is 
\begin{equation}
\lambda'_1 = -\lambda_1
\end{equation}
and we end up with
\begin{equation}
\hat T_1 = \lambda'_1 S_1, \quad 
\hat T_2 = -\lambda_2 S_2,
\end{equation}
with the diagonal sign matrices $S_j$ defined in equation~(\ref{SSS}).
Finally, without changing the finiteness of $G$ we can remove the phase
factors $\lambda'_1$ and $-\lambda_2$.
Thus, the only case we have to investigate is when $G_\ell$ is a Klein
four-group. 

According to the discussion in section~\ref{basic forms},
we know that for every pair of
indices $i,j$ the matrix $U^\dagger S_i U S_j$ has eigenvalues 
$1$, $\sigma^{(ij)}$, $\left( \sigma^{(ij)} \right)^*$, where 
$\sigma^{(ij)}$ is a root of unity. 
So we have the same situation as in section~\ref{basic forms}, with $T$
replaced by $U$. Therefore, instead of $|T|$, $|U|$ itself must be of one of the
five basic forms and we have the solutions
\begin{eqnarray}
\mathcal{CD}_1: &&
|U|^2 = \left( \begin{array}{ccc}
\frac{1}{2} & \frac{1}{4} & \frac{1}{4} \\
\frac{1}{4} & \frac{1}{2} & \frac{1}{4} \\
\frac{1}{4} & \frac{1}{4} & \frac{1}{2} 
\end{array} \right),
\\
\mathcal{CD}_2: &&
|U|^2 = \left( \begin{array}{ccc}
0 & \frac{1}{2} & \frac{1}{2} \\
\frac{1}{2} & \frac{1}{4} & \frac{1}{4} \\
\frac{1}{2} & \frac{1}{4} & \frac{1}{4} 
\end{array} \right),
\\
\mathcal{CD}_3: &&
|U|^2 = \left( \begin{array}{ccc}
\frac{1}{4} & \frac{1}{8}(3-\sqrt{5}) & \frac{1}{8}(3+\sqrt{5}) \\
\frac{1}{8}(3+\sqrt{5}) & \frac{1}{4} & \frac{1}{8}(3-\sqrt{5}) \\
\frac{1}{8}(3-\sqrt{5}) & \frac{1}{8}(3+\sqrt{5}) & \frac{1}{4} 
\end{array} \right),
\\
\mathcal{CD}_4: &&
|U|^2 = \left( \begin{array}{ccc}
\frac{1}{2} & \frac{1}{4} & \frac{1}{4} \\
\frac{1}{4} & \frac{1}{8}(3-\sqrt{5}) & \frac{1}{8}(3+\sqrt{5}) \\
\frac{1}{4} & \frac{1}{8}(3+\sqrt{5}) & \frac{1}{8}(3-\sqrt{5})
\end{array} \right),
\\
\mathcal{CD}_5: &&
|U|^2 = \left( \begin{array}{ccc}
1 & 0 & 0 \\
0 & \cos^2\theta & \sin^2\theta \\
0 & \sin^2\theta & \cos^2\theta
\end{array} \right).
\end{eqnarray}
In the last form, $e^{i\theta}$ must be a root of unity.

Scanning the genuine three-flavour mixing solutions of section~\ref{external}
we find that the $|U|^2$ of $\mathcal{C}_1$ is
of form~2 of equation~(\ref{basic}). 
Naturally, the $|U|^2$ of all solutions
with $T$ of form~5A are of form~5 as well.
But a $|U|^2$ of form~1, 3 or~4 of equation~(\ref{basic}) 
does not occur in the previous section. 
Therefore, $\mathcal{CD}_1$, $\mathcal{CD}_3$ and $\mathcal{CD}_4$ 
provide new mixing matrices.
We note also that, even though 
the previous argument is not enough to
prove that the $\mathcal{CD}_i$ solutions lead to finite groups, it
can be easily checked by explicit construction of the groups generated
by $S_1$, $S_2$, $T_1$, and $T_2$ that they are finite (as are the ones
associated with
the $\mathcal{C}_i$ and
$\mathcal{P}_i$ solutions).

\section{Conclusions}
\label{concl}
In this paper, 
assuming that neutrinos are Majorana particles, 
we have presented a complete classification of all 
possible lepton mixing matrices $U$ such that $|U|^2$, defined as the matrix 
with the elements $|U_{ij}|^2$, is completely determined
by residual symmetries. 
In this model-independent framework, the entries
of $|U|^2$ are obtained as pure numbers, determined by group-theoretical
considerations. Evidently, the resulting matrices $|U|^2$ are independent of
any parameter of a possible underlying theory and of the
lepton masses. In our analysis we used the 
\textit{ad hoc} assumption that the flavour group $G$ is finite, 
which allowed us to use suitable theorems related to sums of roots of unity.

We have found 22 solutions associated with a genuine three-flavour
mixing matrix $U$, \textit{i.e.}\ where all three flavours are mixed:
$\mathcal{C}_{1}-\mathcal{C}_{17}$, $\mathcal{C}_{30}$ 
and $\mathcal{CD}_{1}-\mathcal{CD}_{4}$.
Since four pairs of solutions produce the same 
mixing matrix---$\left(\mathcal{C}_{1},\mathcal{CD}_{2}\right)$,
$\left(\mathcal{C}_{6},\mathcal{C}_{15}\right)$, 
$\left(\mathcal{C}_{8},\mathcal{C}_{17}\right)$ and 
$\left(\mathcal{C}_{11},\mathcal{C}_{13}\right)$---there 
are only 17 sporadic $\left|U\right|^{2}$ patterns and
one infinite series denoted by $\mathcal{C}_{2}$. The precise
groups formed by the generators of the residual symmetries $G_{\nu}$
and $G_{\ell}$ will depend on one or more phases which are free parameters;
for example, the $T$ generator can always be multiplied by an arbitrary
root of unity. Nevertheless, with the help of GAP~\cite{GAP}, 
one can reach the conclusion 
that for the 17 sporadic $\left|U\right|^{2}$ patterns, the minimal
groups\footnote{These are the smallest groups possible. In addition, 
for the cases $\mathcal{C}_{3-17}$, $\mathcal{CD}_{1}$, $\mathcal{CD}_{3}$
and $\mathcal{CD}_{4}$ it can be shown that the full flavour group
must always contain these as subgroups.} 
are 
\begin{itemize}
\item
$S_{4}$ for $\mathcal{C}_{1}/\mathcal{CD}_{2}$,
\item
$PSL\left(2,7\right)$ for 
$\mathcal{C}_{3}$,
$\mathcal{C}_{4}$, 
$\mathcal{C}_{5}$,
$\mathcal{CD}_{1}$, 
\item
$\Sigma\left(360\times3\right)$ for 
$\mathcal{C}_{6}/\mathcal{C}_{15}$,
$\mathcal{C}_{7}$, 
$\mathcal{C}_{8}/\mathcal{C}_{17}$,
$\mathcal{C}_{9}$, 
$\mathcal{C}_{10}$,
$\mathcal{C}_{14}$,
$\mathcal{C}_{16}$,
$\mathcal{CD}_{4}$,
\item
$A_{5}$ for 
$\mathcal{C}_{11}/\mathcal{C}_{13}$,
$\mathcal{C}_{12}$, 
$\mathcal{CD}_{3}$,
\item
$A_{4}$ for 
$\mathcal{C}_{30}$. 
\end{itemize}
At this point a comparison of our results with those in the
literature is in order. 
The list of mixing matrices in~\cite{toorop2} induced by
$PSL(2,7)$ agrees exactly with our list above and the same holds true
for those induced by $A_5$; moreover, this reference contains, among
some mixing matrices of the series $\mathcal{C}_2$, also
the cases $\mathcal{C}_1/\mathcal{CD}_{2}$ and 
$\mathcal{C}_{30}$ with $S_4$ and $A_4$,
respectively. Finally, the list of mixing matrices above associated with 
$\Sigma\left(360\times3\right)$ agrees with those given
in~\cite{hagedorn} 
in table~5 (with labels XIa/XIb) 
and table~6 (with
labels beginning with~K). 
So we conclude that all our sporadic cases have already been discussed in
the literature.

For
the phenomenologically interesting infinite series
denoted by $\mathcal{C}_{2}$ where
\begin{align}\label{C2mixpattern}
\left|U\right|^{2} & =\frac{1}{3}\left(\begin{array}{ccc}
1 & 1+\textrm{Re}\,\sigma & 1-\textrm{Re}\,\sigma\\
1 & 1+\textrm{Re}\left(\omega\sigma\right) &
1-\textrm{Re}\left(\omega\sigma\right) \\
1 & 1+\textrm{Re}\left(\omega^{2}\sigma\right) & 
1-\textrm{Re}\left(\omega^{2}\sigma\right)
\end{array}\right)
\end{align}
for some root of unity $\sigma=\exp\left(2i\pi p/n\right)$ with $p$
coprime to $n$, 
we have discussed minimal groups and their generators in
appendix~\ref{minimal}. Our result is
\begin{itemize}
\item
$\Delta(6m^2)$ with $m = \textrm{lcm}(6,n)/3$ when $9 \nmid n$,
\item
$\left( \mathbbm{Z}_m \times \zz_{m/3} \right) \rtimes S_3$ 
with $m = \textrm{lcm}(2,n)$ when $9 \mid n$.
\end{itemize}

\begin{figure}
\begin{centering}
\includegraphics[scale=0.115]{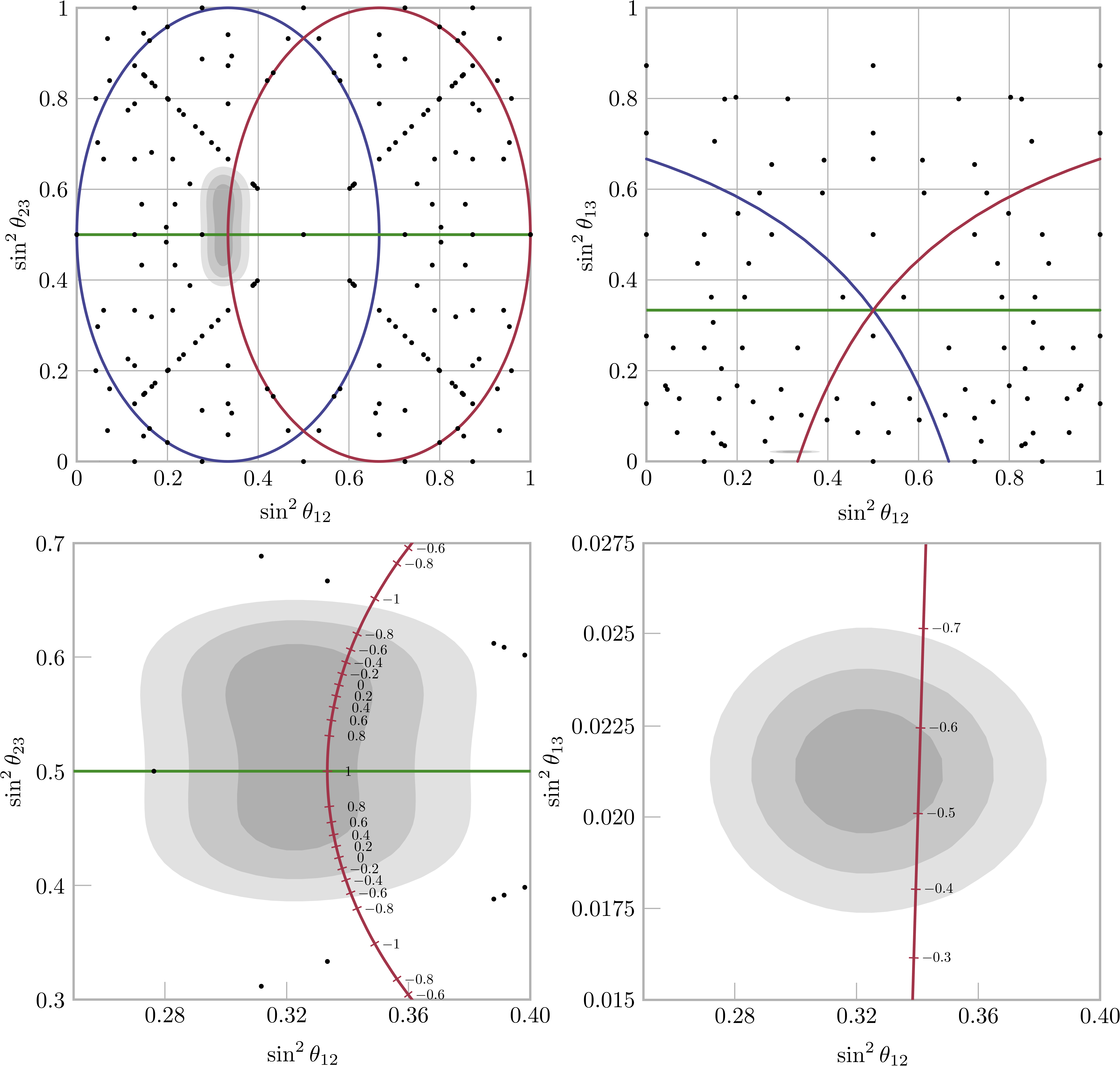}
\par\end{centering}
\caption{\label{fig:Numerics}
Plots with the values of the three lepton mixing
angles in the standard parameterisation, 
for all possible mixing patterns
involving genuine three-flavour mixing (including permutations). 
For each case pertaining to the sporadic mixing matrices, 
a dot has been
placed in these plots (some of them are superimposed).
The three lines (red, blue, green) correspond to different column
permutations of the series of patterns obtained 
from equation~\eqref{C2mixpattern}
by varying the root of unity $\sigma$. 
The lower plots are magnifications of the ones above in the
physically interesting region, with the numbers on the red curves indicating
values of $\mathrm{Re}\,\left(\sigma^6\right)$.
The gray regions mark the 1, 2 and 3 sigma ranges of the mixing angles
calculated by Forero \textit{et al.}, arXiv:1405.7540 in~\cite{fits} for the
normal mass hierarchy.
}
\end{figure}

Once all row and column permutations are considered, 
it turns out that in the four-dimensional space of the 
quadruples $\left\{ \sin^{2}\theta_{12},\sin^{2}\theta_{23},
\sin^{2}\theta_{13},\cos^2\delta\right\}$
the sporadic
$\left|U\right|^{2}$ patterns yield 228 distinct 
values---see figure~\ref{fig:Numerics} 
for a graphical representation
of the mixing angles.\footnote{As for the Dirac-type phase,
$\cos^{2}\delta$ can take a total of 
34 different values associated with the sporadic patterns.
However, note that in some cases the absolute values of the entries of 
the lepton mixing matrix do
not depend on the Dirac phase $\delta$. This happens when 
$\sin\theta_{12}\cos\theta_{12}\sin\theta_{23}\cos\theta_{13}\sin\theta_{13}=0$,
in which case there is at least one null entry in $U$, and consequently
there is no CP-violation associated to $\delta$.}
All of them are excluded at 
3~sigma 
by current neutrino oscillation
data. If one leaves out $\cos^2\delta$, there are still 212
distinct points in the space of the triples
$\left\{ \sin^{2}\theta_{12},\sin^{2}\theta_{23},\sin^{2}\theta_{13}\right\}$.

This leaves us with the infinite series of mixing patterns given
in equation~(\ref{C2mixpattern}) as the only phenomenologically viable
case; taking the 
3~sigma 
range of $\sin^2\theta_{13}$ of Forero
\textit{et al.}, arXiv:1405.7540
in~\cite{fits},
this translates into 
$-0.69\lesssim\textrm{Re}\left(\sigma^{6}\right)\lesssim-0.37$
for roots of unity $\sigma$.
Suitable values of $n$ include, sorted by group size, 
\begin{itemize}
\item
$n=9,\,18$ with $m=18$, 
$G = \left(\zz_{18} \times \zz_6 \right) \rtimes S_3$ and
$\mbox{order}(G) = 648$,
\item
$n=11,\,22,\,33,\,66$ with $m=22$, 
$G = \Delta(6 \times 22^2)$ and $\mbox{order}(G) = 2904$,
\item
$n=28,\,84$ with $m=28$, 
$G = \Delta(6 \times 28^2)$ and $\mbox{order}(G) = 4704$,
\item
$n=32,\,96$ with $m=32$, 
$G = \Delta(6 \times 32^2)$ and $\mbox{order}(G) = 6144$,
\end{itemize}
among others.
However, we note that in the past, with less precise neutrino
oscillation data,  values of $n$ equal to 5 and 16 were valid as
well---see for instance~\cite{holthausen,king1,GL2013}, 
see also~\cite{toorop1,toorop2} for earlier references.
One can easily check that equation~(\ref{C2mixpattern}) gives a
trivial Dirac-type phase $\delta$.
In figure~\ref{fig:Numerics}, the case $\mathcal{C}_2$ is represented
by three lines which are obtained by varying the root of unity~$\sigma$.
The colours correspond to different column permutations:
\begin{itemize}
\item
red: $\cos^2\theta_{13} \sin^2\theta_{12} = 1/3$,
\item
blue: $\cos^2\theta_{13} \cos^2\theta_{12} = 1/3$,
\item
green: $\sin^2\theta_{13} = 1/3$.
\end{itemize}
%%%%
From the two lower plots in figure~1 we can read off that, if
$\mathcal{C}_2$ is realized in nature,
then $\sin^2 \theta_{23}$ must be quite far from $0.5$.

For completeness, we have also presented 
in appendix~\ref{2f}
the solutions where $U$ is
block-diagonal, \textit{i.e.} one flavour does not mix with the other two.
As a byproduct, our analysis has yielded all instances where one
row of $|U|^2$ is fixed.\footnote{Our analysis does \emph{not} produce the
  cases where a column in $U$ is determined.} Actually, if one takes only the
cases without a zero in the row, then there are only two cases,
$\mathcal{P}_1$ and $\mathcal{P}_2$, which have for instance been found
earlier in~\cite{lavoura} with the help of GAP.

The result of our analysis is a bit sobering. Taking up the position that
residual symmetries should be capable of reproducing the results of
fitting $U$ to the data~\cite{fits}, there is only 
the series $\mathcal{C}_2$ which can do the job; some mixing matrices 
of this series were discovered earlier in~\cite{holthausen} by using GAP. 
For the sporadic mixing matrices, the main obstacle is to reproduce 
the small quantity $\sin^2\theta_{13} \simeq 0.021$ 
%% RF: this value comes from the updated data shown in 
%% http://tinyurl.com/q9aln98 which will, I was told, go into
%% the final version of arXiv:1405.7540
obtained from the oscillation data. On the one hand, 
some sporadic mixing matrices $|U|^2$ have a zero entry, which is
ruled out as it would imply $\sin^2\theta_{13} = 0$. On the other hand,
the smallest non-zero entry of all sporadic $|U|^2$ is
$(5-\sqrt{21})/12 \simeq 0.035$ occurring in $\mathcal{C}_5$,
which is significantly larger than the physical $\sin^2\theta_{13}$.

Maybe the result of this paper gives more credibility to models where
mixing angles and phases are related to mass ratios. Such models can be based
on Abelian and non-Abelian flavour groups. Texture zeros are one
possibility---these are, in effect, practically synonymous with Abelian
symmetries~\cite{tanimoto}. One might also try to relax the assumptions of our
analysis, for example allowing one neutrino mass to be 
zero~\cite{joshipura},
adopting the idea that neutrinos are Dirac particles,\footnote{Note
  that, for neutrinos with Dirac nature, $G_\nu$ will in general
  not be the Klein 
  four-group. However, in this is nontheless the case---see~\cite{king},
  then the analysis of this paper still applies.} or giving up the
\textit{ad hoc} assumption of a finite flavour group $G$.
In particular, in the latter case a new mathematical approach would be
necessary. 
It is also worth mentioning that since neither of the 
residual symmetries ($G_\ell$ and $G_\nu$) are symmetries of the full 
Lagrangian, in concrete models there will be radiative corrections to
the mixing patterns presented here.

\vspace{5mm}

\noindent
\textbf{Acknowledgments:} W.G.\ is very grateful to Christoph Baxa for an
illuminating discussion about roots of unity and to Patrick O.\ Ludl
for constant support and valuable help with group theory and GAP. 
We also thank Mariam T\'ortola for providing the up-to-date ranges of
the mixing angles in figure~1, which include data presented at the
Neutrino 2014 conference.
The work of R.F. was supported by the Spanish 
\textit{Ministerio de Econom\'ia y Competitividad} 
through the grants FPA2011-22975 and Multidark Consolider
CSD2009-00064, 
by the \textit{Generalitat Valenciana} through the 
grant PROMETEO/2009/091, and by the 
Portuguese \textit{Funda\c{c}\~ao para a Ci\^encia e a Tecnologia} 
through the grants CERN/FP/123580/2011 and EXPL/FIS-NUC/0460/2013.

\newpage
\appendix
\setcounter{equation}{0}
\renewcommand{\theequation}{A\arabic{equation}}
\section{The possible eigenvalues of $Y^{(ij)}$}
\label{possible eigenvalues-app}
The relations in equation~(\ref{lambda}) have the generic form
\begin{equation}\label{lambda1}
\sum_{k=1}^3 \left( \lambda_k + \lambda_k^* \right) + 2 = 0,
\end{equation}
with roots of unity $\lambda_1$, $\lambda_2$, $\lambda_3$.
This equation can be conceived as a vanishing sum over eight roots of
unity such that the trivial root, 1, occurs in the sum and also
the complex conjugate of every root.

A suitable theorem to deal with such a vanishing sum is theorem~6
of~\cite{conway}, which we have reproduced in this work as 
theorem~\ref{TCJ} in section~\ref{tools}. 
This mathematical result deals with all vanishing sums of at most 9 roots 
of unity, which we have labeled from a) to h). In each sum, all roots 
are different. Let us now go through each of them. 
Sums g) and h) have each 9 roots, so these sums are irrelevant
for a solution of equation~(\ref{lambda1}). On the other hand, the
sums~e) and~f) have 8 distinct roots of unity each, 
but in equation~(\ref{lambda1}) the trivial root~1 
occurs twice, so sums~e) and~f) are also irrelevant for our purpose.
Sums~c) and~d) have 7 roots each, therefore,
adding a further root of unity does not lead to a vanishing
sum. 

Sum~b) has six roots, thus with this vanishing sum we can build a
vanishing sum of eight roots of the form
\begin{equation}\label{s1}
\delta_1 \left( 
- \omega - \omega^2 + \beta + \beta^2 +\beta^3 +\beta^4 \right) + 
\delta_2 \left( 1 -1 \right) = 0,
\end{equation}
where $\omega = e^{2\pi i/3}$, $\beta = e^{2\pi i/5}$,
and $\delta_1$ and $\delta_2$ are further roots of unity yet to be
specified. The trivial root~1 has to occur in both partial sums in
equation~(\ref{s1}). Therefore, without loss of generality, 
$\delta_2 = 1$. Let us first assume $-\delta_1 \omega = 1$. Then
equation~(\ref{s1}) gives 
\begin{equation}
1 + \omega - \omega^2 \left( \beta + \beta^2 +\beta^3 +\beta^4 \right) + 
1 - 1 = 0.
\end{equation}
The root $-1$ should occur two times in this sum, but this is not the
case and we can discard $-\delta_1 \omega = 1$. We can deal similarly
with  $-\delta_1 \omega^2 = 1$. Next we assume $\delta_1 \beta = 1$,
which leads to
\begin{equation}
-\omega \beta^4 - \omega^2 \beta^4 + 1 + \beta + \beta^2 + \beta^3 + 1
-1 = 0.
\end{equation}
Again, $-1$ should occur twice, but it does not. Analogously, we can
exclude all other possibilities like 
$\delta_1 \beta^2 = 1$ and so on.
In summary, we have found that sum~b) of Conway and Jones cannot be part
of the solution of equation~(\ref{lambda1}).

With sum~a) we have the vanishing sum of eight roots of unity
\begin{equation}\label{s2}
\delta_1 \left( 1 + \beta + \beta^2 +\beta^3 +\beta^4 \right) + 
\delta_2 \left( 1 + \omega + \omega^2 \right) = 0.
\end{equation}
Since~1 must occur in both partial sums, we can choose without loss of
generality $\delta_1 = \delta_2 = 1$. Indeed, this represents a
solution of equation~(\ref{lambda1}) with 
$\lambda_1 = \omega$, $\lambda_2 = \beta$, $\lambda_3 = \beta^2$.

According to the theorem of Conway and Jones, the remaining solutions
of equation~(\ref{lambda1}) can only contain powers of $\omega$ and
$-1$. One possibility is
\begin{equation}\label{s3}
\delta_1 \left( 1 + \omega + \omega^2 \right) + 
\delta_2 \left( 1 + \omega + \omega^2 \right) + 
\delta_3 \left( 1 - 1 \right) = 0.
\end{equation}
Suppose the trivial root~1 occurs in each of the partial sums
with $\omega$. 
Then, without loss of generality $\delta_1 = \delta_2 =1$. This
implies $\delta_3^* = -\delta_3$ or $\delta_3 = i$. Thus we have
a second solution of equation~(\ref{lambda1}):
$\lambda_1 = i$, $\lambda_2 = \lambda_3 = \omega$.
On the other hand, we can also assume that the trivial root~1 occurs
in the first and the third partial sum of equation~(\ref{s3}).
Without loss of generality, this is achieved with 
$\delta_1 = \delta_3 =1$. Since, in this case, $-1$ occurs in the
third partial sum, it must occur in the second partial sum too. 
Without loss of generality, this implies $\delta_2 = -1$, leading to
another solution
$\lambda_1 = -1$, $\lambda_2 = \omega$, $\lambda_3 = -\omega$.

The last sum that needs to be discussed is of the form
\begin{equation}\label{s4}
\delta_1 \left( 1 - 1 \right) + 
\delta_2 \left( 1 - 1 \right) + 
\delta_3 \left( 1 - 1 \right) +
\delta_4 \left( 1 - 1 \right) = 0.
\end{equation}
Since the trivial root has to occur twice, we choose, without loss of
generality, $\delta_1 = \delta_2 =1$. Then 
$\delta_3 = \delta_4^* \equiv \lambda$ with an undetermined root $\lambda$.
Thus we have found another solution of equation~(\ref{lambda1}):
$\lambda_1 = -1$, $\lambda_2 = \lambda$, $\lambda_3 = -\lambda$.
The solution of the previous paragraph is special case of this one.

In summary, equation~(\ref{lambda1}) has only three solutions, which
are displayed in equation~(\ref{ABC}).

\setcounter{equation}{0}
\renewcommand{\theequation}{B\arabic{equation}}
\section{The matrix elements $t_{ij} = \left| T_{ij} \right|^2$}
\label{basic forms-app}
Here we determine all possible forms of the matrix
$t = \left( t_{ij} \right)$. As we will see, it lies in the nature of
the method that we can determine $t$ only up to permutations,
\textit{i.e.}\ wherever we find a $t$, then also the matrices 
$P_1 t P_2$, where $P_1$ and $P_2$ are $3 \times 3$ permutation
matrices, are admissible. In the present section, whenever we use ``unique''
for $t$, it is always meant unique up to such permutations.

According to equations~(\ref{t-lambda}) and~(\ref{ABC}),
up to permutations, every line and column of $t$ is identical 
to one of the three following possibilities:
\begin{equation}\label{ABCt}
\renewcommand{\arraystretch}{1.2}
\begin{array}{cc}
\left( \frac{1}{2},\, \frac{1}{4},\, \frac{1}{4} \right)
& \mbox{(A)}, \\
\left( \frac{1}{4},\, \frac{3 + \sqrt{5}}{8},\, \frac{3 - \sqrt{5}}{8} \right)
& \mbox{(B)}, \\
\left( 0,\, \frac{1 + \cos\vartheta}{2},\, \frac{1 - \cos\vartheta}{2} \right)
& \mbox{(C)}.
\end{array}
\end{equation}
Note that
\begin{equation}
\left( \frac{\sqrt{5} \pm 1}{4} \right)^2 = 
\frac{3 \pm \sqrt{5}}{8}.
\end{equation}
The angle $\vartheta$ in (C) is an arbitrary rational
angle,\footnote{Since $\vartheta$ comes from a root of unity, it must be
  $2\pi$ times some rational number, but otherwise it is arbitrary.} 
but if we combine for instance
a column of type (C) with a line of type (A) or (B), then $\cos\vartheta$
can assume only four values:
\begin{equation}\label{CCC}
\renewcommand{\arraystretch}{1.2}
\begin{array}{ccc}
\cos\vartheta = 0: &
\left( 0,\, \frac{1}{2},\, \frac{1}{2} \right)
& \mbox{(C$_1$)}, \\
\cos\vartheta = \frac{1}{2}: &
\left( 0,\, \frac{3}{4},\, \frac{1}{4} \right)
& \mbox{(C$_2$)}, \\
\cos\vartheta = \frac{-1+\sqrt{5}}{4}: &
\left( 0,\, \frac{3 + \sqrt{5}}{8},\, \frac{5 - \sqrt{5}}{8} \right)
& \mbox{(C$_3$)}, \\
\cos\vartheta = \frac{-1-\sqrt{5}}{4}: &
\left( 0,\, \frac{3 - \sqrt{5}}{8},\, \frac{5 + \sqrt{5}}{8} \right)
& \mbox{(C$_4$)}.
\end{array}
\end{equation}
Later we will see in one instance that also the case
\begin{equation}
\cos\vartheta = 1: \quad
\left( 0,\, 1,\, 0 \right)
\quad \mbox{(C$_0$)}
\end{equation}
occurs.

In the following we derive the possible forms of $t$, by exhausting 
all possible combinations of (A), (B) and (C). To find all viable
cases, we must take into account that $t$ stems from a unitary matrix
$T$. Firstly, an admissible $t$ must fulfill
\begin{equation}\label{1}
\sum_{k=1}^3 t_{ik} = \sum_{k=1}^3 t_{kj} = 1 \quad \forall\, i,j = 1,2,3.
\end{equation}
Secondly, for such a $t$, one must also check that it is possible to
find phases $\alpha_{ij}$ such that the scheme of complex numbers
$\sqrt{t_{ij}}\, e^{i\alpha_{ij}}$
fulfills the orthogonality relations of a unitary matrix. It was shown
in~\cite{branco} that for this purpose it is sufficient to check the validity
of the inequality
\begin{equation}\label{ineq}
4 t_{11} t_{22} t_{12} t_{21} \geq
\left( 1 - t_{11} - t_{22} - t_{12} - t_{21} + t_{11} t_{22} +
t_{12} t_{21} \right)^2.
\end{equation}

\subsection{First line in $t$ of type (A)}
Assuming the ordering of the first line as in
equation~(\ref{ABCt}), 
the first column can be either of type (A) or (C$_1$).
\paragraph{First column in $t$ of type (A):}
Then the second line can be of type (A) or (B) or (C$_2$). 
If it is of type (A), we obtain
\begin{equation}
\renewcommand{\arraystretch}{1.2}
t = \left(
\begin{array}{ccc}
\frac{1}{2} & \frac{1}{4} & \frac{1}{4} \\
\frac{1}{4} & \frac{1}{2} & \frac{1}{4} \\
\frac{1}{4} & \frac{1}{4} & \frac{1}{2} 
\end{array} \right)
\quad \mbox{(AAA)}.
\end{equation}
This $t$ is compatible with inequality~(\ref{ineq}).
Here and in the following the capital letters to the right of the
matrix indicate the types of the lines.
If the second line is of type (B), we arrive at 
\begin{equation}\label{ABB}
\renewcommand{\arraystretch}{1.2}
t = \left( \begin{array}{ccc}
\frac{1}{2} & \frac{1}{4} & \frac{1}{4} \\
\frac{1}{4} & \frac{3 + \sqrt{5}}{8} & \frac{3 - \sqrt{5}}{8} \\
\frac{1}{4} & \frac{3 - \sqrt{5}}{8} & \frac{3 + \sqrt{5}}{8}
\end{array} \right)
\quad \mbox{(ABB)}.
\end{equation}
This case is also compatible with inequality~(\ref{ineq}).
Finally, if the second line is of type (C$_2$), we find
\begin{equation}
\renewcommand{\arraystretch}{1.2}
t = \left(
\begin{array}{ccc}
\frac{1}{2} & \frac{1}{4} & \frac{1}{4} \\
\frac{1}{4} & \frac{3}{4} & 0           \\
\frac{1}{4} & 0           & \frac{3}{4} 
\end{array} \right)
\quad \mbox{(AC$_2$C$_2$)}.
\end{equation}
This case is \emph{not} in agreement with inequality~(\ref{ineq}).
\paragraph{First column in $t$ of type (C$_1$):}
Let us assume $t_{21} = 1/2$ and $t_{31} = 0$. Then the second line can
be of type (A) or (C$_1$). The first possibility gives,
upon reordering of the lines, 
\begin{equation}
\renewcommand{\arraystretch}{1.2}
t = \left( \begin{array}{ccc}
          0 & \frac{1}{2} & \frac{1}{2} \\
\frac{1}{2} & \frac{1}{4} & \frac{1}{4} \\
\frac{1}{2} & \frac{1}{4} & \frac{1}{4} 
\end{array} \right)
\quad \mbox{(C$_1$AA)}.
\end{equation}
This is a viable case with respect to inequality~(\ref{ineq}).
If we have type (C$_1$) in the second line, then the third line must
be (C$_2$):
\begin{equation}\label{c}
\renewcommand{\arraystretch}{1.2}
t = \left( \begin{array}{ccc}
\frac{1}{2} & \frac{1}{4} & \frac{1}{4} \\
\frac{1}{2} & \frac{1}{2} & 0           \\
          0 & \frac{1}{4} & \frac{3}{4} 
\end{array} \right)
\quad \mbox{(AC$_1$C$_2$)}.
\end{equation}
However, this case does not comply with inequality~(\ref{ineq}).
Now we have exhausted all possibilities with one or more lines of type (A).

\subsection{First line in $t$ of type (B), but no line of type (A)}
Assuming the ordering of (B) as in equation~(\ref{ABCt}), the first
column can be either of type (A) or (B) or (C$_2$).
\paragraph{First column of type (A):}
With the ordering of the first column as $(1/4,1/4,1/2)$, the
third line must be either of type (A) or of type (C$_1$). 
If it is of type (A), then the second line must be of type (B) and we
are lead to a permutation of (ABB) of equation~(\ref{ABB}).
If it is of type (C$_1$), then the column which has the zero must be
of type (C$_3$) or (C$_4$). But then the second line cannot be one of the
types (A), (B), (C). 
So we come to the conclusion that, if the first column is of
type (A), no new case ensues.
\paragraph{First column of type (B):}
With $t_{21} = (3 - \sqrt{5})/8$, the second line can be
(B) or (C$_4$). In the first case we are lead to the viable matrix
\begin{equation}
\renewcommand{\arraystretch}{1.2}
t = \left( \begin{array}{ccc}
\frac{1}{4} & \frac{3 + \sqrt{5}}{8} & \frac{3 - \sqrt{5}}{8} \\
\frac{3 - \sqrt{5}}{8} & \frac{1}{4} & \frac{3 + \sqrt{5}}{8} \\
\frac{3 + \sqrt{5}}{8} & \frac{3 - \sqrt{5}}{8} & \frac{1}{4}
\end{array} \right)
\quad \mbox{(BBB)}.
\end{equation}
If the second line is of type (C$_4$), the only way to fulfill
equation~(\ref{1}) is
\begin{equation}
\renewcommand{\arraystretch}{1.2}
t = \left( \begin{array}{ccc}
\frac{1}{4} & \frac{3 + \sqrt{5}}{8} & \frac{3 - \sqrt{5}}{8} \\
\frac{3 - \sqrt{5}}{8} & 0 & \frac{5 + \sqrt{5}}{8} \\
\frac{3 + \sqrt{5}}{8} & \frac{5 - \sqrt{5}}{8} & 0
\end{array} \right)
\quad \mbox{(BC$_4$C$_3$)}.
\end{equation}
However, this $t$ is not admissible because inequality~(\ref{ineq}) is
not fulfilled.
\paragraph{First column of type (C$_2$):}
Assuming $t_{21} = 3/4$, then the second line must be of type (C$_2$).
In this case we find 
\begin{equation}
\renewcommand{\arraystretch}{1.2}
t = \left( \begin{array}{ccc}
\frac{1}{4} & \frac{3 + \sqrt{5}}{8} & \frac{3 - \sqrt{5}}{8} \\
\frac{3}{4} & \frac{1}{4} & 0 \\
0 & \frac{3 - \sqrt{5}}{8} & \frac{5 + \sqrt{5}}{8}
\end{array} \right)
\quad \mbox{(BC$_2$C$_4$)}.
\end{equation}
and
\begin{equation}
\renewcommand{\arraystretch}{1.2}
t = \left( \begin{array}{ccc}
\frac{1}{4} & \frac{3 + \sqrt{5}}{8} & \frac{3 - \sqrt{5}}{8} \\
\frac{3}{4} & 0 & \frac{1}{4} \\
0 & \frac{5 - \sqrt{5}}{8} & \frac{3 + \sqrt{5}}{8}
\end{array} \right)
\quad \mbox{(BC$_2$C$_3$)}.
\end{equation}
However, both possibilities must be discarded because
inequality~(\ref{ineq}) is not fulfilled. Now we have exhausted all
cases with lines of type (A) and (B). It remains to discuss matrices
$t$ where \emph{all} lines are of type (C).

\subsection{All lines of type (C)}
Let us assume that the first line is of type (C$_0$). Then we find
\begin{equation}\label{basic5-var}
\renewcommand{\arraystretch}{1.2}
t = \left( \begin{array}{ccc}
1 & 0 & 0 \\
0 & \frac{1 + \cos\vartheta}{2} & \frac{1 - \cos\vartheta}{2} \\
0 & \frac{1 - \cos\vartheta}{2} & \frac{1 + \cos\vartheta}{2} 
\end{array} \right)
\quad \mbox{(C$_0$CC)}.
\end{equation}
All $\cos\vartheta$ are in agreement with inequality~(\ref{ineq}).
If no line of type (C$_0$) occurs in $t$, \textit{i.e.}\ 
$\cos\vartheta \neq \pm 1$,
then one quickly finds that
up to permutations the unique possibility is
\begin{equation}
\renewcommand{\arraystretch}{1.2}
t = \left( \begin{array}{ccc}
0 & \frac{1 + \cos\vartheta}{2} & \frac{1 - \cos\vartheta}{2} \\
\frac{1 - \cos\vartheta}{2} & 0 & \frac{1 + \cos\vartheta}{2} \\
\frac{1 + \cos\vartheta}{2} & \frac{1 - \cos\vartheta}{2} & 0
\end{array} \right)
\quad \mbox{(CCC)}.
\end{equation}
However, application of inequality~(\ref{ineq}) gives 
$\cos^2\vartheta = 1$, which we had already excluded.

In summary, equation~(\ref{ABCt}), which we had derived using the
theorem of Conway and Jones, leads to 11 matrices $t$ which comply with
equation~(\ref{1}). However, checking these matrices 
against inequality~(\ref{ineq}), we are left with only five viable cases:
\[
\mbox{(AAA)}, \; \mbox{(ABB)}, \; \mbox{(C$_1$AA)}, \;
\mbox{(BBB)}, \; \mbox{(C$_0$CC)}.
\]
In the latter case, the angle $\vartheta$ is not restricted, apart
from being a rational angle.

\setcounter{equation}{0}
\renewcommand{\theequation}{C\arabic{equation}}
\section{Details of the derivation of the external phases}
\label{details}
\subsection{Form 2A}
\label{form 2A}
According to the procedure explained in the introduction to
section~\ref{external}, we first compute
\begin{align}
T^{2} & = \left(\begin{array}{ccc}
\frac{1}{2}\left(\kappa_{2}+\kappa_{3}\right) &
-\frac{1}{2\sqrt{2}}\left(\kappa_{2}-\kappa_{3}\right) &
\frac{1}{2\sqrt{2}}\left(\kappa_{2}-\kappa_{3}\right) \\
-\frac{1}{2\sqrt{2}}\left(\kappa_{2}-\kappa_{3}\right) &
\frac{1}{4}\left(2\kappa_{1}+\kappa_{2}+\kappa_{3}\right) &
\frac{1}{4}\left(2\kappa_{1}-\kappa_{2}-\kappa_{3}\right) \\
\frac{1}{2\sqrt{2}}\left(\kappa_{2}-\kappa_{3}\right) &
\frac{1}{4}\left(2\kappa_{1}-\kappa_{2}-\kappa_{3}\right) &
\frac{1}{4}\left(2\kappa_{1}+\kappa_{2}+\kappa_{3}\right) 
\end{array}\right) \hat\kappa.
\end{align}
Since $T^2 \in G$ and having excluded the basic forms~1 and~4, we know that
$|T^2|$ has to be one of the basic forms~2, 3~or~5. The $|T^2|$ above follows
the pattern
\begin{align}
\left|T^{2}\right| & =\left(\begin{array}{ccc}
A & B & B\\
B & D & C\\
B & C & D
\end{array}\right).
\end{align}
Since $B$ occurs four times in $|T^2|$, it follows from equation~(\ref{basic})
that $B = 0$, $1/\sqrt{2}$ or $1/2$, corresponding to 
\begin{equation}
\kappa_2 = \kappa_3, \quad
\kappa_2 = -\kappa_3, \quad
\mbox{or} \quad
\kappa_2 = \pm i\kappa_3,
\end{equation}
respectively. In the latter case one finds $A = 1/\sqrt{2}$, so there could be
one, three or five elements $1/\sqrt{2}$ in $|T^2|$; however, a glance at
equation~(\ref{basic}) reveals that this is impossible, because $1/\sqrt{2}$
appears repeated an even number of times in the basic forms~2,~3 and~5. 
Therefore, the viable possibilities are $\kappa_2 = \pm \kappa_3$.

\subsection{Form 2B}
\label{form 2B}
Proceeding in the same way as in the previous subsection, we find 
\begin{align}
T^{2} & =\left(\begin{array}{ccc}
\frac{\kappa_{1}}{2}+\frac{\kappa_{3}}{2\sqrt{2}} & \frac{\kappa_{3}}{2} & 
\frac{\kappa_{1}}{2}-\frac{\kappa_{3}}{2\sqrt{2}}\\
-\frac{\kappa_{1}}{2\sqrt{2}}-\frac{\kappa_{2}}{2\sqrt{2}}+\frac{\kappa_{3}}{4}
&  
\frac{\kappa_{2}}{2}+\frac{\kappa_{3}}{2\sqrt{2}} &
-\frac{\kappa_{1}}{2\sqrt{2}}+
\frac{\kappa_{2}}{2\sqrt{2}}-\frac{\kappa_{3}}{4}\\ 
\frac{\kappa_{1}}{2\sqrt{2}}-\frac{\kappa_{2}}{2\sqrt{2}}-\frac{\kappa_{3}}{4}
& \frac{\kappa_{2}}{2}-\frac{\kappa_{3}}{2\sqrt{2}} &
\frac{\kappa_{1}}{2\sqrt{2}}+\frac{\kappa_{2}}{2\sqrt{2}}+\frac{\kappa_{3}}{4} 
\end{array}\right) \hat\kappa\,,
\end{align}
and $\left|T^{2}\right|$ follows the pattern
\begin{align}
\left|T^{2}\right| & =\left(\begin{array}{ccc}
A & \frac{1}{2} & B\\
E & C & F\\
G & D & H
\end{array}\right).
\end{align}
Because of
\begin{equation}
\left| \left( T^2 \right)_{ij} \right| \geq \frac{1}{2} 
\left( 1 - \frac{1}{\sqrt{2}} \right)
\quad \mbox{for} \quad (i,j) = (1,1),\,(1,3),\,(2,2),\,(3,2),
\end{equation}
there are, together with  $\left| \left( T^2 \right)_{12} \right|$, 
at least five non-zero elements in $\left|T^{2}\right|$. Moreover, 
of the remaining four elements, not all can be zero at the same time. Thus
there are more than five non-zero elements in $\left|T^{2}\right|$, which
excludes the basic form~5. Thus only the basic forms~2 and~3 with all possible
permutations come into question, with the boundary condition 
$\left| \left( T^2 \right)_{12} \right| = 1/2$. First of all, this gives
\begin{equation}
A = \frac{1}{\sqrt{2}}, \; \frac{1}{2}, \; \frac{\sqrt{5}+1}{4}\;
\mbox{or} \; \frac{\sqrt{5}-1}{4}.
\end{equation}
Secondly, there are eight possible forms of $\left|T^{2}\right|$ which are 
\begin{eqnarray}
&&
\left(\begin{array}{ccc}
\frac{1}{2} & \frac{1}{2} & \frac{1}{\sqrt{2}}\\
\frac{1}{2} & \frac{1}{2} & \frac{1}{\sqrt{2}}\\
\frac{1}{\sqrt{2}} & \frac{1}{\sqrt{2}} & 0
\end{array}\right),
\quad
\left(\begin{array}{ccc}
\frac{1}{2} & \frac{1}{2} & \frac{1}{\sqrt{2}}\\
\frac{1}{\sqrt{2}} & \frac{1}{\sqrt{2}} & 0\\
\frac{1}{2} & \frac{1}{2} & \frac{1}{\sqrt{2}}
\end{array}\right),
\nonumber \\ &&
\left(\begin{array}{ccc}
\frac{1}{\sqrt{2}} & \frac{1}{2} & \frac{1}{2}\\
0 & \frac{1}{\sqrt{2}} & \frac{1}{\sqrt{2}}\\
\frac{1}{\sqrt{2}} & \frac{1}{2} & \frac{1}{2}
\end{array}\right),
\quad
\left(\begin{array}{ccc}
\frac{1}{\sqrt{2}} & \frac{1}{2} & \frac{1}{2}\\
\frac{1}{\sqrt{2}} & \frac{1}{2} & \frac{1}{2}\\
0 & \frac{1}{\sqrt{2}} & \frac{1}{\sqrt{2}}
\end{array}\right),
\nonumber \\ &&
\left(\begin{array}{ccc}
\frac{\sqrt{5}+1}{4} & \frac{1}{2} & \frac{\sqrt{5}-1}{4}\\
\frac{\sqrt{5}-1}{4} & \frac{\sqrt{5}+1}{4} & \frac{1}{2}\\
\frac{1}{2} & \frac{\sqrt{5}-1}{4} & \frac{\sqrt{5}+1}{4}
\end{array}\right),
\quad
\left(\begin{array}{ccc}
\frac{\sqrt{5}+1}{4} & \frac{1}{2} & \frac{\sqrt{5}-1}{4}\\
\frac{1}{2} & \frac{\sqrt{5}-1}{4} & \frac{\sqrt{5}+1}{4}\\
\frac{\sqrt{5}-1}{4} & \frac{\sqrt{5}+1}{4} & \frac{1}{2}
\end{array}\right),
\nonumber \\ &&
\left(\begin{array}{ccc}
\frac{\sqrt{5}-1}{4} & \frac{1}{2} & \frac{\sqrt{5}+1}{4}\\
\frac{\sqrt{5}+1}{4} & \frac{\sqrt{5}-1}{4} & \frac{1}{2}\\
\frac{1}{2} & \frac{\sqrt{5}+1}{4} & \frac{\sqrt{5}-1}{4}
\end{array}\right),
\quad
\left(\begin{array}{ccc}
\frac{\sqrt{5}-1}{4} & \frac{1}{2} & \frac{\sqrt{5}+1}{4}\\
\frac{1}{2} & \frac{\sqrt{5}+1}{4} & \frac{\sqrt{5}-1}{4}\\
\frac{\sqrt{5}+1}{4} & \frac{\sqrt{5}-1}{4} & \frac{1}{2}
\end{array}\right).
\end{eqnarray}
Considering these matrices, we see that always one of the following two
equalities holds: 
\begin{equation}
A = C \quad \mbox{or} \quad A = D.
\end{equation}
Exploiting these equations, we obtain
\begin{equation}
\left| \frac{\kappa_{1}}{2}+\frac{\kappa_{3}}{2\sqrt{2}} \right| =
\left|\frac{\kappa_{2}}{2}\pm\frac{\kappa_{3}}{2\sqrt{2}}\right|
\quad \Rightarrow \quad
\textrm{Re}\left(\kappa_{1}\kappa_{3}^{*}\right) =
\pm\textrm{Re}\left(\kappa_{2}\kappa_{3}^{*}\right).
\end{equation}
Defining $x = \kappa_1 \kappa_3^*$, the external phases are given by 
\begin{equation}
\hat\kappa = \left(x,\pm x,1\right)\kappa_{3}
\quad \mbox{or} \quad
\left(x,\pm x^{*},1\right)\kappa_{3}.
\end{equation}
However, the first relation leads to 
$\left| \left( T^2 \right)_{21} \right| = 1/4$ or 
$\left| \left( T^2 \right)_{31} \right| = 1/4$, which cannot occur in the
basic forms~2 and~3. This leaves us with
\begin{equation}
A = C: \quad \hat\kappa = \left(x,x^*,1\right)\kappa_{3}
\quad \mbox{or} \quad
A = D: \quad \hat\kappa = \left(x,-x^*,1\right)\kappa_{3}.
\end{equation}
Finally, we are in a position to determine $x$ for all cases of $A$:
\begin{equation}\label{Ax}
A = \left|\frac{x}{2}+\frac{1}{2\sqrt{2}}\right| 
\quad \Rightarrow \quad
\left\{
\begin{array}{lll}
x = -\varphi &  \textrm{for} & A=\frac{1}{2}, \\
x = \varphi &   \textrm{for} & A=\frac{1}{\sqrt{2}}, \\
x = \rho_{0} &  \textrm{for} & A=\frac{\sqrt{5}+1}{4}, \\
x = -\rho_{0} & \textrm{for} & A=\frac{\sqrt{5}-1}{4}.
\end{array}
\right.
\end{equation}
The phase factors $\varphi$ and $\rho_0$ are defined in
equation~(\ref{phi-rho0}). Of course, for every $x$ also $x^*$ is a solution
of equation~(\ref{Ax}), but in the light of the discussion
in section~\ref{equivalent} this is a trivial variation.

\subsection{Form 3A}
\label{form 3A}
As before, we consider $T^2$, which in the present case is given by
\begin{align}\label{T3A} 
T^{2} & =
\left(\begin{array}{ccc}
\frac{1}{4}\left(\kappa_{1}-\kappa_{2}-\kappa_{3}\right) &
\frac{1}{8}\left[-a\left(\kappa_{1}+\kappa_{2}\right)+b\kappa_{3}\right] &
\frac{1}{8}\left[-c\left(\kappa_{1}+\kappa_{3}\right)-d\kappa_{2}\right] \\
\frac{1}{8}\left[c\left(\kappa_{1}+\kappa_{2}\right)+d\kappa_{3}\right] &
\frac{1}{4}\left(-\kappa_{1}+\kappa_{2}-\kappa_{3}\right) &
\frac{1}{8}\left[a\left(\kappa_{2}+\kappa_{3}\right)-b\kappa_{1}\right] \\
\frac{1}{8}\left[a\left(\kappa_{1}+\kappa_{3}\right)-b\kappa_{2}\right] &
\frac{1}{8}\left[-d\kappa_{1}-c\left(\kappa_{2}+\kappa_{3}\right)\right] &
\frac{1}{4}\left(-\kappa_{1}-\kappa_{2}+\kappa_{3}\right) 
\end{array}\right)
\hat\kappa,
\end{align}
where we have defined 
\begin{equation}\label{abcd}
a = -1+\sqrt{5}, \quad
b = 3+\sqrt{5}, \quad
c = 1+\sqrt{5}, \quad
d = 3-\sqrt{5}.
\end{equation}
It is easy to show that none of the entries of $|T^2|$ can be as large as one,
which rules out form~5. So it remains to consider forms~2 and~3.
Another observation is that phase factors cannot be 
``aligned,''
\textit{i.e.}\ $\kappa_j = \pm \kappa_k$ 
($j \neq k$), 
because otherwise one or
more entries at the diagonal of $|T^2|$ would be $1/4$ which is forbidden for
all basic forms except form~5 which we have excluded before.

Let us first envisage that $|T^2|$ is of form~3. A specific property of form~3
is that either all elements on the diagonal are the same or all three elements
are different. The latter case is impossible because 
$\left| \left( T^2 \right)_{jj} \right| \leq 3/4$, 
while one element on the diagonal would have to be $(\sqrt{5} + 1)/4$ which is
larger than $3/4$. Since we know that all elements are the same on the
diagonal, we deduce 
\begin{equation}\label{kappa}
\begin{array}{ccc}
\left| \left( T^2 \right)_{11} \right| =
\left| \left( T^2 \right)_{22} \right|
& \Rightarrow & 
\textrm{Re}\left(\kappa_{1}\kappa_{3}^{*}\right) = 
\textrm{Re}\left(\kappa_{2}\kappa_{3}^{*}\right), \\
\left| \left( T^2 \right)_{22} \right| =
\left| \left( T^2 \right)_{33} \right|
& \Rightarrow & 
\textrm{Re}\left(\kappa_{2}\kappa_{1}^{*}\right) = 
\textrm{Re}\left(\kappa_{3}\kappa_{1}^{*}\right).
\end{array}
\end{equation}
Thus we conclude that either 
$\kappa_{1}\kappa_{3}^{*} = \kappa_{2}\kappa_{3}^{*}$ or
$\kappa_{1}\kappa_{3}^{*} = \kappa_{2}^*\kappa_{3}$.
In the first case we would have alignment of phase factors, which we have
excluded before. Therefore, equation~(\ref{kappa}) leads to
the relations
\begin{equation}
\kappa_1\kappa_2 = \kappa_3^2, \quad
\kappa_2\kappa_3 = \kappa_1^2.
\end{equation}
These relations are solved by 
$\hat\kappa = \diag (1,\omega^k, \omega^{2k}) \kappa_1$ with $k =1,2$, since
$k=0$ would lead to alignment of phase factors. 
Since $\kappa_1$ is an arbitrary root of unity and complex conjugation of $T$
leads to a trivial variation, we obtain without loss of generality
\begin{equation}
\hat\kappa = \diag (1,\omega, \omega^2) \kappa_1
\end{equation}
when $|T^2|$ is of form~3. All elements on the diagonal of $|T^2|$ are 
$1/2$ in this case.

Next we discuss the possibility 
that $|T^2|$ is of form~2, 
in which case there 
must be one zero in the matrix. Because of $b > 2a$, we have
\begin{equation}
\left| \left( T^2 \right)_{ij} \right| > 0
\quad \mbox{for} \quad (i,j) = (1,2),\,(2,3),\,(3,1).
\end{equation}
We further observe that
\begin{equation}\label{if}
\left( T^2 \right)_{ij} = 0 
\quad \Rightarrow \quad
\left| \left( T^2 \right)_{ji} \right| = 
\frac{ad+cb}{8c} = \frac{5-\sqrt{5}}{4},
\quad \mbox{if} \quad (i,j) = (2,1),\,(3,2),\,\mbox{or}\,(1,3),
\end{equation}
which is impossible because no such element occurs in form~2.
We conclude that the zero must be on the diagonal of $T^2$:
\begin{itemize}
\item 
$\left( T^2 \right)_{11} = 0$ $\Rightarrow$
$\kappa_{1}=\kappa_{2}+\kappa_{3}$ $\Rightarrow$
$\left|\kappa_{2}+\kappa_{3}\right|=1$. This 
implies $\kappa_{2}\kappa_{3}^{*}=\omega$ or $\omega^{2}$,
so we end up with 
$\hat\kappa = \diag\left(-\omega^{2},\omega,1\right)\kappa_{3}$
or $\diag\left(-\omega,\omega^{2},1\right)\kappa_{3}$.
\item 
$\left( T^2 \right)_{22} = 0$ $\Rightarrow$
$\kappa_{2}=\kappa_{1}+\kappa_{3}$ $\Rightarrow$
$\left|\kappa_{1}+\kappa_{3}\right|=1$. This 
implies $\kappa_{1}\kappa_{3}^{*}=\omega$ or $\omega^{2}$,
so we end up with 
$\hat\kappa = \diag\left(\omega,-\omega^{2},1\right)\kappa_{3}$
or $\diag\left(\omega^{2},-\omega,1\right)\kappa_{3}$.
\item 
$\left( T^2 \right)_{33} = 0$ $\Rightarrow$
$\kappa_{3}=\kappa_{1}+\kappa_{2}$ $\Rightarrow$
$\left|\kappa_{1}+\kappa_{2}\right|=1$. This 
implies $\kappa_{1}\kappa_{2}^{*}=\omega$ or $\omega^{2}$,
so we end up with 
$\hat\kappa = \diag\left(\omega,1,-\omega^{2}\right)\kappa_2$
or $\diag\left(\omega^{2},1,-\omega\right)\kappa_2$.
\end{itemize}
Leaving out cases which emerge from each other by complex conjugation, we end
up with the three solutions
\begin{equation}\label{3A-2}
\hat\kappa = 
\diag\left(-\omega^2,\omega,1\right)\kappa_3, \;
\diag\left(\omega,-\omega^2,1\right)\kappa_3, \;
\diag\left(\omega,1,-\omega^2\right)\kappa_2
\end{equation}
for the external phases, where $\kappa_2$ and $\kappa_3$ are arbitrary roots
of unity.

\subsection{Form 3B}
\label{form 3B}
The matrix $T^{2}$ is as follows:
\begin{align}
T^{2} & =\left(\begin{array}{ccc}
\frac{1}{8}\left(2\kappa_{1}-b\kappa_{2}-d\kappa_{3}\right) &
-\frac{1}{8}\left(c\kappa_{1}+2\kappa_{2}+a\kappa_{3}\right) &
-\frac{1}{8}\left(a\kappa_{1}+c\kappa_{2}-2\kappa_{3}\right)\\ 
\frac{1}{8}\left(c\kappa_{1}+2\kappa_{2}+a\kappa_{3}\right) &
\frac{1}{8}\left(-b\kappa_{1}+d\kappa_{2}+2\kappa_{3}\right) &
\frac{1}{8}\left(-2\kappa_{1}+a\kappa_{2}-c\kappa_{3}\right)\\ 
\frac{1}{8}\left(a\kappa_{1}+c\kappa_{2}-2\kappa_{3}\right) &
\frac{1}{8}\left(-2\kappa_{1}+a\kappa_{2}-c\kappa_{3}\right) &
\frac{1}{8}\left(-d\kappa_{1}+2\kappa_{2}+b\kappa_{3}\right) 
\end{array}\right) 
\hat\kappa,
\end{align}
where $a$, $b$, $c$, $d$ are defined in equation~(\ref{abcd}). The most obvious
observation is that $\left|T^{2}\right|$ is symmetric. 
Therefore, if $|T^2|$ is if form~2 which has one zero, this zero can only be
on the diagonal. However, $b > 2+d$ rules out this possibility. So it
remains to consider $|T^2|$ being of form~3 and of form~5.

If $|T^2|$ is of form~3, then the elements on the diagonal are either
all the same or all different. 
Since the first option leads to a non-symmetric matrix, we are
left with the second option, namely having $1/2$, $(\sqrt{5}+1)/4$ and
$(\sqrt{5}-1)/4$ on the diagonal. Taking into account that
\begin{equation}
\left| \left( T^2 \right)_{jj} \right| \geq\frac{1}{8} (b-2-d) = 
\frac{\sqrt{5}-1}{4},
\end{equation}
we have three cases:
\begin{itemize}
\item 
$\left| \left( T^{2} \right)_{11} \right| = (\sqrt{5}-1)/4$ $\Rightarrow$ 
$\hat\kappa = \diag\left(1,1,-1\right)\kappa_{1}$,
\item 
$\left| \left( T^{2} \right)_{22} \right| = (\sqrt{5}-1)/4$ $\Rightarrow$ 
$\hat\kappa = \diag\left(1,1,1\right)
\kappa_{1}$,
\item 
$\left| \left( T^{2} \right)_{33} \right| = (\sqrt{5}-1)/4$ $\Rightarrow$ 
$\hat\kappa = \diag\left(1,-1,1\right)\kappa_{1}$.
\end{itemize}

If $|T^2|$ is of form~5, we observe that
\begin{equation}
\left| \left( T^{2} \right)_{ij} \right| \leq \frac{1}{8}(2+a+c) =
\frac{\sqrt{5}+1}{4} < 1 \quad \forall\; i \neq j.
\end{equation}
Therefore, the 1 has to be somewhere on the diagonal. Since 
\begin{equation}
\frac{1}{8}(2+b+d) = 1 
\end{equation}
holds, the phase factors necessarily fulfill
$\kappa_{2}=\kappa_{3}=-\kappa_{1}$, leading 
to $\left|T^{2}\right|=\bone$, consistent with $2+a-c=0$ for the off-diagonal
elements. In summary, the external phases are given by
\begin{equation}
\hat\kappa = \left( -1,1,1 \right) \kappa_3.
\end{equation}

\subsection{Form 3C}
\label{form 3C}
This case is quite similar to form~3A. With
\begin{align}
T^{2} & =\left(\begin{array}{ccc}
\frac{1}{8}\left[b\kappa_{1}+a\left(\kappa_{2}-\kappa_{3}\right)\right] &
\frac{1}{8}\left[c\left(\kappa_{2}-\kappa_{1}\right)-d\kappa_{3}\right] &
\frac{1}{4}\left(\kappa_{1}+\kappa_{2}+\kappa_{3}\right)\\ 
\frac{1}{4}\left(-\kappa_{1}+\kappa_{2}+\kappa_{3}\right) &
\frac{1}{8}\left[b\kappa_{2}+a\left(\kappa_{1}+\kappa_{3}\right)\right] &
\frac{1}{8}\left[c\left(\kappa_{2}-\kappa_{3}\right)-d\kappa_{1}\right]\\ 
\frac{1}{8}\left[-c\left(\kappa_{1}+\kappa_{3}\right)+d\kappa_{2}\right] &
\frac{1}{4}\left(\kappa_{1}+\kappa_{2}-\kappa_{3}\right) &
\frac{1}{8}\left[b\kappa_{3}+a\left(\kappa_{2}-\kappa_{1}\right)\right] 
\end{array}\right)
\hat\kappa
\end{align}
and $a$, $b$, $c$, $d$ defined in equation~(\ref{abcd}), we find that none of
its entries can be 
as large as~1, 
which rules out form~5. Furthermore, the
phase factors $\kappa_j$ cannot be aligned, lest one or more entries of
$|T^2|$ be $1/4$. 

First we consider form~3. The entries 
\begin{equation}\label{ee}
\left| \left( T^{2} \right)_{ij} \right|
\quad \mbox{with} \quad
(i,j) = (1,3),\, (2,1),\, (3,2)
\end{equation}
are either 
all the same
or all different. 
However, the latter case cannot occur,
because these entries are all smaller or equal $3/4$, which is smaller than 
$(\sqrt{5}+1)/4$. Evaluating the requirement that the elements of
equation~(\ref{ee}) 
have to be equal, we proceed along the same lines as for
form~3A and obtain the relations
\begin{equation}
\kappa_1^2 = -\kappa_2 \kappa_3, \quad
\kappa_3^2 = -\kappa_1 \kappa_2.
\end{equation}
From this we arrive at the external phases 
\begin{equation}
\hat\kappa = \left( \omega,-\omega^2, 1 \right) \kappa_3.
\end{equation}

It remains to consider form~2, which has one zero in some place. The zero
cannot be on the diagonal of $|T^2|$ because $b-2a = 5 -\sqrt{5} > 0$.
Moreover, proceeding as in equation~(\ref{if}), we find
\begin{itemize}
\item
$\left( T^2 \right)_{12} = 0$ $\Rightarrow$ 
$\left| \left( T^2 \right)_{33} \right| = (5-\sqrt{5})/4$,
\item
$\left( T^2 \right)_{23} = 0$ $\Rightarrow$ 
$\left| \left( T^2 \right)_{11} \right| = (5-\sqrt{5})/4$,
\item
$\left( T^2 \right)_{31} = 0$ $\Rightarrow$ 
$\left| \left( T^2 \right)_{22} \right| = (5-\sqrt{5})/4$,
\end{itemize}
which is impossible for form~2.
Thus we have the following three possibilities:
\begin{itemize}
\item 
$\left( T^2 \right)_{21} = 0$ $\Rightarrow$
$\kappa_{1}=\kappa_{2}+\kappa_{3}$ $\Rightarrow$
$\left|\kappa_{2}+\kappa_{3}\right|=1$. This 
implies $\kappa_{2}\kappa_{3}^{*}=\omega$ or $\omega^{2}$,
so we end up with 
$\hat\kappa = \diag\left(-\omega^{2},\omega,1\right)\kappa_{3}$
or $\diag\left(-\omega,\omega^{2},1\right)\kappa_{3}$.
\item 
$\left( T^2 \right)_{13} = 0$ $\Rightarrow$
$\kappa_{2}=-\kappa_{1}-\kappa_{3}$ $\Rightarrow$
$\left|\kappa_{1}+\kappa_{3}\right|=1$. This 
implies $\kappa_{1}\kappa_{3}^{*}=\omega$ or $\omega^{2}$,
so we end up with 
$\hat\kappa = \diag\left(\omega,\omega^{2},1\right)\kappa_{3}$
or $\diag\left(\omega^{2},\omega,1\right)\kappa_{3}$.
\item 
$\left( T^2 \right)_{32} = 0$ $\Rightarrow$
$\kappa_{3}=\kappa_{1}+\kappa_{2}$ $\Rightarrow$
$\left|\kappa_{1}+\kappa_{2}\right|=1$. This 
implies $\kappa_{1}\kappa_{2}^{*}=\omega$ or $\omega^{2}$,
so we end up with 
$\hat\kappa = \diag\left(\omega,1,-\omega^{2}\right)\kappa_2$
or $\diag\left(\omega^{2},1,-\omega\right)\kappa_2$.
\end{itemize}
Leaving out the complex conjugate solutions, 
we finally obtain
\begin{equation}\label{3C-2}
\hat\kappa = 
\diag\left(-\omega^2,\omega,1\right)\kappa_3, \;
\diag\left(\omega,\omega^2,1\right)\kappa_3, \;
\diag\left(\omega,1,-\omega^2\right)\kappa_2
\end{equation}
for the external phases, with $\kappa_2$ and $\kappa_3$ being arbitrary roots
of unity. 

\subsection{Form 3D}
\label{form 3D}
Also this case is treated analogously to form~3A. With
\begin{align}
T^{2} & =\left(\begin{array}{ccc}
\frac{1}{8}\left[d\kappa_{1}-c\left(\kappa_{2}+\kappa_{3}\right)\right] &
\frac{1}{4}\left(\kappa_{1}-\kappa_{2}+\kappa_{3}\right) &
\frac{1}{8}\left[a\left(\kappa_{3}-\kappa_{1}\right)-b\kappa_{2}\right]\\ 
\frac{1}{8}\left[a\left(\kappa_{2}-\kappa_{1}\right)-b\kappa_{3}\right] &
\frac{1}{8}\left[c\left(\kappa_{3}-\kappa_{1}\right)+d\kappa_{2}\right] &
\frac{1}{4}\left(\kappa_{1}+\kappa_{2}+\kappa_{3}\right)\\ 
\frac{1}{4}\left(\kappa_{1}+\kappa_{2}-\kappa_{3}\right) &
\frac{1}{8}\left[b\kappa_{1}+a\left(\kappa_{2}+\kappa_{3}\right)\right] &
\frac{1}{8}\left[c\left(\kappa_{2}-\kappa_{1}\right)+d\kappa_{3}\right] 
\end{array}\right)
\hat\kappa
\end{align}
and $a$, $b$, $c$, $d$ defined in equation~(\ref{abcd}), we find again
that none of its entries can be as large as one, thus ruling out form~5. 
Furthermore, the phase factors $\kappa_j$ cannot be aligned in order to avoid
one or more entries of $|T^2|$ being $1/4$. 

First we consider form~3. The entries 
\begin{equation}\label{eee}
\left| \left( T^{2} \right)_{ij} \right|
\quad \mbox{with} \quad
(i,j) = (1,2),\, (2,3),\, (3,1)
\end{equation}
are either all the same or all different. However, the latter case cannot occur,
because $3/4$ is an upper bound to these entries and $3/4 < (\sqrt{5}+1)/4$.
Evaluating the requirement that the elements of
equation~(\ref{eee}) 
have to be equal, we obtain the relations
\begin{equation}
\kappa_2^2 = -\kappa_1 \kappa_3, \quad
\kappa_3^2 = -\kappa_1 \kappa_2,
\end{equation}
which give
\begin{equation}
\hat\kappa = \left( -\omega,\omega^2, 1 \right) \kappa_3.
\end{equation}

Assuming now that $|T^2|$ is of form~2, there must be a
zero in the matrix. However, 
\begin{equation}
\left| \left( T^2 \right)_{ij} \right| > 0
\quad \mbox{for} \quad (i,j) = (2,1),\,(3,2),\,(1,3)
\end{equation}
because of $b > 2a$. Moreover, it follows that
\begin{itemize}
\item
$\left( T^2 \right)_{11} = 0$ $\Rightarrow$ 
$\left| \left( T^2 \right)_{32} \right| = (5-\sqrt{5})/4$,
\item
$\left( T^2 \right)_{22} = 0$ $\Rightarrow$ 
$\left| \left( T^2 \right)_{13} \right| = (5-\sqrt{5})/4$,
\item
$\left( T^2 \right)_{33} = 0$ $\Rightarrow$ 
$\left| \left( T^2 \right)_{21} \right| = (5-\sqrt{5})/4$,
\end{itemize}
in the same vein as for form~3A in equation~(\ref{if}),
which is impossible because no such element occurs in form~2.
Thus we have to consider the remaining three possibilities:
\begin{itemize}
\item 
$\left( T^2 \right)_{12} = 0$ $\Rightarrow$
$\kappa_2=\kappa_1+\kappa_3$ $\Rightarrow$
$\left|\kappa_1+\kappa_3\right|=1$. This 
implies $\kappa_1\kappa_3^{*}=\omega$ or $\omega^{2}$,
so we end up with 
$\hat\kappa = \diag\left(\omega,-\omega^2,1\right)\kappa_{3}$
or $\diag\left(\omega^2,-\omega,1\right)\kappa_{3}$.
\item 
$\left( T^2 \right)_{23} = 0$ $\Rightarrow$
$\kappa_2=-\kappa_{1}-\kappa_{3}$ $\Rightarrow$
$\left|\kappa_{1}+\kappa_{3}\right|=1$. This 
implies $\kappa_{1}\kappa_{3}^{*}=\omega$ or $\omega^{2}$,
so we end up with 
$\hat\kappa = \diag\left(\omega,\omega^{2},1\right)\kappa_3$
or $\diag\left(\omega^{2},\omega,1\right)\kappa_3$.
\item 
$\left( T^2 \right)_{31} = 0$ $\Rightarrow$
$\kappa_3=\kappa_{1}+\kappa_{2}$ $\Rightarrow$
$\left|\kappa_{1}+\kappa_{2}\right|=1$. This 
implies $\kappa_{1}\kappa_{2}^{*}=\omega$ or $\omega^{2}$,
so we end up with 
$\hat\kappa = \diag\left(\omega,1,-\omega^{2}\right)\kappa_{2}$
or $\diag\left(\omega^{2},1,-\omega\right)\kappa_{2}$.
\end{itemize}
Dropping the complex conjugate solutions, we arrive at
\begin{equation}\label{3D-2}
\hat\kappa = 
\diag\left(\omega,-\omega^2,1\right)\kappa_3, \;
\diag\left(\omega,\omega^2,1\right)\kappa_3, \;
\diag\left(\omega,1,-\omega^2\right)\kappa_2,
\end{equation}
with $\kappa_2$ and $\kappa_3$ being arbitrary roots of unity.

\setcounter{equation}{0}
\renewcommand{\theequation}{D\arabic{equation}}
\section{Minimal groups for $\mathcal{C}_2$}
\label{minimal}
Here we determine minimal flavour groups $G$ associated with the series
$\mathcal{C}_2$. The generators of the groups of this series are
\begin{equation}
T = \kappa_1 \left( \begin{array}{ccc}
0 & \frac{1}{\sqrt{2}} & \frac{1}{\sqrt{2}} \\
\frac{1}{\sqrt{2}} & -\frac{1}{2} & \frac{1}{2} \\
\frac{1}{\sqrt{2}} & \frac{1}{2} & -\frac{1}{2}
\end{array} \right)
\diag \left( 1, \sigma^{-3}, -\sigma^{-3} \right)
\end{equation}
and $S_j$ ($j=1,2,3$).
We make a basis transformation with
\begin{equation}
V = \left( \begin{array}{ccc}
1 & 0 & 0 \\
0 & -\frac{1}{\sqrt{2}} \sigma^{-2} & -\frac{1}{\sqrt{2}} \sigma^{-2} \\
0 & -\frac{1}{\sqrt{2}} \sigma^{-1} & \frac{1}{\sqrt{2}} \sigma^{-1} 
\end{array} \right),
\end{equation}
leading to 
\begin{equation}\label{TE}
T' = V T V^\dagger = \kappa E^2
\quad \mbox{with} \quad 
E^2 = \left( \begin{array}{ccc}
0 & 0 & 1 \\ 1 & 0 & 0 \\ 0 & 1 & 0
\end{array} \right)
\quad \mbox{or} \quad
E = \left( \begin{array}{ccc}
0 & 1 & 0 \\ 0 & 0 & 1 \\ 1 & 0 & 0
\end{array} \right)
\end{equation}
and $S'_j = V S_j V^\dagger$ given by
\begin{equation}
S'_1 = S_1, 
\quad 
S'_2 = 
\left( \begin{array}{ccc} 
-1 & 0 & 0 \\ 0 & 0 & \sigma^* \\ 0 & \sigma & 0
\end{array} \right),
\quad
S'_3 = 
\left( \begin{array}{ccc} 
-1 & 0 & 0 \\ 0 & 0 & -\sigma^* \\ 0 & -\sigma & 0
\end{array} \right).
\end{equation}

For the time being, we set $\kappa = 1$, which entails $T' = E^2$ and
$\sigma = -\kappa_2^*$. At any rate, $\sigma$ is an abitrary root of unity. 
It is now useful to switch to another set of generators, as described
in appendix~F of~\cite{GL-review}. For this purpose we compute
\begin{equation}\label{EE}
E \left( E S'_3 \right)^2 E^\dagger = 
\diag \left( \sigma^2, \sigma^*, \sigma^* \right) \equiv \tilde F \in G.
\end{equation}
Instead of $E$ it is certainly admissible to
use $\tilde E \equiv \tilde F E$ and $\tilde F$ as generators.
But then, with a further basis change, one can remove the phases from
$\tilde E$ and $S'_3$: 
\begin{equation}
V_1 = \diag \left( \sigma^*,\sigma,1 \right)
\end{equation}
and 
\begin{equation}
V_1 \tilde E V_1^\dagger = E, \quad 
V_1 S'_3 V_1^\dagger \equiv B = \left( 
\begin{array}{ccc}
-1 & 0 & 0 \\ 0 & 0 & -1 \\ 0 & -1 & 0 
\end{array} \right).
\end{equation}

In this final basis, our group $G$ has the generators $S_1$, $\tilde F$,
$E$ and $B$. Clearly, $E$ and $B$ are generators of permutations, so
$G$ has the structure~\cite{GL-review} 
$G = N \rtimes S_3$, where $N$ is the 
normal subgroup consisting of all diagonal matrices and
$S_3$ consists of all $3 \times 3$ permutations matrices. 
This $N$ is generated by $S_1$, $\tilde F$, and matrices
where the elements on the diagonal of $S_1$ and $\tilde F$ are permuted.

We can write $\sigma$ as $\sigma = e^{2\pi ip/n}$ 
with integers $p$ and $n$, and $p$ coprime to $n$.
If $n$ is even, we define
\begin{equation}\label{F1}
E (\tilde F)^{-1} E^\dagger = 
\diag \left(\sigma, \sigma, \sigma^{-2} \right) \equiv F_1.
\end{equation}
If $n$ is odd, we consider $-\sigma$ instead for the definition of
$F_1$. In this case we have
\begin{equation}
-\sigma = \exp \left( 2\pi i \frac{2p + n}{2n} \right).
\end{equation}
Thus it is useful to define $n' = \textrm{lcm}(2,n)$ 
and $p'$, where $p'=p$ for $n$ even and $p'=2p+n$ for $n$ odd. Note
that $p'$ is coprime to $n'$. Since 
\begin{equation} 
F_1^{n'/2} = \diag \left( -1,-1,1 \right) = E^\dagger S_1 E,
\end{equation}
we can dispense with the generator $S_1$.

According to theorem~1 in~\cite{GL-CD}, there is a procedure to find two
generators which reveal the structure of $N$ and, as a
consequence, the structure of $G$. 
We first need to find the element with the highest order in
$N$. Since $p'$ is coprime to $n'$, 
there must be a positive integer $a$ such that 
\begin{equation}
F \equiv F_1^a = \diag \left( \epsilon, \epsilon, \epsilon^{-2}
\right)
\quad \mbox{with} \quad \epsilon = e^{2\pi i/n'}.
\end{equation}
This matrix has the highest possible order in $N$.
In the next step we have to find the generator of the subgroup of
$N$ which has~1 as the first entry. It is easy to see that
in our case every element in this subgroup is a power of
\begin{equation}
g \equiv \diag \left( 1, \epsilon^{-3}, \epsilon^3 \right).
\end{equation}
If $3 \nmid n'$, there is a positive integer $b$ such
that
\begin{equation}
g^b = \diag \left( 1, \epsilon^*, \epsilon \right)
\end{equation}
and we obviously arrive at the conclusion that the flavour group is
$\Delta(6{n'}^2)$. 

For the remaining cases we use again the results of~\cite{GL-CD}.
If $9$ divides $n$, the flavour group is given by
$\left( \zz_{n'} \times \zz_{n'/3} \right) \rtimes S_3$.

Finally, we have to consider $3 \mid n$ but $9 \nmid n$. In
this case we put $\kappa = \omega^\alpha$ with $\alpha = pn/3$. Note
that $\alpha = 1$ or $2\: (\textrm{mod}\,3)$. Now we proceed as before,
but with the replacements $E \to \omega^{2\alpha} E$ in
equation~(\ref{TE}) and, due to equation~(\ref{EE}),
\begin{equation}
\tilde F \to \omega^\alpha \tilde F = 
\diag \left( {\sigma'}^2, {\sigma'}^*, {\sigma'}^* \right)
\quad \mbox{with} \quad \sigma' = \omega^{-\alpha} \sigma.
\end{equation}
Next we have a closer look at
\begin{equation}
\arg \sigma' = 2\pi \left( \frac{p}{n} - \frac{pn}{9} \right) = 
2\pi \, \frac{p}{n} \left( 1 - \left(\frac{n}{3} \right)^2 \right). 
\end{equation}
Given that $n$ is divisible by $3$ but not by $9$, we find that 
$1 - (n/3)^2$ is divisible by $3$. So we end up with
\begin{equation}
\sigma' = \exp\left( 2\pi i\, 
\frac{p \left[1-(n/3)^2 \right]/3}{n/3} \right).
\end{equation}
This should be compared to $\sigma = e^{2\pi ip/n}$ defined before
equation~(\ref{F1}). 
In effect, we replace $n$ by $n/3$ and we proceed as before, starting
from equation~(\ref{F1}). Eventually, we arrive at the group 
$\Delta \left( 6 ( n'/3 )^2 \right)$, 
with $n'$ being defined above.

\setcounter{equation}{0}
\renewcommand{\theequation}{E\arabic{equation}}
\section{Two-flavour solutions}
\label{2f}
We are dealing here with
\begin{equation}
T=
\left(\begin{array}{ccc}
1 & 0 & 0\\
0 & \cos\theta & \sin\theta\\
0 & -\sin\theta & \cos\theta
\end{array}\right) \hat \kappa,
\end{equation}
where $\theta$ is a rational angle, \textit{i.e.}\ 
$e^{i\theta}$ is a root of unity. 
This case clearly involves only two-family mixing and we present it here only
for completeness. Because it is physically irrelevant, we do not discuss
all the details.

The phase factor $\kappa_1$ and the product $\kappa_2\kappa_3$ 
must be roots of unity.
The only non-trivial part of $T$ is the 23-sector, so 
we focus exclusively on this sector. There we have 
the two generators 
\begin{equation}
\mathcal{T} \equiv
\left( \begin{array}{cc}
 \cos\theta \kappa_2 & \sin\theta \kappa_3 \\
-\sin\theta \kappa_2 & \cos\theta \kappa_3
\end{array} \right),
\quad
\mathcal{S} \equiv
\left( \begin{array}{cc}
1 & 0 \\ 0 & -1
\end{array} \right),
\end{equation}
where $\mathcal{S}$ is the 23-block of the generator $S_{2}$ of
equation~(\ref{SSS}). We search for possible values of the pair
$\left(\cos\theta, \kappa_{2}\kappa_{3}^{*}\right)$
which lead to a finite group. 
Let us denote the eigenvalues of $\mathcal{T}$ by $a_{1}$, $a_{2}$ and those
of $\mathcal{TS}$ by $b_{1}$, $b_{2}$. 
Then we find
\begin{equation}
\textrm{Tr}\mathcal{T} = a_{1}+a_{2} =
\cos\theta\left(\kappa_{2}+\kappa_{3}\right),
\quad 
\textrm{Tr}(\mathcal{TS}) 
=b_{1}+b_{2}=\cos\theta\left(\kappa_{2}-\kappa_{3}\right) 
\end{equation}
or 
\begin{equation}
2\cos\theta\kappa_{2} = 
\textrm{Tr}\mathcal{T}+\textrm{Tr}(\mathcal{TS}),
\quad
2\cos\theta\kappa_{3} =
\textrm{Tr}\mathcal{T}-\textrm{Tr}(\mathcal{TS}).
\end{equation}
Taking the square of the absolute values of these relations, 
we obtain consistency only if
$
\left(\textrm{Tr}\mathcal{T}\right)^* \textrm{Tr}(\mathcal{TS})
$
is purely imaginary.
Finally, we can express $\cos^2\theta$ and $\kappa_2\kappa_3^*$ as
\begin{eqnarray}\label{cos}
4\cos^{2}\theta &=& 
\left|\textrm{Tr}\mathcal{T}\right|^{2} + 
\left|\textrm{Tr}(\mathcal{TS})\right|^{2},
\\
\label{kappa23}
\kappa_{2}\kappa_{3}^{*} &=&
\frac{\left|\textrm{Tr}\mathcal{T}\right|^{2} -
\left|\textrm{Tr}(\mathcal{TS})\right|^{2} \pm
2i\left|\textrm{Tr}\mathcal{T}\right|
\left|\textrm{Tr}(\mathcal{TS})\right|}%
{\left|\textrm{Tr}\mathcal{T}\right|^{2} +
\left|\textrm{Tr}(\mathcal{TS})\right|^{2}} 
\end{eqnarray}

We can use equation~(\ref{cos}) to constrain the possible values
$\left|\textrm{Tr}\mathcal{T}\right|^{2}$ and 
$\left|\textrm{Tr}(\mathcal{TS})\right|^{2}$.
Using $4\cos^{2}\theta=e^{-2i\theta}+e^{2i\theta}+2$, we
get the following vanishing sum of roots of unity:
\begin{equation}
0  = 2 - e^{2i\theta} - e^{-2i\theta} +
a_{1}a_{2}^{*} + a_{1}^*a_{2} + 
b_{1}b_{2}^{*} + b_{1}^*b_{2}.
\end{equation}
This sum is exactly of the form of equation~(\ref{lambda2}) of
section~\ref{basic forms}, with its solutions given in
equation~(\ref{ABC}). These solutions imply
\begin{equation}
\left(-e^{2i\theta},\, a_{1}a_{2}^{*},\, b_{1}b_{2}^{*}\right) =
\left(i,\omega,\omega\right),
\quad
\left(\omega,\beta,\beta^{2}\right),
\quad\mbox{or}\quad
\left(-1,\lambda,-\lambda\right),
\end{equation}
up to conjugations and permutations, where $\lambda$ is some arbitrary
root of unity. With this we obtain the seven cases
\begin{eqnarray}
\lefteqn{\left(\left|\textrm{Tr}\mathcal{T}\right|^{2},\,
\left|\textrm{Tr}(\mathcal{TS})\right|^{2}\right) =}
\nonumber \\ &&  
\left(1,1\right),
\;
\left(1,2\right),
\;
\left(1,2+\beta+\beta^{4}\right),
\;
\left(1,2+\beta^{2}+\beta^{3}\right),
\;
\left(2+\beta+\beta^{4},2+\beta^{2}+\beta^{3}\right),
\nonumber \\ && \label{st}
\left(0,2+\lambda+\lambda^{*}\right),
\;
\left(2+\lambda+\lambda^{*},2-\lambda-\lambda^{*}\right)
\end{eqnarray}
or the reversed order.

Note that 
$\left|\textrm{Tr}\mathcal{T}\right|^{2}
\leftrightarrow
\left|\textrm{Tr}(\mathcal{TS})\right|^{2}$
preserves $\cos^{2}\theta$---see equation~(\ref{cos})---and 
$\kappa_{2}\kappa_{3}^{*}$
is transformed to $-\left(\kappa_{2}\kappa_{3}^{*}\right)^{*}$---see
equation~(\ref{kappa23}). While the conjugation of this phase is irrelevant, 
in general this is not true for the sign reversal. So, 
for each of the seven cases in
equation~(\ref{st}), one has to take into account the two signs of
$\kappa_{2}\kappa_{3}^{*}$. 

A related concern would be the signs
of $\cos\theta$ and $\sin\theta$, as equation~(\ref{cos}) only
tells us the value of $\cos^{2}\theta$. However, it is easy to see
that such signs do not affect $\left|U\right|^{2}$.

In the following, we will list the solutions 
for each of the seven cases in equation~(\ref{st}) and 
the two sign variations of $\kappa_{2}\kappa_{3}^{*}$,
without going into detail.
For each of the seven cases we will display
\begin{enumerate}
\item
$\cos^2\theta$ and $\kappa_{2}\kappa_{3}^{*}$,
\item
$\hat\kappa$ and the eigenvalues of $T$ for the positive sign,
\item
$|U|^2$ for the positive sign,
\item
the degenerate case of 
$|U|^2$ for the positive sign,
\item
$\hat\kappa$ and the eigenvalues of $T$ for the negative sign,
\item
$|U|^2$ for the negative sign,
\item
the degenerate case of $|U|^2$ for the negative sign.
\end{enumerate}
\paragraph{First subcase:}
$\left(\left|\textrm{Tr}\mathcal{T}\right|^{2},\,
\left|\textrm{Tr}(\mathcal{TS})\right|^{2}\right) = (1,1)$ \\
This corresponds 
\begin{equation}
\cos^2\theta = \frac{1}{2}, \quad 
\kappa_{2}\kappa_{3}^{*} = \pm i.
\end{equation}
Since here $\kappa_{2}\kappa_{3}^{*}$ is purely imaginary, 
its sign variation is irrelevant. \\ 
\textit{Non-degenerate case:}
\begin{equation}
\hat\kappa = \diag\left( \kappa_1, i\kappa_3, \kappa_3 \right),
\quad
\hat\lambda^{(0)} = \diag\left( 
\kappa_1, -\omega^2 e^{i\pi/4} \kappa_3, -\omega e^{i\pi/4} \kappa_3 \right),
\end{equation}
\begin{equation}
\mathcal{C}_{18}: \quad
|U|^2 = \left(
\begin{array}{ccc}
1 & 0 & 0 \\
0 & \frac{1}{6} \left(3+\sqrt{3}\right) & \frac{1}{6} \left(3-\sqrt{3}\right) \\
0 & \frac{1}{6} \left(3-\sqrt{3}\right) & \frac{1}{6}
   \left(3+\sqrt{3}\right) 
\end{array}
\right).
\end{equation}
\textit{Degenerate case:}
\begin{equation}
\hat\lambda^{(0)} = \kappa_1\, \diag\left( \omega,1,1 \right),
\end{equation}
\begin{equation}
\mathcal{P}_3: \quad
|U|^2 = \left(
\begin{array}{ccc}
0 & \frac{1}{6} \left(3+\sqrt{3}\right) & \frac{1}{6} \left(3-\sqrt{3}\right) \\
\times & \times & \times \\
\times & \times & \times 
\end{array}
\right).
\end{equation}
\paragraph{Second subcase:}
$\left(\left|\textrm{Tr}\mathcal{T}\right|^{2},\,
\left|\textrm{Tr}(\mathcal{TS})\right|^{2}\right) = (1,2)$ \\
This corresponds to
\begin{equation}
\cos^{2}\theta = \frac{3}{4}, \quad
\kappa_{2}\kappa_{3}^{*} =
\pm\frac{1}{3}\left(-1+2i\sqrt{2}\right) \equiv 
\pm\left(F_{1}\right)^{2}.
\end{equation}
This means that we can write the phases as
$\kappa_2 = F_1 \kappa$, $\kappa_3 = F_1^* \kappa$ with a root of unity
$\kappa$. 
\\
\textit{Positive sign, non-degenerate case:} 
\begin{equation}
\hat\kappa = \diag\left( \kappa_1, F_1 \kappa, F_1^* \kappa \right),
\quad
\hat\lambda^{(0)} = \diag\left( 
\kappa_1, -\omega^2 \kappa, -\omega \kappa \right), 
\end{equation}
\begin{equation}
\mathcal{C}_{19}: \quad
|U|^2 = \left(
\begin{array}{ccc}
1 & 0 & 0 \\
0 & \frac{1}{6} \left(3+\sqrt{6}\right) & 
\frac{1}{6} \left(3-\sqrt{6}\right) \\
0 & \frac{1}{6} \left(3-\sqrt{6}\right) & 
\frac{1}{6} \left(3+\sqrt{6}\right) \\
\end{array}
\right).
\end{equation}
\textit{Positive sign, degenerate case:} 
\begin{equation}
\hat\lambda^{(0)} = \kappa_1\, \diag\left( \omega,1,1 \right),
\end{equation}
\begin{equation}
\mathcal{P}_4: \quad
|U|^2 = \left(
\begin{array}{ccc}
0 & 
\frac{1}{6} \left(3+\sqrt{6}\right) & 
\frac{1}{6} \left(3-\sqrt{6}\right) \\
\times & \times & \times \\
\times & \times & \times 
\end{array}
\right).
\end{equation}
\textit{Negative sign, non-degenerate case:} 
\begin{equation}
\hat\kappa = \diag\left( \kappa_1, -F_1 \kappa, F_1^* \kappa \right),
\quad
\hat\lambda^{(0)} = \diag\left( 
\kappa_1, -ie^{-i\pi/4} \kappa, e^{-i\pi/4} \kappa \right) 
\end{equation}
\begin{equation}
\mathcal{C}_{20}: \quad
|U|^2 = \left(
\begin{array}{ccc}
1 & 0 & 0 \\
0 & \frac{1}{4} \left(2+\sqrt{2}\right) & 
\frac{1}{4} \left(2-\sqrt{2}\right) \\
0 & \frac{1}{4} \left(2-\sqrt{2}\right) & 
\frac{1}{4} \left(2+\sqrt{2}\right) \\
\end{array}
\right).
\end{equation}
\textit{Negative sign, degenerate case:} 
\begin{equation}
\hat\lambda^{(0)} = \kappa_1\, \diag\left( i,1,1 \right),
\end{equation}
\begin{equation}
\mathcal{P}_5: \quad
|U|^2 = \left(
\begin{array}{ccc}
0 & 
\frac{1}{4} \left(2-\sqrt{2}\right) & 
\frac{1}{4} \left(2+\sqrt{2}\right) \\
\times & \times & \times \\
\times & \times & \times 
\end{array}
\right).
\end{equation}
\paragraph{Third subcase:}
$\left(\left|\textrm{Tr}\mathcal{T}\right|^{2},\,
\left|\textrm{Tr}(\mathcal{TS})\right|^{2}\right) = (1,2+\beta+\beta^4)$
\\
This corresponds to
\begin{equation}
\cos^{2}\theta = \frac{1}{8}\left(5+\sqrt{5}\right),
\quad
\kappa_{2}\kappa_{3}^{*} = 
\pm\frac{1-2i}{\sqrt{5}} \equiv \pm\left(F_{2}\right)^{2}.
\end{equation}
\textit{Positive sign, non-degenerate case:}
\begin{equation}
\hat\kappa = \diag\left( \kappa_1, F_2 \kappa, F_2^* \kappa \right),
\quad
\hat\lambda^{(0)} = \diag\left( 
\kappa_1, -\beta^3 \kappa, -\beta^2 \kappa \right),
\end{equation}
\begin{equation}
\mathcal{C}_{21}: \quad
|U|^2 = \left(
\begin{array}{ccc}
1 & 0 & 0 \\
0 & 
\frac{1}{20} \left(10-\sqrt{10 \left(5+\sqrt{5}\right)}\right) & 
\frac{1}{20} \left(10+\sqrt{10 \left(5+\sqrt{5}\right)}\right) \\
0 & 
\frac{1}{20} \left(10+\sqrt{10 \left(5+\sqrt{5}\right)}\right) & 
\frac{1}{20} \left(10-\sqrt{10 \left(5+\sqrt{5}\right)}\right) 
\end{array}
\right).
\end{equation}
\textit{Positive sign, degenerate case:}
\begin{equation}
\hat\lambda^{(0)} = \kappa_1\, \diag\left( \beta,1,1 \right),
\end{equation}
\begin{equation}
\mathcal{P}_6: \quad
|U|^2 = \left(
\begin{array}{ccc}
0 & 
\frac{1}{20} \left(10-\sqrt{10 \left(5+\sqrt{5}\right)}\right) & 
\frac{1}{20} \left(10+\sqrt{10 \left(5+\sqrt{5}\right)}\right) \\
\times & \times & \times \\
\times & \times & \times 
\end{array}
\right).
\end{equation}
\textit{Negative sign, non-degenerate case:}
\begin{equation}
\hat\kappa = \diag\left( \kappa_1, -F_2 \kappa, F_2^* \kappa \right),
\quad
\hat\lambda^{(0)} = \diag\left( 
\kappa_1, -i \omega^2 \kappa, -i \omega \kappa \right),
\end{equation}
\begin{equation}
\mathcal{C}_{22}: \quad
|U|^2 = \left(
\begin{array}{ccc}
1 & 0 & 0 \\
0 & 
\frac{1}{12} \left( 6 + \sqrt{3} + \sqrt{15} \right) &
\frac{1}{12} \left( 6 - \sqrt{3} - \sqrt{15} \right) \\
0 & 
\frac{1}{12} \left( 6 - \sqrt{3} - \sqrt{15} \right) & 
\frac{1}{12} \left( 6 + \sqrt{3} + \sqrt{15} \right)
\end{array}
\right).
\end{equation}
\textit{Negative sign, degenerate case:}
\begin{equation}
\hat\lambda^{(0)} = \kappa_1\, \diag\left( \omega,1,1 \right), 
\end{equation}
\begin{equation}
\mathcal{P}_7: \quad
|U|^2 = \left(
\begin{array}{ccc}
0 & 
\frac{1}{12} \left(6 + \sqrt{3} + \sqrt{15} \right) &
\frac{1}{12} \left(6 - \sqrt{3} - \sqrt{15} \right) \\
\times & \times & \times \\
\times & \times & \times 
\end{array}
\right).
\end{equation}
\paragraph{Fourth subcase:}
$\left(\left|\textrm{Tr}\mathcal{T}\right|^{2},\,
\left|\textrm{Tr}(\mathcal{TS})\right|^{2}\right) = (1,2+\beta^2+\beta^3)$
\\
This corresponds to
\begin{equation}
\cos^{2}\theta = \frac{1}{8}\left(5-\sqrt{5}\right),
\quad
\kappa_{2}\kappa_{3}^{*} = 
\pm\frac{1+2i}{\sqrt{5}} \equiv \pm\left(F_{3}\right)^{2}.
\end{equation}
\textit{Positive sign, non-degenerate case:}
\begin{equation}
\hat\kappa = \diag\left( \kappa_1, F_3 \kappa, F_3^* \kappa \right),
\quad
\hat\lambda^{(0)} = \diag\left( 
\kappa_1, -\omega^2 \kappa, -\omega \kappa \right),
\end{equation}
\begin{equation}
\mathcal{C}_{23}: \quad
|U|^2 = \left(
\begin{array}{ccc}
1 & 0 & 0 \\
0 & 
\frac{1}{12} \left( 6 - \sqrt{3} + \sqrt{15} \right) &
\frac{1}{12} \left( 6 + \sqrt{3} - \sqrt{15} \right) \\
0 & 
\frac{1}{12} \left( 6 + \sqrt{3} - \sqrt{15} \right) & 
\frac{1}{12} \left( 6 - \sqrt{3} + \sqrt{15} \right)
\end{array}
\right).
\end{equation}
\textit{Positive sign, degenerate case:}
\begin{equation}
\hat\lambda^{(0)} = \kappa_1\, \diag\left( \omega,1,1 \right), 
\end{equation}
\begin{equation}
\mathcal{P}_8: \quad
|U|^2 = \left(
\begin{array}{ccc}
0 & 
\frac{1}{12} \left(6 - \sqrt{3} + \sqrt{15} \right) &
\frac{1}{12} \left(6 + \sqrt{3} - \sqrt{15} \right) \\
\times & \times & \times \\
\times & \times & \times 
\end{array}
\right).
\end{equation}
\textit{Negative sign, non-degenerate case:}
\begin{equation}
\hat\kappa = \diag\left( \kappa_1, -F_3 \kappa, F_3^* \kappa \right),
\quad
\hat\lambda^{(0)} = \diag\left( 
\kappa_1,-i \beta^4 \kappa, -i \beta \kappa \right),
\end{equation}
\begin{equation}
\mathcal{C}_{24}: \quad
|U|^2 = \left(
\begin{array}{ccc}
1 & 0 & 0 \\
0 & 
\frac{1}{20} \left(10+\sqrt{50-10 \sqrt{5}}\right) &
\frac{1}{20} \left(10-\sqrt{50-10 \sqrt{5}}\right) \\
0 &
\frac{1}{20} \left(10-\sqrt{50-10 \sqrt{5}}\right) &
\frac{1}{20} \left(10+\sqrt{50-10 \sqrt{5}}\right)
\end{array}
\right).
\end{equation}
\textit{Negative sign, degenerate case:}
\begin{equation}
\hat\lambda^{(0)} = 
\kappa_1\, \diag\left( \beta^2,1,1 \right), 
\end{equation}
\begin{equation}
\mathcal{P}_9: \quad
|U|^2 = \left(
\begin{array}{ccc}
0 &
\frac{1}{20} \left(10-\sqrt{50-10 \sqrt{5}}\right) &
\frac{1}{20} \left(10+\sqrt{50-10 \sqrt{5}}\right) \\
\times & \times & \times \\
\times & \times & \times 
\end{array}
\right).
\end{equation}
\paragraph{Fifth subcase:}
$\left(\left|\textrm{Tr}\mathcal{T}\right|^{2},\,
\left|\textrm{Tr}(\mathcal{TS})\right|^{2}\right) = 
(2+\beta+\beta^4,2+\beta^2+\beta^3)$
\\
This corresponds to
\begin{equation}
\cos^{2}\theta = \frac{3}{4},
\quad
\kappa_{2}\kappa_{3}^{*} = 
\pm \frac{1}{3} \sqrt{1+4i\sqrt{5}} \equiv \pm\left(F_{4}\right)^{2}.
\end{equation}
\textit{Positive sign, non-degenerate case:}
\begin{equation}
\hat\kappa = \diag\left( \kappa_1, F_4 \kappa, F_4^* \kappa \right),
\quad
\hat\lambda^{(0)} = \diag\left( 
\kappa_1, -\beta ^3 \kappa, -\beta ^2 \kappa \right),
\end{equation}
\begin{equation}
\mathcal{C}_{25}: \quad
|U|^2 = \left(
\begin{array}{ccc}
1 & 0 & 0 \\
0 & 
\frac{1}{20} \left(10+\sqrt{50-10 \sqrt{5}}\right) &
\frac{1}{20} \left(10-\sqrt{50-10 \sqrt{5}}\right) \\
0 &
\frac{1}{20} \left(10-\sqrt{50-10 \sqrt{5}}\right) &
\frac{1}{20} \left(10+\sqrt{50-10 \sqrt{5}}\right)
\end{array}
\right).
\end{equation}
\textit{Positive sign, degenerate case:}
\begin{equation}
\hat\lambda^{(0)} = 
\kappa_1\, \diag\left( \beta,1,1 \right), 
\end{equation}
\begin{equation}
\mathcal{P}_{10}: \quad
|U|^2 = \left(
\begin{array}{ccc}
0 &
\frac{1}{20} \left(10+\sqrt{50-10 \sqrt{5}}\right) &
\frac{1}{20} \left(10-\sqrt{50-10 \sqrt{5}}\right) \\
\times & \times & \times \\
\times & \times & \times 
\end{array}
\right).
\end{equation}
\textit{Negative sign, non-degenerate case:}
\begin{equation}
\hat\kappa = \diag\left( \kappa_1, -F_4 \kappa, F_4^* \kappa \right),
\quad
\hat\lambda^{(0)} = \diag\left( 
\kappa_1, -i \beta ^4 \kappa, -i \beta \kappa \right),
\end{equation}
\begin{equation}
\mathcal{C}_{26}: \quad
|U|^2 = \left(
\begin{array}{ccc}
1 & 0 & 0 \\
0 & 
\frac{1}{20} \left(10+\sqrt{50+10 \sqrt{5}}\right) &
\frac{1}{20} \left(10-\sqrt{50+10 \sqrt{5}}\right) \\
0 &
\frac{1}{20} \left(10-\sqrt{50+10 \sqrt{5}}\right) &
\frac{1}{20} \left(10+\sqrt{50+10 \sqrt{5}}\right)
\end{array}
\right).
\end{equation}
\textit{Negative sign, degenerate case:}
\begin{equation}
\hat\lambda^{(0)} = 
\kappa_1\, \diag\left( \beta^2,1,1 \right), 
\end{equation}
\begin{equation}
\mathcal{P}_{11}: \quad
|U|^2 = \left(
\begin{array}{ccc}
0 &
\frac{1}{20} \left(10-\sqrt{50+10 \sqrt{5}}\right) &
\frac{1}{20} \left(10+\sqrt{50+10 \sqrt{5}}\right) \\
\times & \times & \times \\
\times & \times & \times 
\end{array}
\right).
\end{equation}
\paragraph{Sixth subcase:}
$\left(\left|\textrm{Tr}\mathcal{T}\right|^{2},\,
\left|\textrm{Tr}(\mathcal{TS})\right|^{2}\right) = 
(0,2+\lambda+\lambda^*)$
\\
With $\lambda = e^{i\vartheta}$ unrestricted, 
except for being a root of unity, this corresponds to 
\begin{equation}
\cos^{2}\theta = \cos^2\frac{\vartheta}{2},
\quad
\kappa_{2}\kappa_{3}^{*} = \pm 1.
\end{equation}
\textit{Positive sign, non-degenerate case:}
\begin{equation}
\hat\kappa = \diag\left( \kappa_1, \kappa_3, \kappa_3 \right),
\quad
\hat\lambda^{(0)} = \diag\left( 
\kappa_1, \kappa_3 e^{i\theta}, \kappa_3 e^{-i\theta} \right), 
\end{equation}
\begin{equation}
\mathcal{C}_{27}: \quad
|U|^2 = \left(
\begin{array}{ccc}
1 & 0 & 0 \\
0 & \frac{1}{2} & \frac{1}{2} \\
0 & \frac{1}{2} & \frac{1}{2} 
\end{array}
\right).
\end{equation}
\textit{Positive sign, degenerate cases:}
\begin{equation}
\hat\lambda^{(0)} = 
\kappa_1\, \diag\left( e^{2i\theta},1,1 \right), 
\end{equation}
\begin{equation}
\mathcal{P}_{12}: \quad
|U|^2 = \left(
\begin{array}{ccc}
0 & \frac{1}{2} & \frac{1}{2} \\
\times & \times & \times \\
\times & \times & \times 
\end{array}
\right),
\end{equation}
\begin{equation}
\hat\lambda^{(0)} = 
\diag\left( \kappa_1, \kappa_3, \kappa_3 \right), 
\end{equation}
\begin{equation}
\mathcal{P}_{13}: \quad
|U|^2 = \left(
\begin{array}{ccc}
1 & 0 & 0 \\
\times & \times & \times \\
\times & \times & \times 
\end{array}
\right).
\end{equation}
\textit{Negative sign, non-degenerate case:}
\begin{equation}
\hat\kappa = \diag\left( \kappa_1, -\kappa_3, \kappa_3 \right),
\quad
\hat\lambda^{(0)} = \diag\left( 
\kappa_1, \kappa_3, -\kappa_3 \right), 
\end{equation}
\begin{equation}
\mathcal{C}_{28}: \quad
|U|^2 = \left(
\begin{array}{ccc}
1 & 0 & 0 \\
0 & \sin^{2}\frac{\theta}{2}  & \cos^{2}\frac{\theta}{2} \\
0 & \cos^{2}\frac{\theta}{2}  & \sin^{2}\frac{\theta}{2} 
\end{array}
\right).
\end{equation}
\textit{Negative sign, degenerate case:}
\begin{equation}
\hat\lambda^{(0)} = \kappa_1\,\diag \left(-1, 1, 1 \right), 
\end{equation}
\begin{equation}
\mathcal{P}_{14}: \quad
|U|^2 = \left(
\begin{array}{ccc}
0 & \cos^{2}\frac{\theta}{2} & \sin^{2}\frac{\theta}{2} \\
\times & \times & \times \\
\times & \times & \times 
\end{array}
\right).
\end{equation}

At this point a note is in order. The only restriction on 
the angle $\theta$ in $\mathcal{C}_{28}$ is that it must
be a rational multiple of $\pi$, \textit{i.e.}\ $\exp i\theta$
must be a root of unity. It is therefore natural to ask whether or
not the $|U|^2$ in $\mathcal{C}_{18}$--$\mathcal{C}_{27}$
and $\mathcal{C}_{29}$ are particular cases
of $\mathcal{C}_{28}$. It turns out that the 
mixing matrices associated with 
$\mathcal{C}_{20}$, $\mathcal{C}_{27}$ and $\mathcal{C}_{29}$ 
are indeed particular cases of the one of
$\mathcal{C}_{28}$, 
with $\theta = 3\pi/4$, $\pi/2$ and $\theta=\pi$, respectively. 
For the remaining cases, theorem~\ref{roots1}
can be used to show that the associated angles $\theta$ are not
rational.
\paragraph{Seventh subcase:}
$\left(\left|\textrm{Tr}\mathcal{T}\right|^{2},\,
\left|\textrm{Tr}(\mathcal{TS})\right|^{2}\right) = 
(2+\lambda+\lambda^*,2+\lambda+\lambda^*)$
\\
This corresponds to 
\begin{equation}
\cos^{2}\theta = 1, \quad 
\kappa_{2}\kappa_{3}^{*} = \pm \lambda.
\end{equation}
Since $\lambda$ is a generic root of unity, the sign is irrelevant here and
$\kappa_1$, $\kappa_2$, $\kappa_3$ are different roots of unity leading to
\begin{equation}
\mathcal{C}_{29}: \quad |U|^2 = 
\left( \begin{array}{ccc}
1 & 0 & 0 \\ 0 & 1 & 0 \\ 0 & 0 & 1 
\end{array} \right).
\end{equation}
The degenerate case has already been treated in $\mathcal{P}_{13}$.


\begin{thebibliography}{99}

\bibitem{review-groups}
P.~Ramond,
\textit{Group theory: a physicist's survey},
(Cambridge University Press, Cambridge, UK, 2010); 
\\
G.~Altarelli and F.~Feruglio,
\textit{Discrete flavor symmetries and models of neutrino mixing},
Rev.\ Mod.\ Phys.\  {\bf 82} (2010) 2701
[arXiv:1002.0211 [hep-ph]];
\\
H.~Ishimori, T.~Kobayashi, H.~Ohki, Y.~Shimizu, H.~Okada and M.~Tanimoto,
\textit{Non-Abelian Discrete Symmetries in Particle Physics},
Prog.\ Theor.\ Phys.\ Suppl.\  {\bf 183} (2010) 1
[arXiv:1003.3552 [hep-th]];
\\
S.~F.~King and C.~Luhn,
\textit{Neutrino mass and mixing with discrete symmetry},
Rept.\ Prog.\ Phys.\  {\bf 76} (2013) 056201
[arXiv:1301.1340 [hep-ph]].

\bibitem{GL-review}
W.~Grimus and P.~O.~Ludl,
\textit{Finite flavour groups of fermions},
J.\ Phys.\ A {\bf 45} (2012) 233001
[arXiv:1110.6376 [hep-ph]].

\bibitem{residual}
C.~S.~Lam,
\textit{Determining horizontal symmetry from neutrino mixing},
Phys.\ Rev.\ Lett.\ {\bf 101} (2008) 121602
[arXiv:0804.2622 [hep-ph]];
\\
C.~S.~Lam,
\textit{The unique horizontal symmetry of leptons},
Phys.\ Rev.\ D {\bf 78} (2008) 073015
[arXiv:0809.1185 [hep-ph]];
\\
W.~Grimus, L.~Lavoura and P.~O.~Ludl,
\textit{Is $S_4$ the horizontal symmetry of tri-bimaximal lepton mixing?},
J.\ Phys.\ G {\bf 36} (2009) 115007
[arXiv:0906.2689 [hep-ph]];
\\
C.~S.~Lam,
\textit{A bottom--up analysis of horizontal symmetry},
arXiv:0907.2206 [hep-ph];
\\
S.-F.~Ge, D.~A.~Dicus, and W.~W.~Repko,
\textit{$\mathbbm{Z}_2$ symmetry prediction for the leptonic Dirac $CP$ phase},
Phys.\ Lett.\ B {\bf 702} (2011) 220
[arXiv:1104.0602 [hep-ph]];
\\
H.-J.~He and F.-R.~Yin,
\textit{Common origin of $\mu$--$\tau$ and $CP$ breaking in neutrino seesaw,
baryon asymmetry,
and hidden flavor symmetry}, 
Phys.\ Rev.\ D {\bf 84} (2011) 033009
[arXiv:1104.2654 [hep-ph]];
\\
S.-F.~Ge, D.~A.~Dicus, and W.~W.~Repko,
\textit{Residual symmetries for neutrino mixing
with a large $\theta_{13}$ and nearly maximal $\delta_D$}, 
Phys.\ Rev.\ Lett.\  {\bf 108} (2012) 041801
[arXiv:1108.0964 [hep-ph]];
\\
H.-J.~He and X.-J.~Xu,
\textit{Octahedral symmetry with geometrical breaking:
New prediction for neutrino mixing angle $\theta_{13}$ and $CP$ violation},
Phys.\ Rev.\ D {\bf 86} (2012) 111301 (R)
[arXiv:1203.2908 [hep-ph]];
\\
D.~Hernandez and A.~Yu.~Smirnov, 
\textit{Lepton mixing and discrete symmetries}, 
Phys.\ Rev.\ D {\bf 86} (2012) 053014 
[arXiv:1204.0445 [hep-ph]];
\\
C.~S.~Lam,
\textit{Finite symmetry of leptonic mass matrices},
Phys.\ Rev.\ D {\bf 87} (2013) 013001 
[arXiv:1208.5527 [hep-ph]];
\\
D. Hernandez and A.~Yu.~Smirnov, 
\textit{Discrete symmetries and model-independent patterns of lepton mixing},
Phys.\ Rev.\ D {\bf 87} (2013) 053005 
[arXiv:1212.2149 [hep-ph]];
\\
B.~Hu,
\textit{Neutrino mixing and discrete symmetries},
Phys.\ Rev.\ D {\bf 87} (2013) 033002 
[arXiv:1212.2819 [hep-ph]];
\\
D.~Hernandez and A.~Yu.~Smirnov,
\textit{Relating neutrino masses and mixings by discrete symmetries},
Phys.\ Rev.\ D {\bf 88} (2013) 093007
[arXiv:1304.7738 [hep-ph]].

\bibitem{toorop1}
R.~de Adelhart Toorop, F.~Feruglio and C.~Hagedorn,
\textit{Discrete flavour symmetries in light of T2K},
Phys.\ Lett.\ B {\bf 703} (2011) 447
[arXiv:1107.3486 [hep-ph]].

\bibitem{toorop2}
R.~de Adelhart Toorop, F.~Feruglio and C.~Hagedorn,
\textit{Finite modular groups and lepton mixing},
Nucl.\ Phys.\ B {\bf 858} (2012) 437
[arXiv:1112.1340 [hep-ph]].

\bibitem{holthausen}
M.~Holthausen, K.~S.~Lim and M.~Lindner,
\textit{Lepton mixing patterns from a scan of finite discrete groups},
Phys.\ Lett.\ B {\bf 721} (2013) 61
[arXiv:1212.2411 [hep-ph]].

\bibitem{hagedorn}
C.~Hagedorn, A.~Meroni and L.~Vitale,
\textit{Mixing patterns from the groups $\Sigma(n\phi)$},
J.\ Phys.\ A {\bf 47} (2014) 055201
[arXiv:1307.5308 [hep-ph]].

\bibitem{king1}
S.~F.~King, T.~Neder and A.~J.~Stuart,
\textit{Lepton mixing predictions from $\Delta(6n^2)$ family symmetry},
Phys.\ Lett.\ B {\bf 726} (2013) 312
[arXiv:1305.3200 [hep-ph]].

\bibitem{lavoura}
L.~Lavoura and P.~O.~Ludl,
\textit{Residual $\zz_2 \times \zz_2$ symmetries and lepton mixing},
Phys. Lett. B {\bf 731} (2014) 331
[arXiv:1401.5036 [hep-ph]].

\bibitem{holthausen1}
M.~Holthausen and K.~S.~Lim,
\textit{Quark and leptonic mixing patterns from the breakdown of a
  common discrete flavor symmetry}, 
Phys.\ Rev.\ D {\bf 88} (2013) 033018
[arXiv:1306.4356 [hep-ph]];
\\
T.~Araki, H.~Ishida, H.~Ishimori, T.~Kobayashi and A.~Ogasahara,
\textit{CKM matrix and flavor symmetries},
Phys.\ Rev.\ D {\bf 88} (2013) 096002
[arXiv:1309.4217 [hep-ph]].

\bibitem{king} % Klein group in quark sector
H.~Ishimori and S.~F.~King,
\textit{A model of quarks with $\Delta(6N^2)$ family symmetry},
arXiv:1403.4395 [hep-ph].

\bibitem{lam2014}
C.~S.~Lam,
\textit{A built-in horizontal symmetry of $SO(10)$},
Phys.\ Rev.\ D {\bf 89} (2014) 095017
[arXiv:1403.7835 [hep-ph]].

\bibitem{grimus}
W.~Grimus,
\textit{Discrete symmetries, roots of unity, and lepton mixing},
J.\ Phys.\ G {\bf 40} (2013) 075008
[arXiv:1301.0495 [hep-ph]].

\bibitem{conway}
J.~H. Conway and A.~J. Jones,
\textit{Trigonometric diophantine equations (On vanishing sums of
  roots of unity)},
Acta Arithmetica {\bf 30} (1976) 229.

\bibitem{washington}
L.~C. Washington,
\textit{Introduction to cyclotomic fields},
(Springer, New York, 1982).

\bibitem{speyer}
D. Speyer, 
\textit{Sums over roots of unity},
Mathematics Stack Exchange, 
URL 
http://math.stackexchange.com/questions/39856/sums-of-roots-of-unity/39864
(version 2011-05-18).

\bibitem{branco}
G.~C. Branco and L. Lavoura,
\textit{Rephasing-invariant parametrization of the quark mixing
  matrix},
Phys.\ Lett. B {\bf 208} (1988) 123.

\bibitem{jarlskog}
C.~Jarlskog,
\textit{Commutator of the quark mass matrices in the Standard
  Electroweak Model and a measure of maximal CP violation},
Phys.\ Rev.\ Lett.\  {\bf 55} (1985) 1039.

\bibitem{GAP}
\textit{Groups, Algorithms, Programming -- A System for Computational
  Discrete Algebra} (GAP),
http://www.gap-system.org.

\bibitem{SGL}
H.~U.~Besche, B.~Eick and E.~A.~O'Brien, \textit{SmallGroups -- A GAP
  package}, 2002, 
\\
http://www.gap-system.org/Packages/sgl.html.

\bibitem{GL2013}
W.~Grimus and L.~Lavoura,
\textit{Double seesaw mechanism and lepton mixing},
JHEP {\bf 1403} (2014) 004
[arXiv:1309.3186 [hep-ph]].

\bibitem{tanimoto}
W.~Grimus, A.~S.~Joshipura, L.~Lavoura and M.~Tanimoto,
\textit{Symmetry realization of texture zeros},
Eur.\ Phys.\ J.\ C {\bf 36} (2004) 227
[hep-ph/0405016].

\bibitem{joshipura}
A.~S.~Joshipura and K.~M.~Patel,
\textit{Horizontal symmetries of leptons with a massless neutrino},
Phys.\ Lett.\ B {\bf 727} (2013) 480
[arXiv:1306.1890 [hep-ph]].

\bibitem{fits}
D.~V.~Forero, M.~T\'ortola, and J.~W.~F.~Valle,
\textit{Global status of neutrino oscillation parameters
after recent reactor measurements},
Phys.\ Rev.\ D {\bf 86} (2012) 073012
[arXiv:1205.4018 [hep-ph]];
\\
G.~L.~Fogli, E.~Lisi, A.~Marrone, D.~Montanino, A.~Palazzo, and A.~M.~Rotunno, 
\textit{Global analysis of neutrino masses, mixings and phases:
Entering the era of leptonic $CP$ violation searches},
Phys.\ Rev.\ D {\bf 86} (2012) 013012
[arXiv:1205.5254 [hep-ph]];
\\
M.~C.~Gonzalez-Garcia, M.~Maltoni, J.~Salvado, and T.~Schwetz,
\textit{Global fit to three neutrino mixing:
Critical look at present precision}, 
J.\ High Energy Phys.\ {\bf 1212} (2012) 123
[arXiv:1209.3023 [hep-ph]];
\\
D.~V.~Forero, M.~T\'ortola, and J.~W.~F.~Valle,
\textit{Neutrino oscillations refitted},
arXiv:1405.7540 [hep-ph].

\bibitem{GL-CD}
W.~Grimus and P.~O.~Ludl,
\textit{On the characterization of the $SU(3)$-subgroups of type C and D},
J.\ Phys.\ A {\bf 47} (2014) 075202 
[arXiv:1310.3746 [math-ph]].

\end{thebibliography}
\end{document}